\input harvmac.tex
 \input epsf.tex
 \input amssym
\input ulem.sty

\def\figin{\epsfcheck\figin}\def\figins{\epsfcheck\figins}
\def\epsfcheck{\ifx\epsfbox\UnDeFiNeD
\message{(NO epsf.tex, FIGURES WILL BE IGNORED)}
\gdef\figin##1{\vskip2in}\gdef\figins##1{\hskip.5in}
\else\message{(FIGURES WILL BE INCLUDED)}%
\gdef\figin##1{##1}\gdef\figins##1{##1}\fi}
\def\DefWarn#1{}
\def\figinsert{\goodbreak\midinsert}
\def\ifig#1#2#3{\DefWarn#1\xdef#1{fig.~\the\figno}
\writedef{#1\leftbracket fig.\noexpand~\the\figno} %
\figinsert\figin{\centerline{#3}}\medskip\centerline{\vbox{\baselineskip12pt
\advance\hsize by -1truein\noindent\footnotefont{\bf
Fig.~\the\figno:} #2}}
\bigskip\endinsert\global\advance\figno by1}



\def\unit{\relax{\rm 1\kern-.26em I}}
\def\nada{\relax{\rm 0\kern-.30em l}}
\def\tilde{\widetilde}


\def \la {\langle}
\def \ra {\rangle}
\def \pa {\partial}

\def \eps {\epsilon}

\def\det{{\rm det}}

\noblackbox
\def\IL{\relax{\rm I\kern-.18em L}}
\def\IH{\relax{\rm I\kern-.18em H}}
\def\IR{\relax{\rm I\kern-.18em R}}
\def\IC{\relax\hbox{$\inbar\kern-.3em{\rm C}$}}
\def\IZ{\relax\ifmmode\mathchoice
{\hbox{\cmss Z\kern-.4em Z}}{\hbox{\cmss Z\kern-.4em Z}} {\lower.9pt\hbox{\cmsss Z\kern-.4em Z}}
{\lower1.2pt\hbox{\cmsss Z\kern-.4em Z}}\else{\cmss Z\kern-.4em Z}\fi}


\def\cO{{\cal O}}

\def\det{{\rm det}\,}

\font\manual=manfnt \def\dbend{\lower3.5pt\hbox{\manual\char127}}

\def\cO{{\cal O}}
\def\<{\langle}
\def\>{\rangle}
\def\oo{\infty}



\lref\LuscherEZ{
  M.~Luscher and G.~Mack,
Commun.\ Math.\ Phys.\  {\bf 41}, 203 (1975)..
}

\lref\FitzpatrickZM{
  A.~L.~Fitzpatrick, E.~Katz, D.~Poland and D.~Simmons-Duffin,
JHEP {\bf 1107}, 023 (2011).
[arXiv:1007.2412 [hep-th]].
}

\lref\CornalbaXK{
  L.~Cornalba, M.~S.~Costa, J.~Penedones and R.~Schiappa,
  ``Eikonal Approximation in AdS/CFT: From Shock Waves to Four-Point Functions,''
JHEP {\bf 0708}, 019 (2007).
[hep-th/0611122].
}

\lref\RobertsISA{
  D.~A.~Roberts, D.~Stanford and L.~Susskind,
JHEP {\bf 1503}, 051 (2015).
[arXiv:1409.8180 [hep-th]].
}

\lref\ELOP{
R.~J.~Eden,  P.~V.~Landshoff, D.~I.~Olive and J.~C.~Polkinghorne,
  ``The analytic S matrix,''
Nucl.\ Phys.\ B {\bf 12}, 281 (1969).
}

\lref\GaryAE{
  M.~Gary, S.~B.~Giddings and J.~Penedones,
  ``Local bulk S-matrix elements and CFT singularities,''
Phys.\ Rev.\ D {\bf 80}, 085005 (2009).
[arXiv:0903.4437 [hep-th]].
}

\lref\EdenTU{
  B.~Eden, P.~Heslop, G.~P.~Korchemsky and E.~Sokatchev,
  ``Constructing the correlation function of four stress-tensor multiplets and the four-particle amplitude in N=4 SYM,''
Nucl.\ Phys.\ B {\bf 862}, 450 (2012).
[arXiv:1201.5329 [hep-th]].
}

\lref\WittenNN{
  E.~Witten,
  ``Perturbative gauge theory as a string theory in twistor space,''
Commun.\ Math.\ Phys.\  {\bf 252}, 189 (2004).
[hep-th/0312171].
}

\lref\PenedonesUE{
  J.~Penedones,
  ``Writing CFT correlation functions as AdS scattering amplitudes,''
JHEP {\bf 1103}, 025 (2011).
[arXiv:1011.1485 [hep-th]].
}

\lref\AldayZY{
  L.~F.~Alday, B.~Eden, G.~P.~Korchemsky, J.~Maldacena and E.~Sokatchev,
  ``From correlation functions to Wilson loops,''
JHEP {\bf 1109}, 123 (2011).
[arXiv:1007.3243 [hep-th]].
}

\lref\PappadopuloJK{
  D.~Pappadopulo, S.~Rychkov, J.~Espin and R.~Rattazzi,
  ``OPE Convergence in Conformal Field Theory,''
Phys.\ Rev.\ D {\bf 86}, 105043 (2012).
[arXiv:1208.6449 [hep-th]].
}

\lref\HogervorstSMA{
  M.~Hogervorst and S.~Rychkov,
  ``Radial Coordinates for Conformal Blocks,''
Phys.\ Rev.\ D {\bf 87}, 106004 (2013).
[arXiv:1303.1111 [hep-th]].
}

\lref\PolyakovGS{
  A.~M.~Polyakov,
  ``Nonhamiltonian approach to conformal quantum field theory,''
Zh.\ Eksp.\ Teor.\ Fiz.\  {\bf 66}, 23 (1974).
}

\lref\FerraraYT{
  S.~Ferrara, A.~F.~Grillo and R.~Gatto,
  ``Tensor representations of conformal algebra and conformally covariant operator product expansion,''
Annals Phys.\  {\bf 76}, 161 (1973).
}

\lref\MackJR{
  G.~Mack,
  ``Duality in quantum field theory,''
Nucl.\ Phys.\ B {\bf 118}, 445 (1977).
}

\lref\BelavinVU{
  A.~A.~Belavin, A.~M.~Polyakov and A.~B.~Zamolodchikov,
  ``Infinite Conformal Symmetry in Two-Dimensional Quantum Field Theory,''
Nucl.\ Phys.\ B {\bf 241}, 333 (1984).
}

\lref\RattazziPE{
  R.~Rattazzi, V.~S.~Rychkov, E.~Tonni and A.~Vichi,
  ``Bounding scalar operator dimensions in 4D CFT,''
JHEP {\bf 0812}, 031 (2008).
[arXiv:0807.0004 [hep-th]].
}

\lref\MooreUZ{
  G.~W.~Moore and N.~Seiberg,
  ``Polynomial Equations for Rational Conformal Field Theories,''
Phys.\ Lett.\ B {\bf 212}, 451 (1988).
}

\lref\HellermanBU{
  S.~Hellerman,
  ``A Universal Inequality for CFT and Quantum Gravity,''
JHEP {\bf 1108}, 130 (2011).
[arXiv:0902.2790 [hep-th]].
}

\lref\ShenkerCWA{
  S.~H.~Shenker and D.~Stanford,
  ``Stringy effects in scrambling,''
JHEP {\bf 1505}, 132 (2015).
[arXiv:1412.6087 [hep-th]].
}

\lref\SonodaI{
  H.~Sonoda,
  ``Sewing conformal field theories I",
  Nucl.\ Phys.\ B {\bf 311}, 401 (1988)
}

\lref\HartmanLFA{
  T.~Hartman, S.~Jain and S.~Kundu,
  ``Causality Constraints in Conformal Field Theory,''
[arXiv:1509.00014 [hep-th]].
}

\lref\SonodaII{
  H.~Sonoda,
  ``Sewing conformal field theories II",
  Nucl.\ Phys.\ B {\bf 311}, 417 (1988)
}

\lref\DolanHV{
  F.~A.~Dolan and H.~Osborn,
  ``Conformal partial waves and the operator product expansion,''
Nucl.\ Phys.\ B {\bf 678}, 491 (2004).
[hep-th/0309180].
}

\lref\RychkovIJ{
  V.~S.~Rychkov and A.~Vichi,
  ``Universal Constraints on Conformal Operator Dimensions,''
Phys.\ Rev.\ D {\bf 80}, 045006 (2009).
[arXiv:0905.2211 [hep-th]].
}

\lref\PolandWG{
  D.~Poland and D.~Simmons-Duffin,
  ``Bounds on 4D Conformal and Superconformal Field Theories,''
JHEP {\bf 1105}, 017 (2011).
[arXiv:1009.2087 [hep-th]].
}

\lref\ElShowkHT{
  S.~El-Showk, M.~F.~Paulos, D.~Poland, S.~Rychkov, D.~Simmons-Duffin and A.~Vichi,
  ``Solving the 3D Ising Model with the Conformal Bootstrap,''
Phys.\ Rev.\ D {\bf 86}, 025022 (2012).
[arXiv:1203.6064 [hep-th]].
}

\lref\BajnokHLA{
  Z.~Bajnok and R.~A.~Janik,
  ``String field theory vertex from integrability,''
JHEP {\bf 1504}, 042 (2015).
[arXiv:1501.04533 [hep-th]].
}

\lref\KosTGA{
  F.~Kos, D.~Poland and D.~Simmons-Duffin,
  ``Bootstrapping the $O(N)$ vector models,''
JHEP {\bf 1406}, 091 (2014).
[arXiv:1307.6856 [hep-th]].
}

 
\lref\HwaZZ{
  R.~C.~Hwa and V.~L.~Teplitz,
  ``Notes On Homology And Its Application To Analytic S Matrix Theory,''
}

\lref\ColemanNorton{
  S.~Coleman and R.~E.~Norton,
  ``Singularities in the physical region,''
Nuovo Cim.\  {\bf 38}, 438 (1965).
}
\lref\GoncharovJF{
  A.~B.~Goncharov, M.~Spradlin, C.~Vergu and A.~Volovich,
  ``Classical Polylogarithms for Amplitudes and Wilson Loops,''
Phys.\ Rev.\ Lett.\  {\bf 105}, 151605 (2010).
[arXiv:1006.5703 [hep-th]].
}
\lref\DixonPW{
  L.~J.~Dixon, J.~M.~Drummond and J.~M.~Henn,
  ``Bootstrapping the three-loop hexagon,''
JHEP {\bf 1111}, 023 (2011).
[arXiv:1108.4461 [hep-th]].
}

\lref\ArkaniHamedJHA{
  N.~Arkani-Hamed and J.~Trnka,
  ``The Amplituhedron,''
JHEP {\bf 1410}, 30 (2014).
[arXiv:1312.2007 [hep-th]].
}

\lref\FidkowskiNF{
  L.~Fidkowski, V.~Hubeny, M.~Kleban and S.~Shenker,
  ``The Black hole singularity in AdS/CFT,''
JHEP {\bf 0402}, 014 (2004).
[hep-th/0306170].
}

\lref\BianchiNK{
  M.~Bianchi, M.~B.~Green, S.~Kovacs and G.~Rossi,
  ``Instantons in supersymmetric Yang-Mills and D instantons in IIB superstring theory,''
JHEP {\bf 9808}, 013 (1998).
[hep-th/9807033].
}

\lref\CaronHuotFEA{
  S.~Caron-Huot,
  ``When does the gluon reggeize?,''
[arXiv:1309.6521 [hep-th]].
}

\lref\MaldacenaKM{
  J.~M.~Maldacena and H.~Ooguri,
  ``Strings in AdS(3) and the SL(2,R) WZW model. Part 3. Correlation functions,''
Phys.\ Rev.\ D {\bf 65}, 106006 (2002).
[hep-th/0111180].
}

\lref\CornalbaXM{
  L.~Cornalba, M.~S.~Costa, J.~Penedones and R.~Schiappa,
  ``Eikonal Approximation in AdS/CFT: Conformal Partial Waves and Finite N Four-Point Functions,''
  Nucl.\ Phys.\ B {\bf 767}, 327 (2007)
  [hep-th/0611123].
}

\lref\HeemskerkPN{
  I.~Heemskerk, J.~Penedones, J.~Polchinski and J.~Sully,
  ``Holography from Conformal Field Theory,''
JHEP {\bf 0910}, 079 (2009).
[arXiv:0907.0151 [hep-th]].
}

\lref\FitzpatrickDM{
  A.~L.~Fitzpatrick and J.~Kaplan,
  ``Unitarity and the Holographic S-Matrix,''
  JHEP {\bf 1210}, 032 (2012)
  [arXiv:1112.4845 [hep-th]].
}

\lref\FitzpatrickZM{
  A.~L.~Fitzpatrick, E.~Katz, D.~Poland and D.~Simmons-Duffin,
  ``Effective Conformal Theory and the Flat-Space Limit of AdS,''
JHEP {\bf 1107}, 023 (2011).
[arXiv:1007.2412 [hep-th]].
}

\lref\DrummondNDA{
  J.~Drummond, C.~Duhr, B.~Eden, P.~Heslop, J.~Pennington and V.~A.~Smirnov,
  ``Leading singularities and off-shell conformal integrals,''
JHEP {\bf 1308}, 133 (2013).
[arXiv:1303.6909 [hep-th]].
}

\lref\ArutyunovPY{
  G.~Arutyunov and S.~Frolov,
  ``Four point functions of lowest weight CPOs in N=4 SYM(4) in supergravity approximation,''
Phys.\ Rev.\ D {\bf 62}, 064016 (2000).
[hep-th/0002170].
}

\lref\BianchiXSA{
  M.~Bianchi, A.~Brandhuber, G.~Travaglini and C.~Wen,
  ``Simplifying instanton corrections to N=4 SYM correlators,''
[arXiv:1312.3916 [hep-th]].
}

\lref\GreenTV{
  M.~B.~Green and M.~Gutperle,
  ``Effects of D instantons,''
Nucl.\ Phys.\ B {\bf 498}, 195 (1997).
[hep-th/9701093].
}

\lref\BanksNR{
  T.~Banks and M.~B.~Green,
  ``Nonperturbative effects in AdS in five-dimensions x S**5 string theory and d = 4 SUSY Yang-Mills,''
JHEP {\bf 9805}, 002 (1998).
[hep-th/9804170].
}

\lref\RubakovVZ{
  V.~A.~Rubakov and M.~E.~Shaposhnikov,
  ``Electroweak baryon number nonconservation in the early universe and in high-energy collisions,''
Usp.\ Fiz.\ Nauk {\bf 166}, 493 (1996), [Phys.\ Usp.\  {\bf 39}, 461 (1996)].
[hep-ph/9603208].
}

\lref\BergUQ{
  B.~Berg and M.~Luscher,
  ``Computation of Quantum Fluctuations Around Multi-Instanton Fields from Exact Green's Functions: The CP**n-1 Case,''
Commun.\ Math.\ Phys.\  {\bf 69}, 57 (1979).
}

\lref\ChangJTA{
  C.~-M.~Chang, Y.~-H.~Lin, S.~-H.~Shao, Y.~Wang and X.~Yin,
  ``Little String Amplitudes (and the Unreasonable Effectiveness of 6D SYM),''
[arXiv:1407.7511 [hep-th]].
}

\lref\ColemanKK{
  S.~R.~Coleman and H.~J.~Thun,
  ``On the Prosaic Origin of the Double Poles in the {Sine-Gordon} S Matrix,''
Commun.\ Math.\ Phys.\  {\bf 61}, 31 (1978).
}

\lref\PolyakovXD{
  A.~M.~Polyakov,
  ``Conformal symmetry of critical fluctuations,''
JETP Lett.\  {\bf 12}, 381 (1970), [Pisma Zh.\ Eksp.\ Teor.\ Fiz.\  {\bf 12}, 538 (1970)].
}

\lref\MaldacenaRE{
  J.~M.~Maldacena,
  ``The Large N limit of superconformal field theories and supergravity,''
Adv.\ Theor.\ Math.\ Phys.\  {\bf 2}, 231 (1998).
[hep-th/9711200].
}
\lref\WittenQJ{
  E.~Witten,
  ``Anti-de Sitter space and holography,''
Adv.\ Theor.\ Math.\ Phys.\  {\bf 2}, 253 (1998).
[hep-th/9802150].
}
\lref\GubserBC{
  S.~S.~Gubser, I.~R.~Klebanov and A.~M.~Polyakov,
  ``Gauge theory correlators from noncritical string theory,''
Phys.\ Lett.\ B {\bf 428}, 105 (1998).
[hep-th/9802109].
}

\lref\PolchinskiYD{
  J.~Polchinski, L.~Susskind and N.~Toumbas,
  ``Negative energy, superluminosity and holography,''
Phys.\ Rev.\ D {\bf 60}, 084006 (1999).
[hep-th/9903228].
}

\lref\BassoZOA{
  B.~Basso, S.~Komatsu and P.~Vieira,
  ``Structure Constants and Integrable Bootstrap in Planar N=4 SYM Theory,''
[arXiv:1505.06745 [hep-th]].
}

\lref\GreenMY{
  M.~B.~Green,
  ``A Gas of D instantons,''
Phys.\ Lett.\ B {\bf 354}, 271 (1995).
[hep-th/9504108].
}

\lref\GrossKZA{
  D.~J.~Gross and P.~F.~Mende,
  ``The High-Energy Behavior of String Scattering Amplitudes,''
Phys.\ Lett.\ B {\bf 197}, 129 (1987).
}

\lref\AldayHR{
  L.~F.~Alday and J.~M.~Maldacena,
  ``Gluon scattering amplitudes at strong coupling,''
JHEP {\bf 0706}, 064 (2007).
[arXiv:0705.0303 [hep-th]].
}

\lref\BargheerFAA{
  T.~Bargheer, J.~A.~Minahan and R.~Pereira,
  ``Computing Three-Point Functions for Short Operators,''
JHEP {\bf 1403}, 096 (2014).
[arXiv:1311.7461 [hep-th]].
}

\lref\VenezianoYB{
  G.~Veneziano,
  ``Construction of a crossing - symmetric, Regge behaved amplitude for linearly rising trajectories,''
Nuovo Cim.\ A {\bf 57}, 190 (1968).
}
  
\lref\AlessandriniJY{
  V.~Alessandrini, D.~Amati and B.~Morel,
  ``The asymptotic behaviour of the dual pomeron amplitude,''
Nuovo Cim.\ A {\bf 7}, 797 (1972).
}

\lref\MendeWT{
  P.~F.~Mende and H.~Ooguri,
  ``Borel Summation of String Theory for Planck Scale Scattering,''
Nucl.\ Phys.\ B {\bf 339}, 641 (1990).
}

\lref\GreenTV{
  M.~B.~Green and M.~Gutperle,
  ``Effects of D instantons,''
Nucl.\ Phys.\ B {\bf 498}, 195 (1997).
[hep-th/9701093].
}

\lref\ArkaniHamedKY{
  N.~Arkani-Hamed, S.~Dubovsky, A.~Nicolis, E.~Trincherini and G.~Villadoro,
  ``A Measure of de Sitter entropy and eternal inflation,''
JHEP {\bf 0705}, 055 (2007).
[arXiv:0704.1814 [hep-th]].
}

\lref\GaryAE{
  M.~Gary, S.~B.~Giddings and J.~Penedones,
  ``Local bulk S-matrix elements and CFT singularities,''
Phys.\ Rev.\ D {\bf 80}, 085005 (2009).
[arXiv:0903.4437 [hep-th]].
}

\lref\ErdoganBGA{
  O.~Erdogan,
  ``Coordinate-space singularities of massless gauge theories,''
Phys.\ Rev.\ D {\bf 89}, no.\ 8, 085016 (2014), [Erratum-ibid.\ D {\bf 90}, no.\ 8, 089902 (2014)].
[arXiv:1312.0058 [hep-th]].
}

\lref\HwaZZ{
  R.~C.~Hwa and V.~L.~Teplitz,
  ``Notes On Homology And Its Application To Analytic S Matrix Theory,''
}

\lref\ColemanNorton{
  S.~Coleman and R.~E.~Norton,
 ``Singularities in the physical region,''
Nuovo Cim.\  {\bf 38}, 438 (1965).
}

\lref\ZamolodchikovIE{
  Al.~B.~Zamolodchikov,
  ``Conformal Symmetry In Two-dimensions: An Explicit Recurrence Formula For The Conformal Partial Wave Amplitude,''
Commun.\ Math.\ Phys.\  {\bf 96}, 419 (1984).

}
\lref\ZamolodchikovXX{
Al.~B.~Zamolodchikov, 
``Conformal Symmetry in Two-Dimensional Space: Recursion Representation of Conformal Block,''
Theor.\ Math.\ Phys.\ {\bf 73}, 103-110 (1987).
}

\lref\HartmanOAA{
  T.~Hartman, C.~A.~Keller and B.~Stoica,
  ``Universal Spectrum of 2d Conformal Field Theory in the Large c Limit,''
[arXiv:1405.5137 [hep-th]].
}

\lref\HeemskerkPN{
  I.~Heemskerk, J.~Penedones, J.~Polchinski and J.~Sully,
  ``Holography from Conformal Field Theory,''
JHEP {\bf 0910}, 079 (2009).
[arXiv:0907.0151 [hep-th]].
}

\lref\ChangJTA{
  C.~M.~Chang, Y.~H.~Lin, S.~H.~Shao, Y.~Wang and X.~Yin,
  ``Little String Amplitudes (and the Unreasonable Effectiveness of 6D SYM),''
[arXiv:1407.7511 [hep-th]].
}

\lref\ColemanKK{
  S.~R.~Coleman and H.~J.~Thun,
  ``On the Prosaic Origin of the Double Poles in the {Sine-Gordon} S Matrix,''
Commun.\ Math.\ Phys.\  {\bf 61}, 31 (1978).
}

\lref\MaldacenaHW{
  J.~M.~Maldacena and H.~Ooguri,
  ``Strings in AdS(3) and SL(2,R) WZW model 1.: The Spectrum,''
J.\ Math.\ Phys.\  {\bf 42}, 2929 (2001).
[hep-th/0001053].
}

\lref\MaldacenaKV{
  J.~M.~Maldacena, H.~Ooguri and J.~Son,
  ``Strings in AdS(3) and the SL(2,R) WZW model. Part 2. Euclidean black hole,''
J.\ Math.\ Phys.\  {\bf 42}, 2961 (2001).
[hep-th/0005183].
}

\lref\MaldacenaKM{
  J.~M.~Maldacena and H.~Ooguri,
  ``Strings in AdS(3) and the SL(2,R) WZW model. Part 3. Correlation functions,''
Phys.\ Rev.\ D {\bf 65}, 106006 (2002).
[hep-th/0111180].
}

\lref\PakmanZZ{
  A.~Pakman, L.~Rastelli and S.~S.~Razamat,
  ``Diagrams for Symmetric Product Orbifolds,''
JHEP {\bf 0910}, 034 (2009).
[arXiv:0905.3448 [hep-th]].
}

\lref\CaronHuotFEA{
  S.~Caron-Huot,
  ``When does the gluon reggeize?,''
[arXiv:1309.6521 [hep-th]].
}

\lref\CostaCB{
  M.~S.~Costa, V.~Goncalves and J.~Penedones,
  ``Conformal Regge theory,''
JHEP {\bf 1212}, 091 (2012).
[arXiv:1209.4355 [hep-th]].
}

\lref\BianchiXSA{
  M.~Bianchi, A.~Brandhuber, G.~Travaglini and C.~Wen,
  ``Simplifying instanton corrections to N = 4 SYM correlators,''
JHEP {\bf 1404}, 101 (2014).
[arXiv:1312.3916 [hep-th]].
}

\lref\DoreyPD{
  N.~Dorey, T.~J.~Hollowood, V.~V.~Khoze, M.~P.~Mattis and S.~Vandoren,
  ``Multi-instanton calculus and the AdS/CFT correspondence in ${\cal N}=4$ superconformal field theory,''
Nucl.\ Phys.\ B {\bf 552}, 88 (1999).
[hep-th/9901128].
}

\lref\RubakovVZ{
  V.~A.~Rubakov and M.~E.~Shaposhnikov,
  ``Electroweak baryon number nonconservation in the early universe and in high-energy collisions,''
Usp.\ Fiz.\ Nauk {\bf 166}, 493 (1996), [Phys.\ Usp.\  {\bf 39}, 461 (1996)].
[hep-ph/9603208].
}

\lref\GreenTV{
  M.~B.~Green and M.~Gutperle,
  ``Effects of D instantons,''
Nucl.\ Phys.\ B {\bf 498}, 195 (1997).
[hep-th/9701093].
}

\lref\LandauFI{
  L.~D.~Landau,
  ``On analytic properties of vertex parts in quantum field theory,''
Nucl.\ Phys.\  {\bf 13}, 181 (1959).
}

\lref\PenedonesUE{
  J.~Penedones,
  ``Writing CFT correlation functions as AdS scattering amplitudes,''
JHEP {\bf 1103}, 025 (2011).
[arXiv:1011.1485 [hep-th]].
}

\lref\OkudaYM{
  T.~Okuda and J.~Penedones,
  ``String scattering in flat space and a scaling limit of Yang-Mills correlators,''
Phys.\ Rev.\ D {\bf 83}, 086001 (2011).
[arXiv:1002.2641 [hep-th]].
}

\lref\LuscherEZ{
  M.~Luscher and G.~Mack,
  ``Global Conformal Invariance in Quantum Field Theory,''
Commun.\ Math.\ Phys.\  {\bf 41}, 203 (1975).
}

\lref\MackGY{
  G.~Mack,
  ``D-dimensional Conformal Field Theories with anomalous dimensions as Dual Resonance Models,''
Bulg.\ J.\ Phys.\  {\bf 36}, 214 (2009).
[arXiv:0909.1024 [hep-th]].
}

\lref\FitzpatrickVUA{
  A.~L.~Fitzpatrick, J.~Kaplan and M.~T.~Walters,
  ``Universality of Long-Distance AdS Physics from the CFT Bootstrap,''
JHEP {\bf 1408}, 145 (2014).
[arXiv:1403.6829 [hep-th]].
}

\lref\FitzpatrickYX{
  A.~L.~Fitzpatrick, J.~Kaplan, D.~Poland and D.~Simmons-Duffin,
  ``The Analytic Bootstrap and AdS Superhorizon Locality,''
JHEP {\bf 1312}, 004 (2013).
[arXiv:1212.3616 [hep-th]].
}

\lref\MaldacenaRE{
  J.~M.~Maldacena,
  ``The Large N limit of superconformal field theories and supergravity,''
Adv.\ Theor.\ Math.\ Phys.\  {\bf 2}, 231 (1998).
[hep-th/9711200].
}
\lref\WittenQJ{
  E.~Witten,
  ``Anti-de Sitter space and holography,''
Adv.\ Theor.\ Math.\ Phys.\  {\bf 2}, 253 (1998).
[hep-th/9802150].
}
\lref\GubserBC{
  S.~S.~Gubser, I.~R.~Klebanov and A.~M.~Polyakov,
  ``Gauge theory correlators from noncritical string theory,''
Phys.\ Lett.\ B {\bf 428}, 105 (1998).
[hep-th/9802109].
}

\lref\GiddingsGJ{
  S.~B.~Giddings and R.~A.~Porto,
  ``The Gravitational S-matrix,''
Phys.\ Rev.\ D {\bf 81}, 025002 (2010).
[arXiv:0908.0004 [hep-th]].
}

\lref\SiversIG{
  D.~Sivers and J.~Yellin,
  ``Review of recent work on narrow resonance models,''
Rev.\ Mod.\ Phys.\  {\bf 43}, 125 (1971).
}

\lref\JacksonNLA{
  S.~Jackson, L.~McGough and H.~Verlinde,
  ``Conformal Bootstrap, Universality and Gravitational Scattering,''
[arXiv:1412.5205 [hep-th]].
}

\lref\AlmheiriLWA{
  A.~Almheiri, X.~Dong and D.~Harlow,
  ``Bulk Locality and Quantum Error Correction in AdS/CFT,''
[arXiv:1411.7041 [hep-th]].
}

\lref\ArkaniHamedJHA{
  N.~Arkani-Hamed and J.~Trnka,
  ``The Amplituhedron,''
JHEP {\bf 1410}, 30 (2014).
[arXiv:1312.2007 [hep-th]].
}

\lref\ArkaniHamedKCA{
  N.~Arkani-Hamed and J.~Trnka,
  ``Into the Amplituhedron,''
JHEP {\bf 1412}, 182 (2014).
[arXiv:1312.7878 [hep-th]].
}

\lref\PerlmutterIYA{
  E.~Perlmutter,
  ``Virasoro conformal blocks in closed form,''
[arXiv:1502.07742 [hep-th]].
}

\lref\HogervorstSMA{
  M.~Hogervorst and S.~Rychkov,
Phys.\ Rev.\ D {\bf 87}, 106004 (2013).
[arXiv:1303.1111 [hep-th]].
}

\lref\DolanHV{
  F.~A.~Dolan and H.~Osborn,
  ``Conformal partial waves and the operator product expansion,''
Nucl.\ Phys.\ B {\bf 678}, 491 (2004).
[hep-th/0309180].
}

\lref\PolchinskiTT{
  J.~Polchinski and M.~J.~Strassler,
  ``Hard scattering and gauge / string duality,''
Phys.\ Rev.\ Lett.\  {\bf 88}, 031601 (2002).
[hep-th/0109174].
}

\lref\GiddingsGJ{
  S.~B.~Giddings and R.~A.~Porto,
  ``The Gravitational S-matrix,''
Phys.\ Rev.\ D {\bf 81}, 025002 (2010).
[arXiv:0908.0004 [hep-th]].
}

\lref\EdenMV{
  B.~Eden, C.~Schubert and E.~Sokatchev,
  ``Three loop four point correlator in N=4 SYM,''
Phys.\ Lett.\ B {\bf 482}, 309 (2000).
[hep-th/0003096].
}

\lref\AdamsSV{
  A.~Adams, N.~Arkani-Hamed, S.~Dubovsky, A.~Nicolis and R.~Rattazzi,
  ``Causality, analyticity and an IR obstruction to UV completion,''
JHEP {\bf 0610}, 014 (2006).
[hep-th/0602178].
}
 
\lref\MaldacenaWAA{
  J.~Maldacena, S.~H.~Shenker and D.~Stanford,
  ``A bound on chaos,''
[arXiv:1503.01409 [hep-th]].
}

\lref\CornalbaXK{
  L.~Cornalba, M.~S.~Costa, J.~Penedones and R.~Schiappa,
  ``Eikonal Approximation in AdS/CFT: From Shock Waves to Four-Point Functions,''
JHEP {\bf 0708}, 019 (2007).
[hep-th/0611122].
}

\lref\CornalbaXM{
  L.~Cornalba, M.~S.~Costa, J.~Penedones and R.~Schiappa,
  ``Eikonal Approximation in AdS/CFT: Conformal Partial Waves and Finite N Four-Point Functions,''
Nucl.\ Phys.\ B {\bf 767}, 327 (2007).
[hep-th/0611123].
}

\lref\CornalbaZB{
  L.~Cornalba, M.~S.~Costa and J.~Penedones,
  ``Eikonal approximation in AdS/CFT: Resumming the gravitational loop expansion,''
JHEP {\bf 0709}, 037 (2007).
[arXiv:0707.0120 [hep-th]].
}

\lref\CornalbaQF{
  L.~Cornalba, M.~S.~Costa and J.~Penedones,
  ``Eikonal Methods in AdS/CFT: BFKL Pomeron at Weak Coupling,''
JHEP {\bf 0806}, 048 (2008).
[arXiv:0801.3002 [hep-th]].
}

\lref\CornalbaFS{
  L.~Cornalba,
  ``Eikonal methods in AdS/CFT: Regge theory and multi-reggeon exchange,''
[arXiv:0710.5480 [hep-th]].
}

\lref\CamanhoAPA{
  X.~O.~Camanho, J.~D.~Edelstein, J.~Maldacena and A.~Zhiboedov,
  ``Causality Constraints on Corrections to the Graviton Three-Point Coupling,''
[arXiv:1407.5597 [hep-th]].
}
\lref\PoppitzNZ{
  E.~Poppitz, T.~Sch{\"a}fer and M.~{\"U}nsal,
  ``Universal mechanism of (semi-classical) deconfinement and theta-dependence for all simple groups,''
JHEP {\bf 1303}, 087 (2013).
[arXiv:1212.1238 [hep-th]].
}

\lref\FitzpatrickYX{
  A.~L.~Fitzpatrick, J.~Kaplan, D.~Poland and D.~Simmons-Duffin,
  ``The Analytic Bootstrap and AdS Superhorizon Locality,''
JHEP {\bf 1312}, 004 (2013).
[arXiv:1212.3616 [hep-th]].
}

\lref\DateUN{
  G.~Date,
  Ph.D.\ thesis, SUNY, Stony Brook, 1982 ``Factorization Theorems In Perturbative Quantum Field Theory,''
  (Report No.\ UMI-83-07385).
} 
\lref\ErdoganGHA{
  O.~Erdog�an and G.~Sterman,
  ``Ultraviolet divergences and factorization for coordinate-space amplitudes,''
Phys.\ Rev.\ D {\bf 91}, no.\ 6, 065033 (2015).
[arXiv:1411.4588 [hep-ph]].
} 

\lref\KomargodskiEK{
  Z.~Komargodski and A.~Zhiboedov,
  ``Convexity and Liberation at Large Spin,''
JHEP {\bf 1311}, 140 (2013).
[arXiv:1212.4103 [hep-th]].
}

\lref\HogervorstSMA{
  M.~Hogervorst and S.~Rychkov,
  ``Radial Coordinates for Conformal Blocks,''
Phys.\ Rev.\ D {\bf 87}, 106004 (2013).
[arXiv:1303.1111 [hep-th]].
}

\lref\DrummondNDA{
  J.~Drummond, C.~Duhr, B.~Eden, P.~Heslop, J.~Pennington and V.~A.~Smirnov,
  ``Leading singularities and off-shell conformal integrals,''
JHEP {\bf 1308}, 133 (2013).
[arXiv:1303.6909 [hep-th]].
}

\lref\EdenMV{
  B.~Eden, C.~Schubert and E.~Sokatchev,
  ``Three loop four point correlator in N=4 SYM,''
Phys.\ Lett.\ B {\bf 482}, 309 (2000).
[hep-th/0003096].
}

\lref\AldayOTA{
  L.~F.~Alday and A.~Zhiboedov,
  ``Conformal Bootstrap With Slightly Broken Higher Spin Symmetry,''
[arXiv:1506.04659 [hep-th]].
}

\lref\LangKP{
  K.~Lang and W.~Ruhl,
  ``The Critical O(N) sigma model at dimension 2 < d < 4 and order 1/n**2: Operator product expansions and renormalization,''
Nucl.\ Phys.\ B {\bf 377}, 371 (1992).
}

\lref\HattaKN{
  Y.~Hatta, E.~Iancu, A.~H.~Mueller and D.~N.~Triantafyllopoulos,
  ``Jet evolution from weak to strong coupling,''
JHEP {\bf 1212}, 114 (2012).
[arXiv:1210.1534 [hep-th]].
}

\lref\CutkoskySP{
  R.~E.~Cutkosky,
  ``Singularities and discontinuities of Feynman amplitudes,''
J.\ Math.\ Phys.\  {\bf 1}, 429 (1960).
}

\lref\MackMI{
  G.~Mack,
  ``D-independent representation of Conformal Field Theories in D dimensions via transformation to auxiliary Dual Resonance Models. Scalar amplitudes,''
[arXiv:0907.2407 [hep-th]].
}

\lref\PenedonesUE{
  J.~Penedones,
  ``Writing CFT correlation functions as AdS scattering amplitudes,''
JHEP {\bf 1103}, 025 (2011).
[arXiv:1011.1485 [hep-th]].
}

\lref\FitzpatrickIA{
  A.~L.~Fitzpatrick, J.~Kaplan, J.~Penedones, S.~Raju and B.~C.~van Rees,
  ``A Natural Language for AdS/CFT Correlators,''
JHEP {\bf 1111}, 095 (2011).
[arXiv:1107.1499 [hep-th]].
}

\lref\StreaterVI{
  R.~F.~Streater and A.~S.~Wightman,
  ``PCT, spin and statistics, and all that,''
Princeton, USA: Princeton Univ. Pr. (2000) 207 p.
}




\Title{
\vbox{\baselineskip6pt
}}
{\vbox{\centerline{Looking for a bulk point} 
 \centerline{     }
}}

\centerline{  Juan Maldacena,$^1$ David Simmons-Duffin,$^1$ Alexander Zhiboedov$^2$   }
\bigskip
\centerline{\it $^1$ School of Natural Sciences, Institute for Advanced Study, Princeton, NJ, USA }
\centerline{\it $^2$ Center for the Fundamental Laws of Nature, Harvard University, Cambridge, MA, USA}

\vskip .3in \noindent

 We consider Lorentzian correlators of local operators. In perturbation theory, 
singularities occur when we can draw a position-space Landau diagram with 
null lines. In theories with gravity duals, we can also draw Landau diagrams in the bulk. 
We argue that certain singularities can arise only from bulk diagrams, not from boundary diagrams. 
As has been previously observed, these singularities are a clear diagnostic of bulk locality. 
We analyze some properties of these perturbative singularities and discuss their relation to the OPE and the
dimensions of double-trace operators. 
In the exact nonperturbative theory, we expect no singularity at these locations. We prove this
statement in 1+1 dimensions by CFT methods.

 \Date{ }

 \listtoc\writetoc
\vskip .5in \noindent

\vfil \break

\newsec{Introduction}

In Euclidean signature, correlators of local operators are analytic 
for non-coincident points.  However in Lorentzian signature, singularities can arise when ``something happens.'' 
These Lorentzian singularities correspond, in weakly coupled theories, to Landau diagrams consisting of a set of null particles interacting at local vertices in an energy-momentum conserving fashion. 
  We will derive these rules for a generic perturbative quantum field theory (see also \refs{\LandauFI\ColemanNorton\CutkoskySP\DateUN\ErdoganBGA-\ErdoganGHA}).\foot{For a discussion of analytic properties of correlation functions of local operators in a generic QFT see \StreaterVI , the case of a CFT is considered in \LuscherEZ . }
  
  In theories that have gravity duals, singularities can arise from Landau diagrams in the bulk. 
  In some cases, these occur at positions where there is no Landau diagram on the boundary
  \refs{\PolchinskiYD\GaryAE\HeemskerkPN\PenedonesUE-\OkudaYM}. 
  Such singularities are a probe of bulk locality. We call them ``bulk-point singularities.'' 
  We will display examples in 1+1 and 2+1 dimensions. 
  
 The emergence of the bulk is intimately related to the development of these singularities as the boundary
 theory becomes strongly coupled. 
   In this paper we analyze some properties of bulk-point singularities, but we do not give a satisfactory explanation for
   their emergence. 
  There were several previous studies of these interesting singularities including  \refs{\PolchinskiYD\GaryAE\HeemskerkPN\PenedonesUE-\OkudaYM}.
   Some articles (see e.g.\ \refs{\GaryAE,\HeemskerkPN})
assumed the singularity is present and showed how it could be used to extract the flat space scattering 
amplitude.  
  We are simply  adding a few comments to those previous papers.

 We first  review the origin of bulk-point singularities using the local bulk theory. 
  We  argue that finite $\alpha'$ effects  remove the singularity \OkudaYM. 
  We then comment that D-instanton effects are again  singular at this location. 
 Finally, we expect that at finite $G_N$ this singularity should not be present since, in some sense, there was no bulk point to start with in the boundary theory. 
  
  In 1+1 dimensions,  using general CFT arguments, we show explicitly that the singularity is not present in the exact answer. The only singularities of the four-point function are the light-cone singularities.
  
   This paper is organized as follows. In section~2, we 
   derive the position-space Landau rules for correlators. These are analogous to the 
   well-known momentum-space Landau rules \refs{\LandauFI\ColemanNorton-\CutkoskySP}. 
   In section~3, we consider singularities arising from a local bulk. We argue that these singularities
   do not arise from boundary Landau diagrams in 1+1 and 2+1 dimensions. 
   In section~4, we consider stringy and instanton corrections to the gravity formulas, and then we discuss some aspects of the exact answer. 
   In section~5, we discuss the singularities of the four-point function in $d>2$ and discuss how there can be 
   both a bulk UV and IR contribution to the singularity. 
   In section~6, we review the relation between the singularity and the OPE expansion, clarifying the 
   applicability of the OPE for this analysis and also for the Regge limit. 
   In section~7, for 1+1 dimensional CFTs,  we prove that there are no bulk-point singularities in the exact answer.
In appendix~A, we prove a bound on the coefficients of the low energy expansion of a causal flat-space tree level 
four-point scattering amplitude. Other appendices give more details on the discussion in the main body.

\newsec{Singularities of perturbative correlation functions}

  Let us consider a weak coupling expansion of a local quantum field theory. 
    We study time-ordered correlation functions of local operators. 
    At each order in perturbation theory, these are functions of the spacetime positions
    of the operators. 
   In this section, we describe their  possible  singularities. 
  In other words, we have 
   \eqn\refsci{ 
    \langle O(x_1) \cdots O(x_n) \rangle = \sum_{k} g^{k} F_k( x_1, \cdots,x_n),
    }
  and we want to find the spacetime locations where $F_k$ has singular behavior. 
  Previous discussion of this topic includes \refs{\DateUN\ErdoganBGA-\ErdoganGHA}. 
  
  There is a conceptually 
  similar problem   involving  the singularities of perturbative scattering amplitudes, viewed as functions of the momenta. 
  In that case, the singularities are at locations where one can draw a Landau diagram    \refs{\LandauFI\ColemanNorton-\CutkoskySP}. 
  For correlation functions, the situation is similar, and the singularities are at the momenta where one can draw a position-space Landau 
  diagram with on-shell  massless particles interacting at local  vertices with momentum conserved at the vertices. 
  In this section, we derive this rule. First we consider a simple example. 
  
  \subsec{A four-point function example}
  
  Consider a massless field $\phi$ in four dimensions  with an interaction $\int d^4 x \lambda \phi^4$. 
  The leading order correction to the four-point function is given by the  integral 
  \eqn\inteagor{  
  \langle  \phi(x_1) \cdots  \phi(x_4) \rangle \propto    \lambda I , ~~~~I \equiv \int d^4 w { 1 \over \prod_{i=1}^4 [ ( w - x_i)^2 + i \varepsilon ] }, 
  }
 \eqn\inteag{\eqalign{ 
 I = & - { 2 \pi^2 i z \bar z   \over x_{12}^2 x_{34}^2 } \left[ 2 {\rm Li_2}( z ) - 2 {\rm Li}_2(\bar z)  + \log(z \bar z ) \log \left(  {1-z \over 1-\bar z}  \right) \over z - \bar z \right],
  }}
  \eqn\crossratios{
   ~~~~ u = { x_{12}^2 x_{34}^2 \over x_{13}^2 x_{24}^2 } = z \bar z ,~~~~~~~v={ x_{14}^2 x_{23}^2 \over x_{13}^2 x_{24}^2 } =(1- z) (1-\bar z).
   }
  The last line defines the variables $z$ and $\bar z$. 
  In Euclidean space $\bar z = z^*$, and the only possible singularities arise 
   when  points coincide, at $z=\bar z = 0,~1,~\infty$. 
  In particular, there is no singularity when $z = \bar z$ for generic $z$ since the numerator  in  \inteag\ has a zero at that location. 
  Now we can move to Lorentzian signature, where $z$ and $\bar z$ become independent real variables. We have a singularity 
  when  $z=0$, with any $\bar z$. These are the light-cone singularities, arising when two points  are lightlike separated. We can continue past this singularity 
  by using the appropriate $i\varepsilon$ prescription.  Now some of the points are timelike separated and some are spacelike separated. In this regime, we might
  encounter new singularities. In our example, this occurs when $z = \bar z$. What has happened is that we have analytically continued \inteag , going 
   through branch points such that when we set $z = \bar z$ the numerator no longer cancels the denominator. 
   
          \ifig\exampleLand{Landau diagram for the four-point function. There is a point $w$ null separated from the insertion points of the operators. 
          We can put physical massless particles along these null lines so that momentum is conserved at the vertex.  }
          {\epsfxsize1.6in \epsfbox{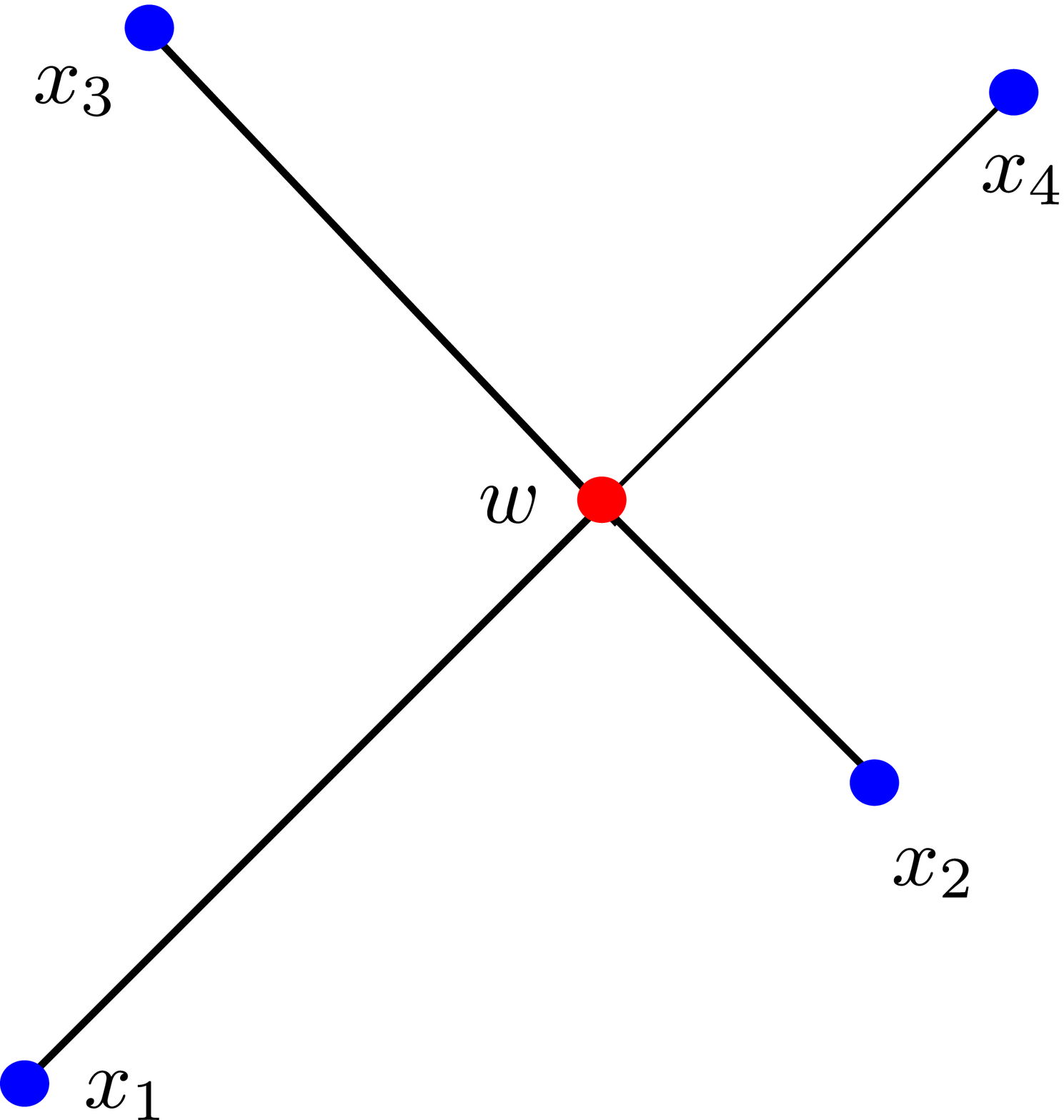} } 

   Let us see this more explicitly. We choose     
   \eqn\fourpc{ \eqalign{ 
& x_1 = (-t,0,1,0) ,~~~~ x_2 = (-t,0,-1,0),~~~~~x_3 = (t,1,0,0) ,~~~~ x_4 = (t,-1,0,0)
\cr
&{1 \over z} =   {1 \over 2} + \sqrt{ t^2 (1-t^2 ) } ,~~~~~~~~~ {1 \over \bar z} = {1 \over 2 } - \sqrt{ t^2 (1-t^2 ) }.
} }
The $i\varepsilon$ prescription corresponds to setting $t = -i \varepsilon + \tilde t$ with real $\tilde t$ in these formulas. 
As we change $\tilde t$ from zero to one, we go from the Euclidean region to the Lorentzian region described above. In doing so, 
$  z$ goes around the branch point at $z=1$, while $\bar z$ goes around $\bar z = \infty$.
 Therefore, when we return to ${1 \over z} = {1 \over \bar z} = 1/2 $, at $t=1$, we pick up a contribution from going to the other branches in \inteag , and now the numerator no longer vanishes when $z=\bar z$. Namely, 
   the term in brackets in \inteag\ becomes 
   $ { ( 2\pi)^2 \over z - \bar z } $ plus terms that are regular at $z= \bar z$. 

It is actually  not necessary to know the explicit answer to find the singularity. One can start with the original integral  \inteagor. In Lorentzian signature,
the $i\varepsilon$ prescription completely defines the integral by stating how we should pick the integration contour. A singularity can only arise if
we cannot deform the contour to avoid zeros in the denominator of the integrand in \inteagor. 
That is, singularities arise when 
   \eqn\bdylandau{
        (x_a - w_0)^2 =0 ,~~~a=1,\cdots 4,
        }
      and we cannot deform the $w$ integral away from $w_0$.  If the  $w$ integral can be deformed by shifting $w^\mu  \to w^\mu  + i v^\mu$ so that 
      \eqn\vcond{ 
       (x_a^\mu - w_0^\mu) v^\mu > 0  ,~~~~~~~~{\rm for ~all~}a,
       }
     then there is no singularity, despite \bdylandau. This is expected to be the generic situation, since
      it is generically possible to solve the four equations \bdylandau\ for four variables. 
      The singularity can be present only when we fail to find a $v^\mu$ obeying \vcond, which happens if and only if the following condition holds: 
    there exist four numbers $\alpha_a$, such that 
      \eqn\neeco{ 
       \sum_{a=1}^4 \alpha_a ( x_a^\mu - w_0^\mu) = 0  ,~~~~~~~~~~ \alpha_a \geq 0 ,~~~~~~~{\rm not ~all~}\alpha_a ~ {\rm zero}.
       }
       It is clear that if \neeco\ holds then \vcond\ cannot hold, since we can simply multiply \neeco\ by $v^\mu$ to find an inconsistency. It is also true that 
       if it is not possible to solve \vcond\ then it is possible to solve \neeco\ (we present the argument in the next subsection).

  The conclusion is that the singularity is present when both \bdylandau\ and \neeco\ hold. The first could hold for generic $x_a$ but the second can only hold 
  for special cases, since, in particular, it requires that $\det ( x_a^\mu - w_0^\mu) =0$, which imposes one more condition beyond \bdylandau. 
  This can be interpreted as follows. We are demanding that there exists a point $w_0$ such that we can send on-shell massless particles from the  position 
  of the operators,  $x_a$,  with momenta $k^\mu_a = \alpha_a (x^\mu_a - w_0^\mu)$ so that momentum is conserved at the point $w_0$.
  The positivity of $\alpha_i$ ensures that the energies are all positive.  We call such a configuration 
  a position-space
   Landau diagram. A Landau diagram is like a Feynman diagram, except that the lines are all null, and we can associate null momenta to all the lines so that they
  obey momentum conservation at the vertices.\foot{A momentum-space Landau diagram  \refs{\LandauFI\ColemanNorton-\CutkoskySP} 
  is a momentum-space Feynman diagram, where we can assign 
  {\it positions}  to the vertices such that the difference in positions between two vertices joined by a line with momentum $p^\mu$ is proportional to $p^\mu$ with
  a positive coefficient. Here the momenta obey $p^2 = -m^2$, with possibly non-zero values of $m^2$.}
  
  After this introductory case, let us consider a general case. 
  
  \subsec{General conditions for singularities of correlators in local quantum field theories }

We consider a time-ordered Lorentzian correlation function in a perturbative field theory. The theory can be massive or massless. It can be in any dimension. 
We will analyze the location of possible singularities at a given order in perturbation theory. 
Some four-dimensional cases were considered previously in \refs{\DateUN\ErdoganBGA-\ErdoganGHA}.  
 
We consider a correlation function of local operators. At a fixed order in perturbation theory, we get an expression of the form  
\eqn\locap{ 
\langle O(x_1) \cdots O(x_n) \rangle =  \sum_k g^{k} 
\int \prod_{a=1}^k d^d w_a   \langle L_{\rm int}(w_1) \cdots L_{\rm int}(w_k)  O(x_1) \cdots O(x_n) \rangle, 
}
where the interaction Lagrangian is a product of local fields. Therefore the integrand is 
  given by a product of propagators. There can be derivatives acting on the propagators. 
We do not expect any singularity arising from infinity. This can be seen by 
deforming the integration contours into the Euclidean direction as soon as the integration variables are larger
than the largest time appearing in the correlator. 
The propagators are functions of distances $G( d_{ij}^2 + i \varepsilon)$, which are regular for positive values of the argument 
but have a singularity when the argument is zero. 
This singularity can be a pole or a branch cut depending on the dimensionality and whether the field is massive or massless.

Let us lump all the integration variables into a big vector $w^M$. 
We consider a region in the integration domain, centered on a $w_0^M$ where any number of these distances is becoming zero 
\eqn\zerodis{ 
 d_{ij}^2(w_0^M) =0 ,~~~~~~~~{\rm for}~~~{i,j \in  S },
 }
 where $S$ is the set of distances that are zero.  
Now let us define 
\eqn\varby{ 
x_{ij}^M =  \left. { \partial d^2_{ij} \over \partial w^M} \right|(w_0). 
}
Next consider shifting the integration contour, which runs over real $w^M$, to the imaginary region $w^M \to w^M + i v^M $. 
The distances change to  
$d_{ij}^2 + i \varepsilon  \to  d^2_{ij} + i ( x_{ij}^M v^M + \varepsilon )  $. We can move away from the dangerous region if 
\eqn\condic{ 
{\rm there ~exists~a}~~~  v^M  ~~~{\rm such~ that} ~~~~ \sum_M x_{ij}^M v^M > 0 ,~~~~~~~~{\rm for } ~~~~ {i,j \in S}.
} 
If we can not do this, then we will have a pinch singularity, where the integration contour lies between 
two singularities that are approaching each other. In this case, the final correlator will generically be singular. 
(Of course, it is possible to have cancellations between different diagrams).

Now, let us denote the indices $ij$ in $S$ collectively by the letter $J$.  
Farkas' lemma states\foot{We thank N.\ Arkani-Hamed for pointing this 
out to us.} that either  \condic\  or the following is true:
\eqn\codc{ 
{\rm There ~exists}~\alpha_J \geq 0,~{\rm not ~all } ~\alpha_J ~{\rm zero}, ~~~~~~\sum_{J} \alpha_{J} x_{J}^M = \sum_{ij\in S} \alpha_{ij} x_{ij}^M  =0.
}
In other words, when we can solve \codc, we have a singularity of the correlator. As in the previous subsection, the condition \codc\ says that we can assign momenta to the null lines so that momentum is conserved. 
  
 We could make the formulas look more uniform by defining $\alpha_{ij} =0$ for all $ij$ that are not in the set of distances, $S$. 
 Then the conditions for a singularity are that we can solve  
 \eqn\twoex{ 
 \alpha_{ij} d_{ij}^2 =0 ~~~{\rm (no ~sum)}, ~~~~~~ \sum_{ij} \alpha_{ij} x_{ij}^M =0, ~~~~~
\alpha_{ij} \geq 0 ~~~{\rm not~all} ~\alpha_{ij} ~{\rm zero}. 
 }

Note that we have massless particles even in a massive theory because the singularity comes from 
the propagation of a very high energy particle. For such high energy particles, we can neglect the mass. 
  
The conclusion is that the only possible singularities of perturbative Lorentzian correlators arise when we can 
draw a Landau diagram with on-shell massless particles that start from the external points and 
undergo collisions. We can assign a momentum along the direction of motion of each massless 
particle in such a way that momentum is conserved at the interaction points, but not at the external 
points.  

  Landau's original equations are similar, but they are related to singularities of Feynman diagrams
in momentum  space \LandauFI.       In the case of momentum-space Feynman diagrams, the Landau diagrams   
 also involve on-shell particles  interacting at localized positions in an energy-momentum conserving fashion. 
But now they represent propagation  for very long distances in spacetime, and
their energy and momentum are finite. For this reason the masses do not drop out.  
The difference in position between two vertices connected by an on-shell line $l$  with momentum $p_l^\mu$ is 
$\Delta x^\mu = \alpha_l  p_l^\mu$. The condition that this defines a consistent set of positions is that 
$\sum_l \alpha_l p_l^\mu =0 $ for each loop. This is analogous to \codc. 

If we consider a quantum field theory in curved space, the same reasoning tells us that we should consider massless 
geodesics in the curved spacetime with momentum locally conserved at each vertex, and redshifted appropriately when we go from one vertex
to the next. We have 
not checked this explicitly. 

\subsec{From Minkowski space to the cylinder}

Notice that besides the momentum  all other conformal charges are preserved at the 
interaction vertex. 
 In order to see this, let us first introduce the angular 
momentum, dilatation, and special conformal charges for any classical  massless particle as 
\eqn\angmo{ 
 J^{\mu \nu} = x^\mu p^\nu  - x^\nu p^\mu  ,~ ~~~~~~D = x^\mu p_\mu ,~~~~~~~~~ K^\mu = - 2 x^\mu  x^\nu p_\nu + 
x^2 p^\mu.
 }
Note that the fact that $p^2=0$ implies that we can evaluate $x^{\mu}$ along any point on the trajectory. 
At a collision point, the fact that momentum is conserved implies that  the charges in 
 \angmo\ are conserved since the position $x^{\mu}$ can be chosen as the interaction point for all 
of the particles coming to a given vertex. 

Now let us consider a CFT and the same process on the cylinder, $R\times S^{d-1}$. We can view the on-shell particles as living on the cylinder and impose that they preserve the natural charges on  the
cylinder, namely energy and angular momentum. These charges are a combination of the various conformal generators on the plane. Given that they are conserved on the plane, we will find that they are also conserved on the cylinder.

\subsec{Applications to symbology}

In computations of perturbative correlators in scale-invariant theories, it
has proven   useful to introduce a ``symbol'' that captures some of the 
singularities of the integrals \GoncharovJF. The symbol contains arguments that are functions
of the kinematic invariants. The zeros of the functions appearing in the symbols should 
correspond to solutions of the Landau equations. This is due to the fact that such zeros 
represent branch point singularities of the answer as a function of the kinematic invariants. 
These branch points might appear only after analytically continuing the answer through some previous 
branch cuts. So, they are singularities of the analytically continued function. Since, a priori, we do not know 
whether such analytic continuation arises from going to Lorentzian signature or from a more formal operation,\foot{E.g., complexifying the coordinates or considering other orderings of the operators.}
we can consider 
 solutions of the Landau equations without insisting on the positivity (or reality)  of
  the $\alpha_{ij}$ parameters (we still demand that they are not all zero). Similarly, we can also allow on-shell three-point vertices.  
On the other hand, solutions of the Landau equations could also appear as prefactors of the symbols or
as zeros inside the symbol. 
In some computations, a guess has been made for the possible arguments of the symbol at a given 
order in perturbation theory (e.g.\ \DixonPW). Here we are simply giving a guiding principle for what these 
arguments can be. Their zeros should correspond to solutions of the Landau equations. At a finite 
order in perturbation theory there is only a finite number of possible Landau graphs.
These remarks suggest that it would be very interesting to find the explicit  locations in kinematic 
space where Landau graphs are possible, but we will not attempt that here.

\newsec{Bulk versus boundary singularities}

We have seen that perturbative singularities of local QFTs correspond to Landau diagrams. This logic applies when either the boundary or the bulk is local and weakly coupled. In the next subsection, we give an example of a perturbative ``bulk-point singularity'' arising from a bulk Landau diagram.  Subsequently, we show that this bulk-point singularity could never arise from perturbation theory on the boundary, at least in 1+1 and 2+1 dimensions. (This is not a contradiction because we do not expect the bulk to be local and perturbative simultaneously with the boundary.)

  \subsec{Bulk-point singularities from a local bulk} 
  
 Consider a $d$-dimensional  CFT with an $AdS_{d+1}$ dual. We view $AdS_{d+1}$ as the universal cover of the hyperboloid in ${\Bbb R}^{2,d}$ given by
 \eqn\adsmetric{
 P^I P_I=-(P^{-1})^2-(P^0)^2+\sum_{I=1}^d (P^I)^2 = -1.
 }
  Consider a $(d+2)$-point correlation function with generic boundary points given by 
   null $X_{a}^I\in {\Bbb R}^{2,d}$ ($a=1,\dots,d+2$). The singularity we are interested in happens 
   when  
 $  x  \equiv \det X^I_a  =0$.
  Notice that $x=0$ defines a codimension-one subspace in the space of cross-ratios.
  
A null vector $X$ representing a boundary point is defined modulo rescaling $X\sim \lambda X$, $\lambda \in {\Bbb R}^+$.  Hence, $x=\det X_a^I$ itself is not a well-defined cross-ratio, but instead transforms with weight $1$ under rescaling of each of the individual $X_a^I$'s.  To define a cross-ratio, we can divide by appropriate factors to form a projective invariant, e.g., 
\eqn\xhatdef{
\hat x = {x \over ((-2X_1\cdot X_2)(-2X_2\cdot X_3)\cdots (-2X_{d+2}\cdot X_1))^{1/2}}.
}
Objects with nonzero weight appear throughout the discussion in this section, but physical results are always projective invariants.  For $d=2$, we have
\eqn\xhatintermsofzzb{
\hat x = - {z-\bar z \over 4z(1-z)} + O((z-\bar z)^2)
}
so that $x=0$ is the same as $z-\bar z=0$, as described in section~2.1.
 
 \ifig\LightonN{An example arrangement of boundary points that leads to a singularity.  $X_1$ and $X_2$ are at time $-\pi/2$ on diametrically opposite sides of the Lorentzian cylinder. The remaining points $X_3,\dots,X_{d+2}$ are at time $\pi/2$ and arranged at generic directions on $S^{d}$.  In this configuration, lightlike particles can propagate into the bulk from $X_1,X_2$, scatter at $P$, and propagate out to $X_3,\dots,X_{d+2}$.}
 {\epsfxsize1.2in \epsfbox{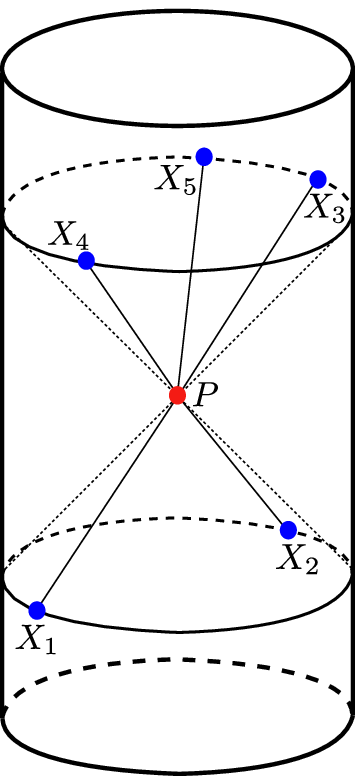} } 
 
  When $x$ vanishes, $X_a^I$ is a singular $(d+2)\times (d+2)$ matrix, so it has a zero-eigenvector $P^I$.  $P$ represents a bulk point that is lightlike separated from the others, $X_a\cdot P=0$, such that we can draw null lines 
 from the boundary points $X_a$ to $P$.
 Using $AdS$ isometries, we may assume $P=(1,0,\dots,0)$.  The $X_a$ then take the form $X_a=(0,n_a)$, where $n_a$ are null $d+1$-vectors.  The $n_a$ represent the direction of the null lines to the boundary, in local coordinates near $P$.
 
 If $\det X_a^I=0$, then $X_a^I$ also has a left zero-eigenvector $k_a$ such that $\sum_a k_a X_a=0$. Explicitly, we may choose
 \eqn\leftzeroeignevector{
 k_a=(-1)^{a-1}\det'_a(n_b^\mu),
 }
 where $\det'_a(\cdot)$ denotes a determinant with the $a$-th column removed.
   Thus we can assign momenta $k_a n_a$ to each null line so that momentum conservation holds at $P$, see \LightonN. 
 
 A singularity arises as a consequence of a local interaction in the 
 bulk. Let us imagine that we have the $d+2$ fields interacting through a local interaction ${\lambda\over (d+2)!}\phi^{d+2}$. 
 The leading perturbative correction to our correlation function is given by the Witten diagram\foot{This $i\varepsilon$ prescription is appropriate when $Q$ lies in the three Poincar\'e patches nearest $X_a$ \CornalbaXK.  Other patches do not contribute to the singularity we are interested in, so we will ignore them.  We leave $\varepsilon$ explicit because it will be important in the following discussions.}
 \eqn\wittendiagramone{\eqalign{
I &= \int_{AdS_{d+1}} dQ  (-i \lambda) \prod_{a=1}^{d+2 }{C_\Delta^{1/2} \over  ( -2Q\cdot X_a + i \varepsilon) ^{\Delta }} \cr
&= 
\left({C_\Delta^{1/2} e^{-i{\pi\over 2}\Delta} \over 2^\Delta\Gamma(\Delta)}\right)^{d+2}\int_{ AdS_{d+1}} dQ \int_0^\infty \left(\prod_a{d\omega_a \over \omega_a} \omega_a^\Delta\right) (-i \lambda) e^{-i Q \cdot \sum_a \omega_a X_a-\varepsilon \sum_a \omega_a},
 }}
 where
 \eqn\bulkboundaryprefactor{
 C_\Delta \equiv {\Gamma(\Delta) \over 2 \pi^{{d\over 2}} \Gamma(\Delta - {d \over 2} + 1)}.
 }
 Suppose that $X_a=(\sigma_a, n_a)$, where the $\sigma_a$ are small.  Note that 
 \eqn\detintermsofepsilons{
 x=\sum_a k_a \sigma_a,
 }
 where the $k_a$ are given by \leftzeroeignevector.
 
 Let us consider the integral \wittendiagramone\ near $Q=P$ in the limit $x\to 0$.  $I$ is singular if the integration contour cannot be deformed away from $Q=P$.  We claim that this occurs if and only if all $k_a$ have the same sign. In  this case we can reorder the $X_a$ so that all the $k_a$ are positive.  Indeed, suppose $k_a>0$ and consider deforming $Q$ in the imaginary direction, $Q \to P-i\delta Q$.  To avoid singularities in the propagators, we must have $\delta Q\cdot X_a > 0$ for all $a$.  However, then $0=\sum_a k_a X_a\cdot \delta Q > 0$, a contradiction.  This shows that positivity is sufficient to have a singularity.  As we explained in 
 section~2.3, it is also necessary.  For the remainder of this section, we will assume that $k_a>0$. This gives an additional constraint on the $X_a$. (For example, they cannot all be in the past (or future) of $P$.)
 
 The integral \wittendiagramone\ will be dominated near the point $Q=P$.  In this region, we can approximate $AdS$ by flat $(d+1)$-dimensional Minkowski space, $Q=(1,y)$, $y\in {\Bbb R}^{1,d}$, so that the wavefunctions $e^{-iQ\cdot \omega_aX_a}$ become plane waves $e^{i\omega_a \sigma_a-iy\cdot\omega_a n_a}$ with momenta $\omega_a n_a$. The integral over $y$ produces a momentum-conserving delta function,
 \eqn\wittendiagramtwo{
 I \approx (2\pi)^{d+1}\left({C_\Delta^{1/2} e^{-i{\pi\over 2}\Delta} \over 2^\Delta\Gamma(\Delta)}\right)^{d+2}\int_0^\infty \left(\prod_a{d\omega_a \over \omega_a} \omega_a^\Delta\right) \delta^{d+1}\left(\sum_a \omega_a n_a\right)(-i\lambda) e^{i\sum_a\omega_a (\sigma_a+i \varepsilon)}.
 }
 This approximation is valid up to subleading terms in the limit $x\to 0$.
 The $\delta$-function constraint is solved by $\omega_a = \omega k_a$ with an arbitrary overall energy $\omega$,
 \eqn\wittendiagramthree{\eqalign{
 I &\approx  (2\pi)^{d+1}\left({C_\Delta^{1/2} e^{-i{\pi\over 2}\Delta} \over 2^\Delta\Gamma(\Delta)}\right)^{d+2} \left(\prod_a k_a^{\Delta-1}\right) \int_0^\infty d\omega\,\omega^{(d+2)(\Delta-1)} (-i\lambda) e^{i\omega (x+i \varepsilon )},
 }}
 Finally, integrating over $\omega$ gives rise to a singularity at $x=0$,\foot{This singularity has projective weight $-\Delta$ in each of the $X_a$, matching \wittendiagramone. To express it in terms of projectively invariant cross-ratios, we can multiply by appropriate powers of $X_a\cdot X_b$.} 
 \eqn\finalsingularity{
 I \propto {\prod_a k_a^{\Delta-1} \over  (-ix)^{(d+2)(\Delta-1)+1}}.
 }

 For a general interaction given by a local amplitude ${\cal A}(p_a)$, the constant $-i\lambda$ gets replaced by the amplitude evaluated at momenta $\omega p_a$, where $p_a=k_a n_a$ (no sum),
 \eqn\wittendiagramfour{ 
   I \approx  (2\pi)^{d+1}\left(\prod_a {C_{\Delta_a}^{1/2} e^{-i{\pi\over 2}\Delta_a}k_a^{\Delta_a-1} \over 2^{\Delta_a}\Gamma(\Delta_a)}\right) \int_0^\infty d\omega\,\omega^{\sum_a(\Delta_a-1)} {\cal A}(\omega p_a) e^{i\omega ( x+i \varepsilon)}.
 }
When $x$ is small, the behavior of $I$ is controlled by the fixed angle scattering amplitude at high energies $\sim 1/\hat x$. (Recall that $\hat x$, defined in \xhatdef, is a projectively invariant version of $x$.)   For local interactions, ${\cal A}$ is polynomial in $\omega$, giving a singularity at $x=0$.

We expect this computation to be reliable for $\hat x \gg 1/M_*$, where $M_*$ is the scale that suppresses higher-dimensional interactions in the bulk.  When the bulk is a string theory, this is $\hat x \gg \ell_{s}$, and in M-theory this is $\hat x\gg \ell_{Pl}$, where $\ell_s$ and $\ell_{Pl}$ are, respectively, the string and Planck lengths in units of $R_{AdS}$.  We discuss what happens near these scales in section~4.

     \subsec{Landau diagrams in 1+1 dimensions} 
        
The bulk-point singularity described above cannot arise at any order in boundary perturbation theory in 1+1 dimensions.
        To see why, consider a  weakly coupled 2d sigma model. 
        The Landau diagrams are very simple in this case. The lines are   lightlike and momentum conservation at the 
        interactions implies that the left- and right-moving momenta are conserved. Therefore it is as if we did not have any 
        interactions. In other words, the Landau diagrams with and without interactions look the same, and the only singularities are 
        light-cone singularities. 
         
        In particular, this implies  that there is no singularity at $x \propto z - \bar z =0$, for generic values of $z$.

        \ifig\nolandauTwoD{Lorentzian cylinder drawn in the plane. The vertical lines at 
         $0$ and $2 \pi$ are identified. Red dots correspond to operator insertions. Black dots stands for the points of first interactions of on-shell particles emitted/absorbed by external operators. Clearly, it is not possible to draw a Landau graph  for $\phi \neq 0, \pi$.  }
          {\epsfxsize3.2in \epsfbox{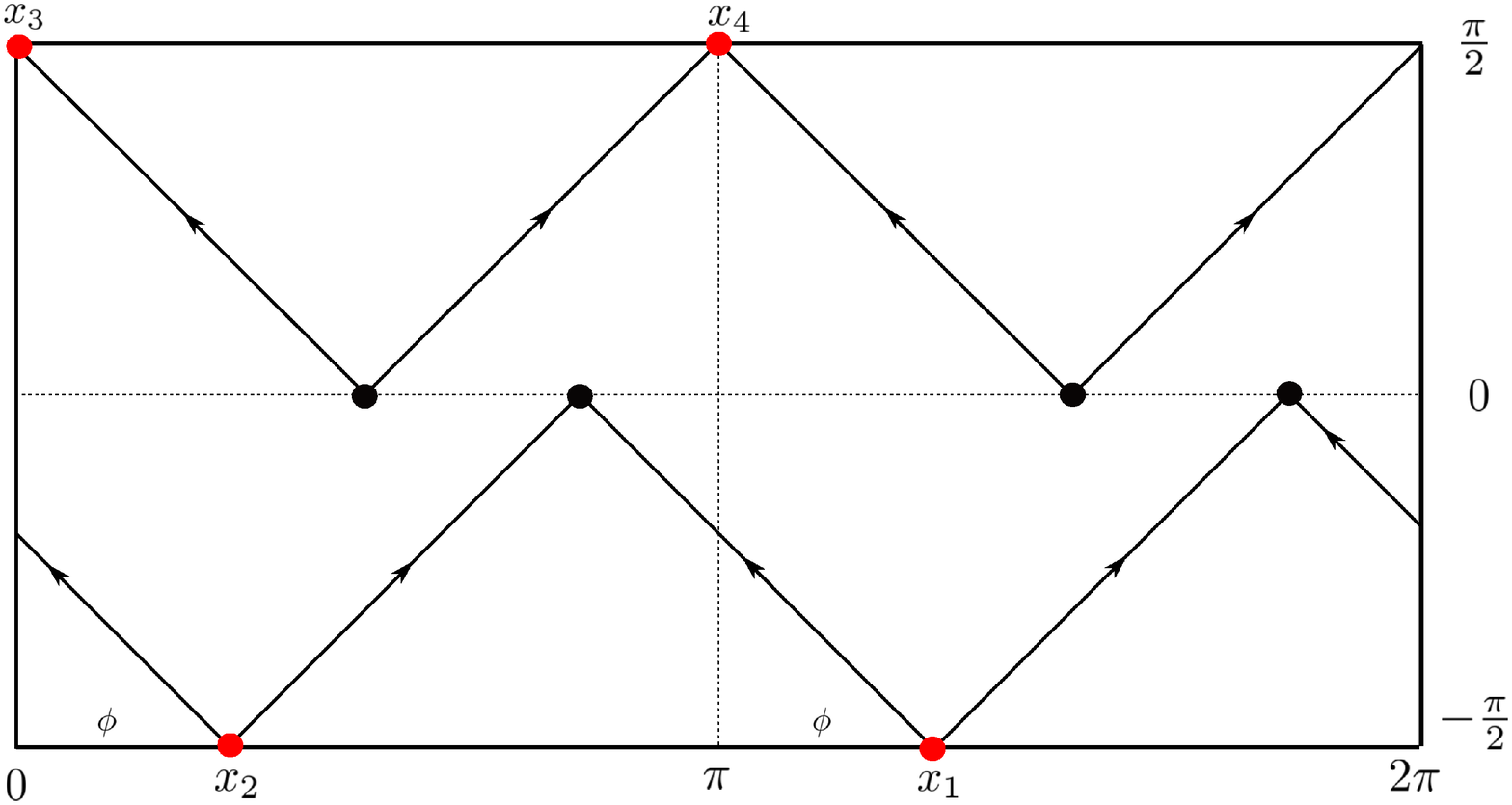} } 
        
        By performing conformal transformations, the points $x_3,x_4$ can be set at $\tau={\pi \over 2}$, and $\varphi =0, \pi$ and the 
         points $x_{1},x_{2}$ at $\tau$ and $\varphi = \phi+\pi, ~\phi$, see \nolandauTwoD. 
        Then $x=0$ corresponds to $\tau =-{\pi \over 2}$ with any value of $\phi$. These points can be 
        joined by a bulk Landau graph but not by any boundary Landau graph. 

        A popular starting point for the sigma model that is dual to $AdS_3 \times S^3 \times M_4$, with $M_4 = K3$ or $T^4$, is the symmetric product 
        $ M_4^k/S_k$,  where $S_k$ is the permutation group of $k$ elements. One then deforms this theory by a twist operator mixing two of the factors at a time. 
        It is interesting to wonder whether this type of perturbation theory can produce the bulk-point singularities at finite order in perturbation theory. 
         At finite order in perturbation theory a connected correlator corresponds to a computation in a theory with an order one value of $k$ (that grows with the order
         of perturbation theory). In a theory with a small value of $k$, we will prove in section~7 that the correlators do not have any singularity except for the ordinary 
         light-cone ones. Since we are expanding around a regular point in the conformal manifold (the space of coupling constants of the theory), 
          we expect that each term in the 
         expansion should also be analytic at $z= \bar z$.

        \subsec{Landau diagrams in 2+1 dimensions} 
        
\ifig\nolandau{We consider the five-point correlation function on the Lorentzian 3d cylinder. The configuration is chosen such that all external points lie on the geodesics emanating from a point in $AdS_4$. In a) the position of five points on the spatial $S^2$ is shown. Red points correspond to operators inserted at $\tau = - {\pi \over 2}$. Blue points correspond to operators inserted at $\tau = {\pi \over 2}$. In b) the moment of first interaction at $\tau = 0$ is shown. Excitations created at $\tau = - {\pi \over 2}$ could potentially interact at $\tau = 0$ along the sphere equator.}
 {\epsfxsize3.2in \epsfbox{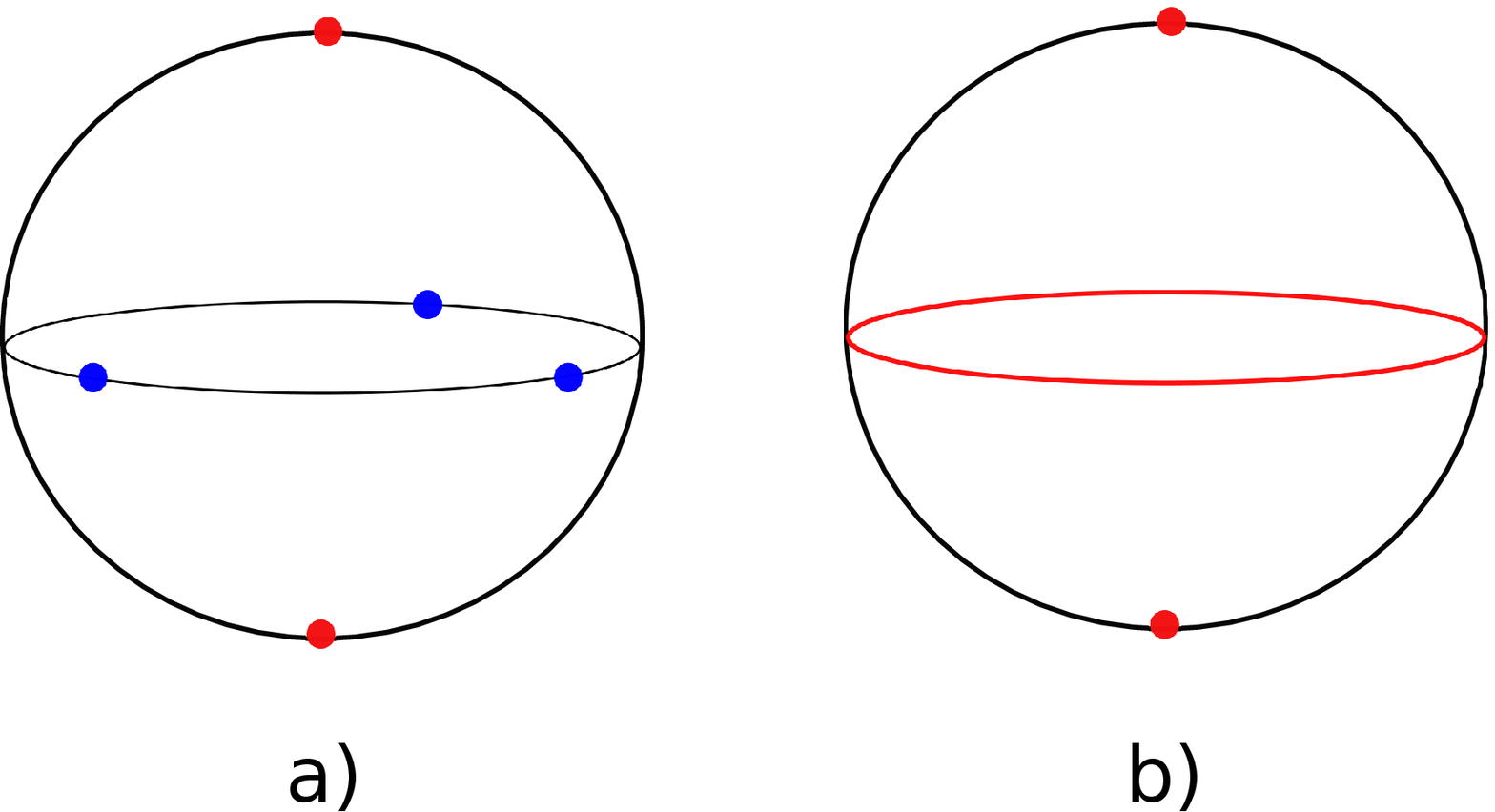} } 

We now argue that the bulk-point singularity $x=0$ cannot arise in perturbation theory in a three-dimensional CFT. In this case $d+2 = 5$, and we are dealing with the five-point function. 
We label the points in terms of $(\tau, \theta, \phi)$, where $\tau$ is time and 
$\theta$ and $\phi$ are standard coordinates on $S^2$. 
 We choose two points to be at $(  - {\pi \over 2},  0, \phi)$, $(  - {\pi \over 2}, \pi, \phi)$ in the past; these points evolve into
the three points 
 $(  {\pi \over 2}, {\pi \over 2}, \phi_1)$, $(  {\pi \over 2}, {\pi \over 2}, \phi_2)$, $(  {\pi \over 2}, {\pi \over 2}, \phi_3)$ in the future, see   \nolandau.

Let us do the same thing as before: starting from initial points, we follow the light-cone to the first possible interaction and then run time backwards from the final points to the last interaction. 

Starting from points $1$ and $2$ (located at the north and south pole respectively), we evolve in time from $\tau = - {\pi \over 2}$ to $\tau = 0$ --- the moment when the light cones first meet at the equator of the sphere. We conclude that the earliest  interactions can  occur at $\tau=0$ and should occur at 
the equator. 
  
Now we can run time backwards. Starting from the final points we conclude that at the moment $\tau = {\pi \over 2}^-$, all particles that are 
away from the equator must move towards it, since the final points are at the equator.

After the initial light-cones meet at the equator, at a time $\tau=0$, there are two options: either the particles remain up to $\tau=\pi/2$ on the equator, or
they leave the equator. In the second case, they must eventually return to the equator. Let us consider the case where there are particles going towards the north pole. 
These particles cannot reach very high latitudes since they have to get back to the equator at time $\pi/2$. Let us consider the highest latitude collision. 
If there are several collisions at this latitude, consider the one having incoming particles going north.  It will have 
 no incoming particles moving south. But it has no outgoing particles going north, therefore momentum along the lines of longitude would not be conserved. 
 This contradiction means that particles remain on the equator from $\tau =0$ until $\tau = \pi/2$. 
 Since the equator has only one spatial dimension, we now have Landau diagrams in 1+1 dimensions. As we have remarked above, these Landau diagrams are
 the same as the ones we would have in the free theory. So we can draw a pair of null lines emanating from an arbitrary point on the equator, one left-moving and one
 right-moving. These lines must end on two separate operators. But this is impossible if the positions of the operators are generic, namely when 
 $|\phi_i - \phi_j |$ is not a multiple of $\pi$.

        \subsec{Landau diagrams in $d\geq 4$}
        
In 3+1 dimensions, we have not been able to find a set of points with $x=0$ where we can prove nonexistence of a boundary Landau diagram.   For some symmetric configurations of points, we can actually find Landau diagrams on 
 the boundary (see appendix~C), but this does not mean that they exist for generic configurations.  It would be interesting to understand further when they do or do not exist. 
 The conditions that determine the existence of Landau diagrams seem reminiscent of the positivity conditions studied in the context of the amplituhedron 
 \ArkaniHamedJHA , and it would be interesting to see if there is any relation.

\newsec{Singularities beyond perturbation theory} 

In this section, we study the effect of various types of corrections. First we discuss
the effect of string worldsheet corrections for bulk-point singularities. At finite $\alpha'$ these corrections should
make bulk-point singularities disappear  \OkudaYM. We then discuss the effect of D-instantons. We claim that the 
D-instanton correction contains a singularity. It turns out that at weak boundary coupling, we can also get similar singularities from 
field theory instantons. We will provide a rationale for the agreement between these two computations. 
Finally, we argue that with a finite gravitational coupling constant $G_N$, we expect no bulk-point singularities. 

While so far we have focused primarily on $d+2$-point functions in $d$-dimensions, it is also interesting to consider four-point functions in $d$-dimensions. We end with a discussion of their expected singularities.

      \subsec{Stringy corrections} 
   
      Here we consider bulk-point singularities in $d+2$-point functions, arising from an interaction localized at a point in the bulk. 
      We consider a theory in the large $N$ expansion, and we examine the correlators at fixed order in this expansion. Each term is a function of
      the 't~Hooft coupling parameter $\lambda $. We recover the discussion in section~3.1 in the $\lambda \to \infty $ regime where the bulk string theory can 
      be approximated by a local field theory. Here, we discuss what happens when $\lambda$ is finite but large. This corresponds to a bulk theory with 
      a finite and small string length in units of the $AdS$ radius,  $\ell_{s} \ll 1 $.  
         In the small $x$ limit,  we 
  expect a correction to the formulas in section~3.1,  which arises as follows. 
We know that the flat space, fixed angle, string scattering amplitude at 
large energy goes as 
\eqn\fixedflat{ 
{\cal A}_{\rm flat}(\omega p_a ) \propto e^{ - \omega^2 h },
}
where $h$ is a positive function of the angles  \refs{\VenezianoYB\AlessandriniJY-\GrossKZA}. 
We saw that in the local regime, described in section~3.1, 
the energy scales as $\omega \sim 1/ x$. We have a range of energies where we can 
approximate the amplitude by the high energy flat space amplitude. In this regime, the 
correlator has the form 
\eqn\correfo{ 
I \propto \int_{o(1)}^\infty  d \omega \, \omega^c e^{ - \ell_s^2 \omega^2 } e^{ i \omega  x - \varepsilon \omega }.
}
The integral is now convergent at large $\omega$ for all $ x$. 
To understand what happened with the singularity it is useful to consider the explicit toy model 
\eqn\toyint{ 
I(\ell_s, x) = \int_0^\infty d \omega \, e^{ - \ell_s^2 \omega^2} e^{ i x \omega}.
}
For $\ell_s =0$, we get $I(0, x ) = i/x $. 
When $\ell_s$ is non-zero, $I(\ell_s,x)$ is an analytic function of $x$ (an error function).
 For  large $x$ with  Im$(x)>0$, 
we have that  $I  \sim i/x$, agreeing with the result  when $\ell_s =0$. However, as $x$ becomes
comparable to $\ell_s$, we find deviations from this behavior. Interestingly, when the imaginary part of
$x$ is very negative, say $x = -i y$ with large positive $y$, then the function grows like $\exp(y^2/( 4 \ell_s^2))$. 
In this region, even though $|x|$ is large, the value of the function is completely different from $i/x$, which
is the  analytic continuation of the $\ell_s=0$ answer to the region where the original integral was not convergent.  
Notice that the explicitly convergent region of the integral is in the direction of the $i \varepsilon$ 
prescription, which corresponds to performing a bit of Euclidean evolution that damps the contribution
of high energy states. If we go in the opposite direction, we enhance high energy contributions, and it 
is not surprising that we get a very different answer.  
One might have naively expected that the pole at $x=0$ would move in the complex
plane when $\ell_s$ becomes finite. The complete disappearance of the singularity is related to the fact that the function is
not analytic in $\ell_s$ at $\ell_s=0$.\foot{We thank A.\ Zamolodchikov for discussion on this point.}

 Of course, if   we expand the amplitude in powers of $\ell_s^2$, then each individual 
term seems to have a singularity at $x=0$. In fact, higher orders in the expansion give 
higher order singularities. However, the full function is perfectly regular at $x=0$. 
 
The conclusion is then that finite $\alpha'$ effects remove the $x=0$ singularity from the string tree-level correlator. This  had been previously observed in \OkudaYM .

When the   energy of the collision is very large, we expect corrections to the 
flat space scattering formula coming from the curvature of $AdS$. High energy scattering amplitudes
in $AdS$ were considered in \AldayHR, where it was found that the amplitude behaves as $
\exp( - ( \log \omega )^2 {\hat h} ) $  for large energies, where $\hat h$ is a positive function of the angles. This also leads to a convergent amplitude at large $\omega$. 

Note that once we have this strong suppression at large $\omega$, it is no longer true that 
$\omega \sim 1/x$. In fact, if we evaluate \correfo\ by saddle-point approximation, we see that even 
for $x =0$, we have a finite value for $\omega$. 
In the gravity regime, we can picture the $x \to 0$ limit as a kind of microscope, or collider, that lets us explore the 
local bulk degrees of freedom. However, this microscope is blurred at the bulk string scale where it 
ceases to explore higher energies.

It would be interesting to evaluate the correlation function of heavy operators using classical strings, 
as in \BargheerFAA ,
to further check that the result is completely regular at $x=0$. 
  
These arguments suggest  that there will be no singularity at $x=0$ 
at any order in string perturbation
theory.

As a final point,  notice that the emergence of a local theory in the bulk is something that can be explored
using the planar approximation of the gauge theory. As we go from small to large 't~Hooft coupling $\lambda$, we 
should get an enhancement of the connected correlator when $x \sim 0$.  This could be done explicitly if integrability techniques 
are developed to compute planar correlation functions. For recent progress in this direction see \refs{\BassoZOA,\BajnokHLA}.

       \subsec{Instanton corrections} 

Here we consider the spacetime dependence of instanton or D-instanton corrections to correlation functions. 
Such corrections are exponentially small, due to the action of the instanton. However, there are cases
where instantons or D-instantons lead to a dependence on one parameter that is invisible in perturbation theory
(e.g., the theta angle in four-dimensional gauge theories). 
In these cases, we can consider the derivative of the correlator with respect to this parameter, whose leading contribution comes
from the one-instanton correction, and we can explore its dependence on the positions in a clear way.\foot{Of course, a singularity in the 
prefactor of the one-instanton correction does not mean that the function after we sum over all instantons is singular  at this 
location.}   
Below, we will show that such corrections typically display a singularity at the same location as the bulk-point singularities.

 { \it D-instantons   }  
   
As argued in \refs{\GreenMY,\GreenTV},  D-instanton corrections to scattering amplitudes  are not exponentially 
 suppressed at high momenta.\foot{In the flat space case, the instanton correction never dominates over the tree amplitude
$e^{ - \ell_{s}^2 p^2}$ or the  $e^{ - \ell_{s} p}$  that we get  after resumming over genera \MendeWT. 
The black hole formation threshold at $\ell_{s} p \sim 1/g$ happens 
first as we increase the energy.}
So if we consider the one-instanton correction to the correlator, we expect singular behavior 
at $x=0$. This is in line with the idea that D-instantons explore points in the bulk.

{ \it Singularities from field theory instantons }

Now we consider singularities due to instantons in the boundary quantum field theory. 
It turns out that instanton corrections often give rise to singularities that are similar to bulk-point singularities. 
Let us explain the mechanism in  a   simple case. 
 Consider a theory in  1+1 dimensions. Imagine a non-linear sigma model that contains 
a non-contractible $S^2$ in its target space. This has instantons that correspond to wrapping the $S^2$, which can be
described as follows. 
Consider first the Euclidean field theory and parametrize space by the coordinates $z$ and $\bar z$. 
We parametrize the target space sphere using the complex coordinate $w$, with  
\eqn\swritx{
w = \tan \theta/2 e^{ i\phi} ,~~~~~~~~ d\theta^2 + \sin^2\theta d\phi^2  = 4 { |d w|^2 \over (1 + |w|^2)^2 }.
}
 The instanton is given by 
\eqn\instaco{ 
 w = { a z + b \over c z + d } ,~~~~~~~~~~~ m \equiv \pmatrix{ a & b \cr c & d } ,~~~~~~ \det\, m =1,
 }
 where the matrix $m$ is related to parameters of the instanton. 
 The moduli space is given by $SL(2,{\Bbb C})/SU(2)$, which can be viewed as $H_3$ since the metric is  $SL(2,{\Bbb C})$-invariant. 
 We now consider operators that change the radius of the sphere,
 \eqn\opera{ 
 O(z,\bar z) =  { \partial_\alpha w \partial_\alpha \bar w \over  (1 + |w|^2)^2 } .
 }
 Evaluated on the holomorphic instanton \instaco, this  operator has the expectation value 
 \eqn\expecva{ 
 \langle O(z,\bar z) \rangle = { 1 \over \left(   (\bar z , 1) m^\dagger m \pmatrix{z \cr 1} \right)^2 }.
 }
 We see that the answer depends on the hermitian matrix 
 $P  = m^\dagger m $, with $\det P =1$,
 which  parametrizes   $SL(2,{\Bbb C})/SU(2)$. 
Writing
 \eqn\ppar{ 
 P = \pmatrix{ Y_{-1} + Y_2 & Y_0^E + i Y_1 \cr 
 Y^E_0 - i Y_1 & Y_{-1} - Y_2 }  ,~~~~~~~\det P =1 = Y_{-1}^2 - ( Y_0^E)^2 - Y_1^2 - Y_2^2  ,
 }
 we recognize the $Y_I$ as embedding coordinates for $H_3$. Furthermore the expression for the operator in \expecva\ is simply given 
 by $ \langle  O \rangle = ( X \cdot Y)^{-2} $, with $X$ the embedding coordinate for the boundary point.   
 We see then that the instanton contribution to the four-point function can be computed as 
 \eqn\insco{ 
 \langle O_1 O_2 O_3 O_4 \rangle \propto  e^{ - I} \int_{H^3}  d^3 Y {1\over  \prod_{i=1}^4  (Y \cdot X_i)^2 }.
 }
 This has precisely the same form as that of a local bulk interaction. Here, $I$ is the action of the instanton. 
 In particular, after continuing \insco\ to Lorentzian signature, we will find precisely the same bulk-point singularity that gravity produces (except for the 
 small factor $e^{-I}$). 
 
 We should remark that the full computation in the sigma model is more involved, since we need to include the fermions and the fluctuations
 in the extra dimensions in target space that make it a full conformal sigma model. If we only had the $S^2$ and nothing else, then the theory would not 
 be conformal. Similarly, the operators might contain some dependence on those dimensions, which would complicate some of the details while 
 retaining the same $x=0$ singularity, at least at leading order in perturbation theory. 
 
 Notice that the presence of a singularity here does not contradict the argument that, in perturbation theory, 
we only get singularities on the light cone in 1+1 dimensions. Here we are talking about a non-perturbative effect. 
 
 In 3+1 dimensions, one obtains a similar result  \refs{\BianchiNK\DoreyPD-\BianchiXSA }.
  The moduli space of Yang-Mills instantons can be viewed as $H_5$, where the radial direction is
 the instanton size. Furthermore, the correction they produce for a correlation function can be found by evaluating the operator in the instanton 
 background, as above. This leads again to functions that are identical to the ones produced by local bulk interactions.\foot{For example,  
 see equation (34) in \BianchiNK.}

  \subsec{General argument for the position dependence of one-instanton corrections} 
   
  Here, we explain the position dependence of the one-instanton correction to correlation functions. 
  We imagine an instanton whose only moduli  are those given by the breaking of the symmetries: conformal symmetries as well as supersymmetries. 
  The main point is that the one-instanton correlation function at leading order in the $1/N$-expansion factorizes as 
  \eqn\ecorcaa{ 
   \langle O(x_1) \cdots O(x_n) \rangle \propto e^{ - I } \int DY \langle O(x_1) \rangle_{{\rm inst},Y} \cdots \langle O(x_n) \rangle_{{\rm inst},Y},
   }
   where $\langle O(x_i) \rangle_{{\rm inst},Y}$ is the expectation value of an operator in the presence of the instanton with fixed values of the zero modes. 
   This one-point function is completely fixed by conformal invariance: it is given by the corresponding bulk-to-boundary propagator. 
   We saw this explicitly in the example above when we only had the conformal moduli. We did not check it explicitly, but we expect that the same symmetry 
   arguments extend to the fermionic moduli, when these arise from broken supersymmetries.
    This then fixes the one-point functions and consequently the spacetime dependence of the correlation function. 
   The important point to notice is that in this argument we did not use the value of the 't~Hooft coupling, so it applies both for strong and weak coupling. 
   In fact, the spacetime dependence of the instanton correction was observed to be the same at both weak and strong
    coupling \refs{\BianchiNK\DoreyPD-\BianchiXSA }. This argument does not put any constraint on the $\lambda$ dependence of the constant
     prefactor.

        \subsec{Exact answer} 
       
If we now consider a theory with a finite Planck scale, then we expect that the fixed angle 
scattering is suppressed exponentially, since the amplitude would be given by a classical gravity solution
whose action is proportional to $G_N \omega^{d-1}$, leading to the corresponding 
exponential suppression $\exp( - G_N \omega^{d-1} ) $.  This is also sometimes explained by saying 
that high energy scattering will typically make a black hole that evaporates into a large number of
particles, so that producing just two particles in the final state would be highly unlikely \refs{\ArkaniHamedKY,\GiddingsGJ}.   
Of course,  we could consider a theory where the string scale and the Planck scale coincide, where 
these strong gravity effects are the first to remove the singularity. 
This suppression leads us to expect that the answer should be analytic at $x=0$. 
We can also say that the production of black holes implies that we cannot explore arbitrarily short 
distances, therefore removing the $x=0$ singularity. Of course, the boundary theory had no bulk points to start with. 

In fact, we will later  prove that in a $1+1$ dimensional CFT, there is indeed no singularity at $x=0$. 

\subsec{Summary}

\ifig\bumpcartoon{An illustration of the $(d+2)$-point function near $\hat x=0$ for different bulk theories. The top curve (dashed light-blue) shows a $\hat x^{-\beta}$ divergence from a tree-level Witten diagram. The bottom curve (dark blue) shows this divergence cut off at $\ell_{s}$, when the bulk is a string theory. The middle curve (medium blue) shows the case $\ell_{s}\to \ell_{Pl}$, where the divergence is cut off by further corrections in the exact gravity theory. The dotted gray curve shows the effect of $D$-instantons, which (though formally singular at $\hat x =0$) are suppressed and can only be trusted down to a scale between $\ell_{Pl}$ and $\ell_{s}$.}
 {\epsfxsize4.6in \epsfbox{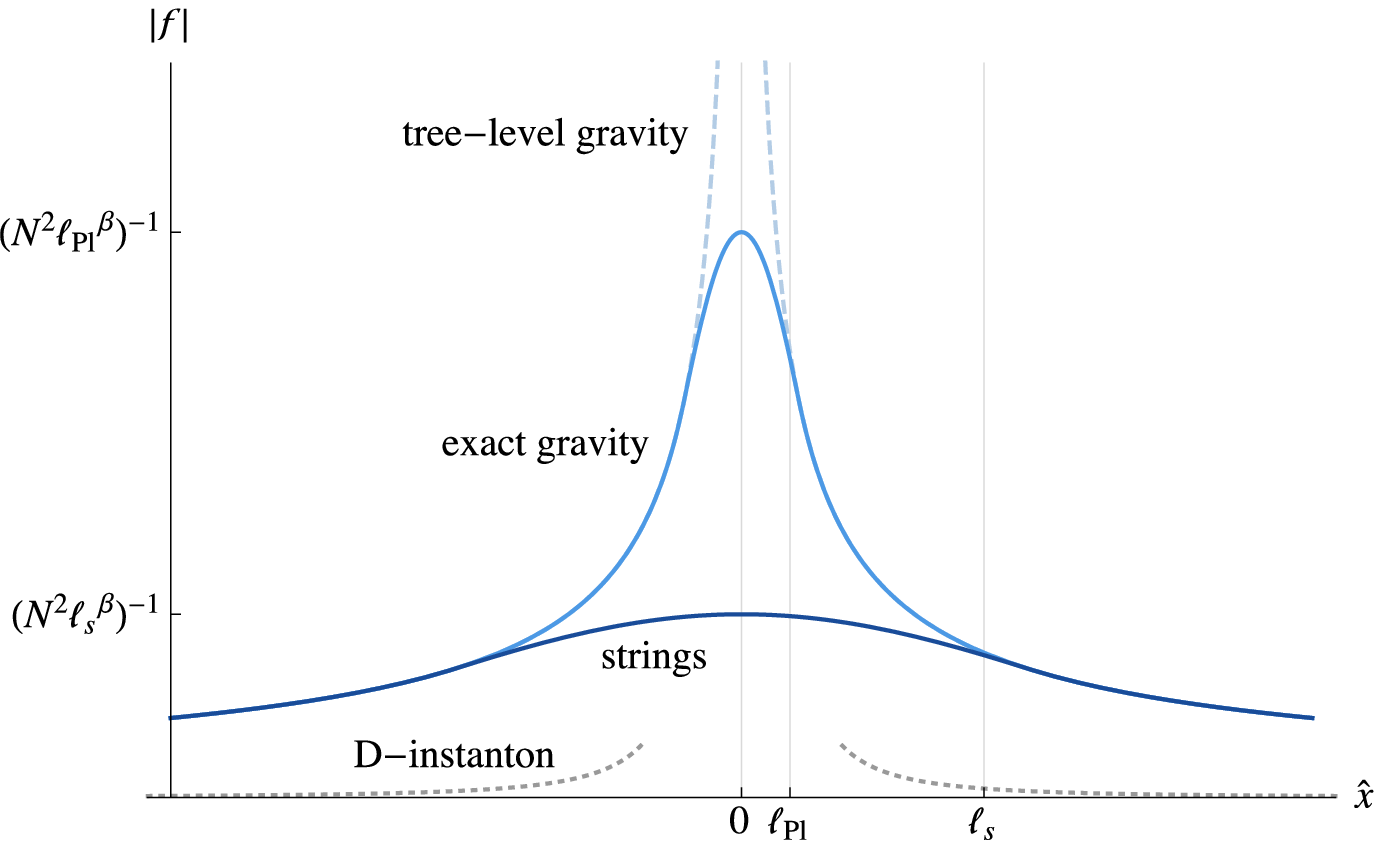} } 

These considerations suggest the following picture for the $(d+2)$-point function near $x=0$, illustrated in \bumpcartoon.  In this discussion, it is useful to refer to the projectively invariant cross-ratio $\hat x$ \xhatdef.  For a bulk amplitude growing like ${\cal A}\propto s^{k}$, a tree-level Witten diagram gives a singularity $f\sim (N^2\hat x^{\beta})^{-1}$, where $\beta=(d+2)\Delta-d-1+2k$.  In the case of an ordinary local gravity theory, we expect that $k=1$. 
 If the bulk is a string theory, the singularity is cut off at the scale $\hat x \sim \ell_{s}$, resulting in a bump in the correlator of height $(N^2 \ell_{s}^\beta)^{-1}$.  When the bulk is not described by weakly-coupled string theory (for instance in M-theory, or in the limit $\ell_{s} \to \ell_{Pl}$), our tree-level computation breaks down near $\hat x\sim \ell_{Pl}$, and we expect gravitational corrections to smooth out the singularity, resulting in a bump of height $(N^2 \ell_{Pl}^\beta)^{-1}\sim N^{{2\beta \over d-1}-2}$.\foot{Here, we assume $c_T \propto N^2$, as for an adjoint theory. In a vector-like theory, the $N$-dependence will be different.}  Note that in this latter case, even though the bump comes from a $1/N^2$ correction, it actually {\it grows} with $N$ when $(d+2)\Delta- 2 d +2k > 0$.  This is because the breakdown of bulk perturbation theory occurs when the effective energy scale $\omega\sim \hat x^{-1}$ becomes of order $\ell_{Pl}^{-1}$, and the breakdown does not depend on whether the correlator itself is small or large.

\newsec{Singularities of the four-point function}

Previous papers on this subject have focused on the four-point function  \refs{\PolchinskiYD\GaryAE\HeemskerkPN\PenedonesUE-\OkudaYM}. 
In $d\geq 3$ dimensions, the CFT four-point function can have a singularity at $ z = \bar z$  already in perturbation theory (see \inteag ) 
because we can draw a Landau diagram on the boundary.  
In addition,  we can also draw Landau diagrams in the bulk. 
In this section,  we discuss in more detail the four-point function, emphasizing the various possible contributions to the singularity at $z =\bar z$. 
 
Let us start with some kinematics. We have four points $X_a^M$ in embedding coordinates. We   form 
 the $4\times 4$ matrix $X_a \cdot X_b$.  Generically its determinant is non-zero. We want to focus on situations where its determinant,  
 $  \det( X_a \cdot X_b) \propto (z-\bar z)^2$,  vanishes. It turns out that there are distinct configurations with zero determinant that cannot be distinguished
 by the cross-ratios of the four points \GaryAE . They can be distinguished by the type of manifold generated by the four vectors  $X_a^M$. This manifold can be either 
 a four-dimensional null manifold or a three-dimensional manifold. As an example, the first case arises when the four points can be located in the interior of the Poincar\'e patch, such as the points in 
 \fourpc\ with $t=1$. 
 An example of the second configuration is the points $X^M = (X^{-1},X^0,X^{1} \cdots , X^{d})$ given by\foot{We use $X^M_i = (\cos \tau_i, \sin \tau_i, \cos \phi_i , \sin \phi_i, \vec 0)$.}  
 \eqn\fullnull{
 X^M_{1,2} = ( 0,-1,\mp \cos \phi, \mp \sin \phi , \vec 0 ), ~~~~~X^M_{3,4} = ( 0,1,\pm 1 ,0, \vec 0) .
  }
 Both types of  configurations have $z=\bar z$ or zero determinant. But they cannot be transformed into each other by a finite 
 conformal transformation.  This type of situation is common in Lorentzian signature. For example, 
 consider the two-point function, which is a function of the proper distance $x^2$. We can have a singularity at $x^2=0$ when points are either null-separated or coincident. By analogy with this situation, we will 
 call the first type of configuration the ``null $z=\bar z$ singularity'' and the second the ``full $z=\bar z$ singularity.'' 
 Since the four-point function depends only on the cross-ratios, we will have the same approach to the singularity in both cases, even though the configurations look rather different. Of course, if we had a 
 higher point function, where only four of the points are approaching the $z=\bar z$ configuration, then these two cases can behave very differently, and
 can be distinguished by looking at other cross-ratios.

 In the case of the  null $z=\bar z$ configuration, we cannot find a bulk point $P$ that is null separated from all other boundary points. The only point that is null separated from the four points lies also  on the boundary. 
 
 A richer situation arises for the  full $z=\bar z$ configuration.  In this case, we can find a point $P$ that is null separated from all the boundary points, $X_a \cdot P =0$. 
 Furthermore, it is also possible to draw bulk 
 Landau diagrams with a vertex at $P$.\foot{This is clear because we can find a point $P$ in the $AdS_3$ subspace  given by $0=X^3 = X^4 =\cdots =X^d$, as in section~3.1.}
 In fact, we get a full family of such points that span an $H_{d-2}$ subspace of $AdS_{d+1}$. 
 For example, in the configuration in \fullnull , this subspace is given by $ -(X^{-1})^2 + (X^{3})^2 + \cdots (X^{d})^2  = -1$. 
 We can understand the symmetries of the configuration as follows. A generic set of four points is only invariant under an $SO(d-2)$ subgroup of the conformal group. However, the 
 Lorentzian full $z=\bar z$ configuration is invariant under a $SO(1,d-2)$ subgroup.\foot{In Euclidean space, configurations with $z=\bar z$ are invariant under $SO(d-1)$.} This subgroup acts 
 on the bulk hyperboloid. This bulk hyperboloid intersects the boundary at an $S^{d-3}$, so that even on the boundary we can have more than one
  Landau diagram.\foot{For $d=3$ we have two points.}
 
  When we perform the bulk computation, we can have contributions to the $z-\bar z$ singularity that come from two sources. We can have a bulk UV contribution that comes from high energy particles 
  colliding at a particular point $P$ in the hyperboloid. In addition, we can have a bulk IR contribution that comes from integrating  the interaction point $P$ over the hyperboloid. 
  Let us discuss first the bulk IR contribution.   
When we go away from $ z = \bar z $, we find that this  integral over hyperbolic space gets cut off at a distance $e^\rho \propto 1/|z-\bar z|$, which 
results in a singularity of the form 
\eqn\signir{ 
\langle { \cal O}(x_1)  { \cal O}(x_2)  { \cal O}(x_3)  { \cal O}(x_4) \rangle \propto { 1 \over (z - \bar z)^{d-3} } 
}
(for $d=3$ we get a logarithm). This singularity does not involve short distances
 in the bulk. From the boundary point of view, it involves short distances near the $S^{d-3}$ of possible interaction points for  boundary Landau diagrams.

Let us now discuss the contribution from the bulk UV singularity. For that purpose, it 
 is convenient to choose the following $AdS_{d+1}$ coordinates: 
\eqn\embdcaI{
Q^M_{AdS_{d+1} }  = \cosh \rho\, \tilde Q^M_{AdS_3} + \sinh \rho\, \vec n_{S^{d-3} }, 
}
where $\tilde Q^M_{AdS_3}$ are embedding coordinates of $AdS_3$ and $\vec n$ is a unit vector on $S^{d-3}$. 
 Then the bulk diagram corresponding to a contact ${\lambda \over 4!} \phi^4$ interaction has the form 
\eqn\diagrin{ 
I = {\rm Vol}_{S^{d-3} } \int_0^\infty d\rho \sinh^{d-3} \rho (\cosh \rho )^3 \int_{AdS_3} d\tilde Q { -i\lambda \over \prod_{a=1}^4 ( - 2 \tilde Q \cdot X_a \cosh \rho + i \varepsilon )^\Delta }.
}
The last term has the same form as the $AdS_3$ problem, except that the $X_a$ have been rescaled. Therefore, we can repeat the derivation in section~3.1
 to go from \wittendiagramone\ to \wittendiagramthree. The only difference is that we replace $x$ by $x \cosh \rho$,
\eqn\witdiagro{ \eqalign{ 
I &\approx  (2\pi)^{3}\left({C_\Delta^{1/2} e^{-i{\pi\over 2}\Delta} \over 2^{\Delta}\Gamma(\Delta)}\right)^4\left(\prod_a k_a^{\Delta-1}\right)  {\rm Vol}_{S^{d-3} }  \cr
&~~~~\times \int_0^\infty d\rho \sinh^{d-3} \rho  \int_0^\infty d\omega\,\omega^{4(\Delta-1)} (-i\lambda) e^{i\omega x \cosh \rho} \cr 
&\approx 2(2\pi)^{{d+3 \over 2}}\left({C_\Delta^{1/2} e^{-i{\pi\over 2}\Delta} \over 2^{\Delta}\Gamma(\Delta)}\right)^4\left(\prod_a k_a^{\Delta-1}\right) \cr
&~~~~\times 
\int_0^\infty d\omega\ \omega^{4(\Delta-1)} (-i \lambda) (-i \omega x)^{{3-d \over 2}} K_{d-3 \over 2}(-i \omega x) \cr
 &\propto {\prod_a k_a^{\Delta-1} \over (-ix)^{4\Delta-3}},
 }}
 where the $(\cosh \rho)^3$ term in \diagrin\ was cancelled by three-dimensional energy momentum conservation in $AdS_3$. In the second line,
 we have done the 
 integral over $\rho$ in order to get the Bessel function.

The integral \witdiagro\ contains both the UV and IR contributions. To separate them, it is useful to replace the contact interaction by a string amplitude. 
We can model this by writing an expression analogous to \wittendiagramfour \ where we replace $\lambda$ by 
${\cal A }( \omega p_a )$. Here, $\omega $ sets the overall energy scale of the process. If the amplitude vanishes rapidly for large $\omega$, then
the UV contribution cancels and we get the result predicted by the general discussion around \signir. Namely, if the amplitude is suppressed beyond 
$\omega_0$, then we get $ e^{i w_0 x \cosh \rho }$, which cuts off the $\rho$ integral at a constant value of $ x e^\rho  $, producing the desired result \signir. 
This discussion is similar to the one in \PolchinskiTT\ for deep inelastic processes. The high energy regions in the bulk are highly 
suppressed and the contribution arises from the regions where a redshift factor effectively lowers the energy to proper energies below the string scale.

For similar reasons, in the exact gravity theory we do not expect a bulk UV contribution. But we do expect the bulk IR contribution. 
In fact, we can argue that we expect a contribution of the form \signir\ for general theories, even those that do not have a gravity dual. 
This should be intuitively clear from the discussion of symmetries above. Namely, the singularity arises from the action of the non-compact symmetry group $SO(1,d-2 )$. 
In fact,  the $1/(z - \bar z)^{d-3}$ singularity is present for each individual conformal block. Here it is clear that it can only arise from the non-compact nature of this symmetry group. 
 The fact that conformal blocks have this singularity is reviewed in appendix~B and can be seen 
by looking at the equation  (2.21) in \HogervorstSMA\ (see also \DolanHV ). It is also clear from the explicit expressions in $d=4,6$.   

Note that a free field theory does not have any singularity at $z=\bar z$. The singularities from each individual conformal block cancel out. 

It is tempting to conjecture that any non-free theory will have a singularity at $ z-\bar z$ of the form \signir. 
As a simple example, consider the large $N$ critical $O(N)$ model, which in a sense is close to a free theory. In that case, the $1/N$ correction to the 
four-point function of the spin fields ($O(N)$ vectors) has a singularity of the form $\Gamma({d-3 \over 2}) x^{3-d}$,  as expected.\foot{This is obtained from the  function $\bar D_{{d \over 2} -1 , 1 , {d \over 2} -1 , 1}$, see   \AldayOTA .}
It is important for this conjecture that we consider the four-point function of the smallest dimension (or twist) operator in the theory. 
  For example, in a product of two independent field theories, where each individual one is interacting within itself, 
 we can consider products of operators in each of the two theories, and we can have a higher order singularity. 
Similarly, in a large $N$ theory, we can have higher order singularities if we have double-trace  external operators.\foot{We thank Douglas Stanford for discussion on this point.} 

An apparent counterexample arises when we consider   the four-point function of scalar half-BPS operators ${\rm Tr[\phi^2]}$ in ${\cal N}=4$ SYM. At one loop, we get ${\lambda \over (z - \bar z)}$ coming from a $\phi^4$-type interaction, and at two loops we get ${\lambda^2 \over (z - \bar z)^2}$ from two copies of that diagram \EdenMV. 
This double pole arises because we have two separate vertices.  In other words, we get a singularity similar to what we would get in two separate field theories. We expect that at higher 
loops, there should be $\log (z- \bar z)$ corrections that ultimately remove this higher power. In fact, at three loops there is such a contribution  \DrummondNDA. 
It is also interesting to try to give a hand-waving bulk understanding of this higher order singularity. Since it should correspond to a string theory with very small string tension, we do not expect a bulk UV singularity. 
However, a very low string tension can give rise to a very big intermediate string that stretches between two very distant bulk points, so that we end up integrating over two separate points on the hyperboloid.  
 But we expect that at higher orders we will get contributions that suppress the separate points as $e^{ - \lambda ( \log (z- \bar z) )^2 }$.

\newsec{Approaching singularities using the OPE}

In this section, we apply the OPE to understand the behavior of Lorentzian correlators. 
We are primarily interested in the bulk-point singularity and the Regge limit.  
One must analytically continue away from the Euclidean sheet to reach these configurations.
Nevertheless, we would like to emphasize that they can be approached in such a way that the OPE remains valid.
  
The general idea is easy to formulate. Consider operators $\cO(\tau_L,\phi)$ on the Lorentzian cylinder parameterized by a time $\tau_L$ and an angle $\phi$ on a great circle of $S^{d-1}$.  We arrange them in the following four-point function:
\eqn\lorcylfourp{
f(u,v) = {\< \cO(- \pi, \phi + \pi) \cO(-\pi, \phi )  \cO(0, 0 )\cO( 0, \pi) \>_{R\times S^{d-1}} \over  \<\cO(0,0)\cO(0,\pi)\>^2},
}
with generic $\phi$.\foot{Throughout the paper we write the correlator $$\<  {\rm T} \left\{ \cO(x_1)\cO(x_2) \cO(x_3)\cO(x_4) \right\} \> = \< \cO(x_1)\cO(x_2) \cO(x_3)\cO(x_4) \>,$$ which makes time ordering implicit.}${}^{,}$\foot{We have shifted all operators by $-\pi/2$ in $\tau_L$ relative to previous sections. This is for convenience when discussing analytic continuation below.} To approach this configuration, we evolve the operators at time $\tau_L = - \pi$ by $- \eps$ in Euclidean time, corresponding to $ - \pi \to  - \pi + i \eps$. Next, we consider the OPE in the $s$-channel $\cO(-\pi + i \eps, \phi )\cO(-\pi + i \eps, \phi + \pi)$.  This OPE converges for $\eps>0$. The region of interest lies at the boundary $\eps \to 0$.

The $s$-channel OPE takes the following form:
\eqn\OPEsMORE{
f(u,v) = \sum_{\Delta, \ell} e^{- i \pi \Delta} c_{\Delta, \ell }^2 g_{\Delta, \ell}(\eps, \phi),
} 
which should be compared with the Euclidean correlator 
\eqn\OPEsMOREeucl{
f_{E}(u,v) = \sum_{\Delta, \ell}c_{\Delta, \ell}^2 g_{\Delta, \ell}(\eps, \phi),
} 
which corresponds to $\< \cO(- \eps, \phi + \pi) \cO(-\eps ,  \phi )  \cO(0, 0 )\cO(0, \pi) \>_{E}$.
The difference between the two is in the phase factor $e^{- i \pi \Delta} $ in \OPEsMORE .
 As we explain below, this can lead to different effects. First, it is easy to see that it can lead to the emergence of singularities in large $N$ theories. In this case the emergence of a singularity is related to infinitely many operators having dimensions such that they sum up ``in phase.'' Second, it can lead to damping of the correlator when the phases are ``random.'' This is the chaos phenomenon.

\subsec{A simple bound on the bulk-point singularity}

\ifig\Rhoradialquantization{Any four points in $\Bbb R^d$ can be brought into the above configuration using conformal transformations.  The four points lie in a two-plane, and $\rho=re^{i\phi}$ is a complex coordinate on that plane.  The quantities $r$ and $\phi$ are alternative parameterizations of the two nontrivial conformal cross-ratios.  Radial quantization around the origin gives an expansion for the four-point function in $r=|\rho|$. (Figure from \HogervorstSMA.)}
 {\epsfxsize2.8in \epsfbox{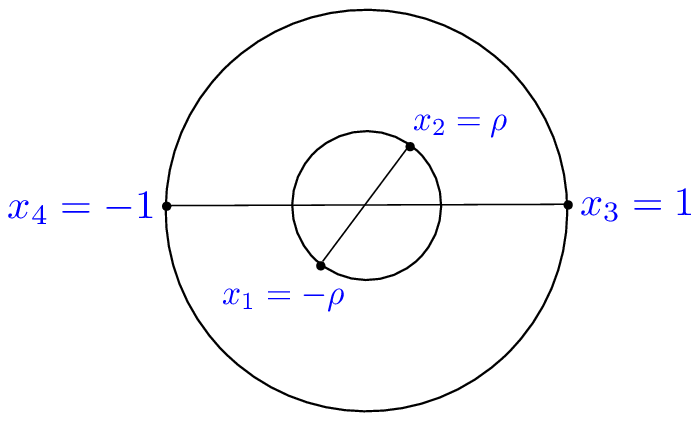} } 

\ifig\nolandau{The map from $z$-plane to the $\rho$-space is shown. The whole $z$-plane minus the $[1,\infty)$ cut is mapped to the 
$|\rho| < 1 $ region. The $[1,\infty)$ cut in the $z$-plane is mapped to the $|\rho| = 1$ locus.}
 {\epsfxsize4.4in \epsfbox{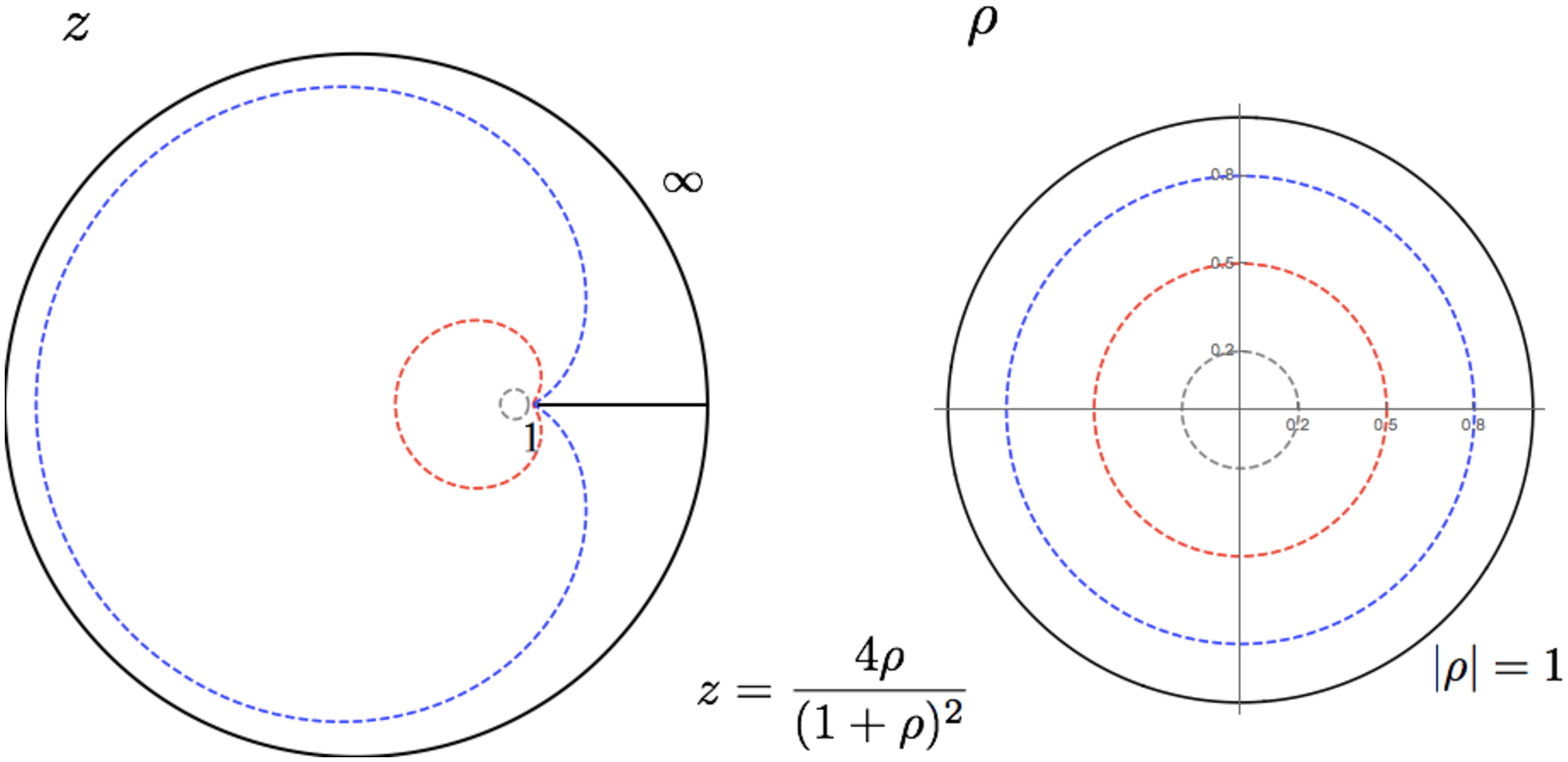} } 

To understand more clearly the relationship between the Lorentzian and Euclidean correlator, let us start on the Euclidean cylinder, 
\eqn\fourptconfigurationtwo{
f_E(u,v) = {\< \cO(\tau_E,\phi+\pi)\cO(\tau_E,\phi) \cO(0,0)\cO(0,\pi)  \>_{E} \over \<\cO(0,0)\cO(0,\pi)\>^2},
}
with $\tau_E < 0$.  In radial quantization, this corresponds to the configuration shown in \Rhoradialquantization, where $\rho=e^{\tau_E+i\phi}$ is the radial coordinate introduced in \refs{\PappadopuloJK,\HogervorstSMA}.  The standard cross-ratios $z,\bar z$ are related to $\tau_E$ and $\phi$ by
\eqn\rhovariabledefinition{\eqalign{
\rho &= r e^{i\phi} = e^{\tau_E+i\phi} = {z \over (1+\sqrt{1-z})^2},\cr
\bar\rho &= r e^{-i\phi} = e^{\tau_E-i\phi} = {\bar z \over (1+\sqrt{1-\bar z})^2}.
}}
In Euclidean space, $\bar \rho$ and $\bar z$ are the complex conjugates of $\rho$ and $z$, respectively. 
The OPE around $r=0$ converges for $|r|< 1$, which translates to the whole cut plane in the $z$ variables, see \nolandau. 

\ifig\zequalzbregions{There are several different regions of the Lorentzian cylinder where $z=\bar z$, but they are easy to distinguish using the variables $\tau_L$ and $\phi$.  Firstly, when $\tau_L=0$ with $\phi$ arbitrary (blue), the correlator is in a Euclidean regime: all operators lie on the unit circle in \Rhoradialquantization.  When $\phi=0,\pi$ with $\tau_L$ arbitrary (green), we have $z=\bar z$.  (Moving along the green line, taking $\tau_L\to\pi$, we approach the Regge limit, discussed below.)  Finally if $\tau_L=\pi$ with $\phi$ arbitrary (red), the correlator is in the bulk-point region.  Although all of these loci have $z=\bar z$, they require different analytic continuations from the Euclidean regime, and hence can have different physics. For example, if we start at the blue line and increase $\tau_L$, then each operator crosses one light cone (dashed line) to get to the green line, while each operator crosses two light cones to reach the red line. One can move around the light cones and connect these different regimes by moving in Euclidean time (making $\epsilon$ finite) as we do below.}
 {\epsfxsize3.8in \epsfbox{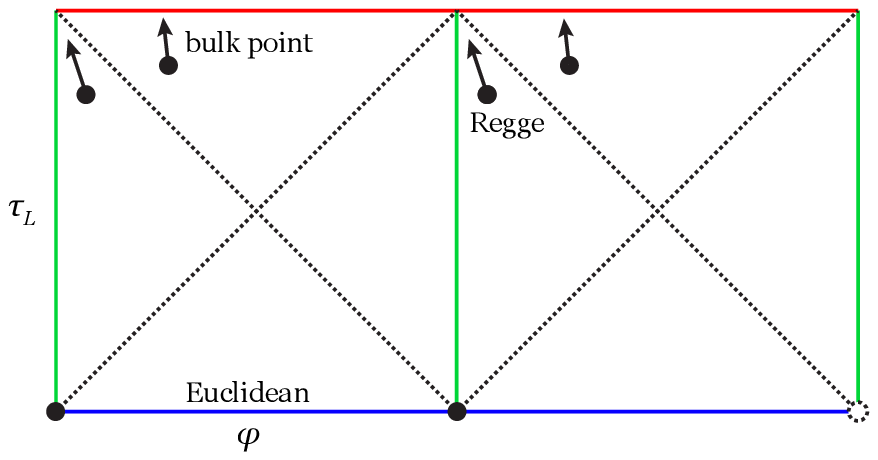} } 

The Lorentzian correlator corresponds to analytically continuing $\tau_E\to i \tau_L$. 
Note that $\rho= e^{ i \tau_L + i \phi}$ and $\bar\rho=e^{ i \tau_L - i \phi } $ are no longer complex-conjugates after this continuation.
  As we approach the Lorentzian region, we maintain convergence of the OPE by taking $\tau_E=i \tau_L - \epsilon$ where $\epsilon $ is small and positive.  The Lorentzian version of the correlator \fourptconfigurationtwo\ can then be written
\eqn\cylinderfourpt{
f(e^{i\tau_L-\epsilon},\phi) = \<\psi|e^{(i\tau_L-\epsilon) D + i \phi R}|\psi\> = \<\psi|e^{-\epsilon D/2} e^{i\tau_L D+i\phi R} e^{-\epsilon D/2}|\psi\>,
}
where $D$ is the Hamiltonian on the cylinder, $R$ is the generator of rotations in the $\phi$ direction and $|\psi\>\equiv   \cO(0,\pi)\cO(0,0)|0\>/\<\cO(0,\pi)\cO(0,0)\>$.  Since $D$ and $R$ are Hermitian, the Cauchy-Schwarz inequality implies
\eqn\cauchyschwartz{
|f(e^{i\tau_L-\epsilon},\phi)|\leq \<\psi|e^{-\epsilon D/2} e^{-\epsilon D/2}|\psi\>=f(e^{-\epsilon},0).
}
This implies that for $\epsilon>0$ we have a strictly convergent OPE expansion. Of course we could still have a divergence as $\epsilon \to 0$. 
We can derive a simple bound on the rate of divergence. 
The configuration $\tau_E=-\epsilon, \phi=0$ is now Euclidean, and we can use the crossed-channel OPE to compute the four-point function,
\eqn\crosschannel{
f(e^{-\epsilon},0) \sim {1 \over {(1-e^{-\epsilon})^{4\Delta} }} \sim {1 \over \epsilon^{4\Delta}} \qquad (\epsilon \ll 1).
}
Combining {\cauchyschwartz} and {\crosschannel}, we see that the correlator  is bounded by $\epsilon^{-4\Delta}$ when approached from the Euclidean direction $\tau_E=i\tau_L-\epsilon$.

More generally, we may be interested in a $d+2$ function of the form 
\eqn\fourpc{ 
 f(e^{\tau_E},  \{ \vec n_\alpha \}  ) =  \left\langle\cO(\tau_E,\vec n_n)\cO(\tau_E,\vec n_s)  \left[ \prod_{i=1}^{d} \cO(0, \vec n_i )  \right]  \right\rangle_{R\times S^{d-1}},
}
where $\vec n_i$ are points on the $S^{d-2}$ sphere at the equator of $S^{d-1}$ and $\vec n_{n,s}$ are on the north and south pole of the $S^{d-1}$. 
This configuration is characterized by several cross-ratios.  However, we can keep all the points on the sphere fixed and change only $\tau_E$. The OPE in the north+south channel is convergent as long as $\tau_E<0$. Furthermore, all points at $\tau_E = i \tau_L - \epsilon$ are at the boundary 
of this OPE convergence region. This includes the locus $\tau_L = - \pi$, where the special cross-ratio $x$ vanishes. 

We can derive a bound similar to \crosschannel\ for this correlator.\foot{The right-hand side of the bound is  $\epsilon^{- (d+2) \Delta } $.}
  In particular, it applies as we approach the singular locus $x=0$ from the Im$(x)>0$ direction.
However, this bound is not very constraining for bulk-point singularities that arise as coefficients in the 
$1/N^2$ expansion. As discussed in section~4.5, the singularity is smoothed out when $\hat x\sim \ell_{Pl}$. At this location, it does not violate the Cauchy-Schwarz inequality.

\subsec{Origin of the singularity from the dimensions of  operators in the OPE }

The OPE in Euclidean \OPEsMOREeucl\ and Lorentzian  \OPEsMORE\ signature  look very similar. It is instructive to understand why 
the Lorentzian one can develop a bulk-point singularity in the  gravity regime, while the Euclidean one remains finite (for generic values of the angle). 
  This question was addressed in \HeemskerkPN, and it will be useful to briefly review it.  Contact interactions in $AdS$ generate an anomalous dimension for double-trace operators $\cO_{n,l}\equiv \cO \pa^{2 n} \pa_{\mu_1}\dots\pa_{\mu_l} \cO$ and also a shift in their three-point coefficients,
\eqn\doubletrace{\eqalign{
\Delta_{n,\ell} &\equiv \dim\cO_{n,\ell} = 2\Delta + 2n + \ell + \gamma^{(1)}_{n,\ell} +\dots\cr
p_{n,\ell} &\equiv f_{\cO\cO\cO_{n,\ell}}^2 = p^{(0)}_{n,\ell} + p^{(1)}_{n,\ell} + \dots,
}}
where $p^{(0)}_{n,\ell}$ are the values in Mean Field Theory (see appendix~B) and $\gamma^{(1)}_{n,\ell}$ and $p^{(1)}_{n,\ell}$ are of order $1/N^2$.  

The four-point function has a  conformal block expansion
\eqn\mftconformalblockexpansion{
f(r,\phi) = 1+\sum_{n=0}^\oo\sum_{\ell=0,2,\dots} p_{n,\ell} g_{\Delta_{n,\ell},\ell}(r,\phi) + {\rm other\ operators},
}
where ``other operators" are single- and multi-trace operators that contribute at first and higher order in the $1/N^2$ expansion.
Under evolution by $-\pi$ in Lorentzian time, each block acquires an overall phase $g_{\Delta,\ell}(e^{-i\pi}r,\phi)=e^{-i\pi\Delta} g_{\Delta,\ell}(r,\phi)$
(for even $\ell $). 
Thus, the correlator becomes
\eqn\fnearlightup{
f(e^{-i\pi-\epsilon},\phi) = 1+\sum_{n=0}^\oo \sum_{\ell=0,2,\dots} p_{n,\ell} e^{-i\pi \Delta_{n,\ell}} g_{\Delta_{n,\ell},\ell}(e^{-\epsilon},\phi) + {\rm other\ operators}.
}
Let us denote the leading correction in the $1/N^2$ expansion by $\delta^{(1)} f$.  We have
\eqn\correctiontofnearlightup{\eqalign{
\delta^{(1)}f(e^{-i\pi - \epsilon},\phi)&= e^{-2\pi i \Delta}\delta^{(1)} f(e^{-\eps},\phi) \cr
&\quad+ e^{-2i\pi\Delta} \sum_{n=0}^\oo \sum_{\ell=0,2,\dots} \left(-i\pi \gamma^{(1)}_{n,\ell}\right) p^{(0)}_{n,\ell} g_{2\Delta + 2n + \ell,\ell}(e^{-\eps},\phi)\cr
&\quad+ {\rm single\,\hbox{-}trace}.
}}
In the first line, we have packaged together contributions $\delta^{(1)} f(e^{-\eps},\phi)$ that are present in Euclidean space. The second line contains extra double-trace terms that arise in Lorentzian signature from the expansion of $e^{-i\pi \gamma_{n,\ell}^{(1)}}$. The third line contains  single-trace terms that we will ignore for now.

The Euclidean terms cannot contribute to a singularity as $\epsilon \to 0$ for generic $\phi$ because no such singularity exists for Euclidean correlators (which have only OPE singularities).
Thus, a singularity as $\eps\to 0$ must come from the second set of terms in {\correctiontofnearlightup}, proportional to the anomalous dimensions $\gamma^{(1)}_{n,\ell}$.  Because double-trace dimensions are nearly equal to $2\Delta$ plus an even integer, these contributions sum up ``in phase," allowing us to pull out the overall factor of $e^{-2\pi i \Delta}$.

As an example, let us reproduce the singularity in $\phi^4$ theory from a sum over blocks.  When $\epsilon$ is small, the sum in {\correctiontofnearlightup} is dominated by terms with $n\epsilon$ of order 1.  The scalar blocks in this limit (see appendix~B) are given by
\eqn\correctlimitforblockspinzero{
g_{2\Delta + 2n,0}(e^{-\epsilon},\phi) \approx {2^{2-d\over 2} \sqrt{n} \over \sqrt \pi |\sin\phi|} \epsilon^{3-d \over 2} K_{d-3\over 2}(2n \epsilon),\qquad n\gg 1,\quad n\epsilon=O(1).
}
To compare to the variables in section~3.1, let us choose boundary points $X_i=(\cos\tau_i,\sin\tau_i,\cos\phi_i,\sin\phi_i)$ with $\tau_{1,2}=-\pi+i\epsilon$ and $\tau_{3,4}=0$ and the $\phi_i$ as before. (For convenience, we are gauge-fixing the rescaling of the $X_i$).
In the limit $\eps\to 0$, we have\foot{In deriving this relation, we must take care to reorder the operators so that the $k_a$ in \leftzeroeignevector\ are positive. The correct ordering depends on the sign of $\sin\phi$, leading to the absolute value.}
\eqn\epstox{\eqalign{
x &\approx 4i \eps|\sin\phi|,\qquad(\eps\ll 1),\cr
k_a &\approx 2|\sin\phi|, \qquad(\eps\ll 1,\ a=1,2,3,4)
}}

A ${\lambda \over 4!}\phi^4$ contact interaction in $AdS$ generates anomalous dimensions only for scalar double-trace operators at leading order in $1/N^2$.  For large $n$, these are given by \refs{\HeemskerkPN,\FitzpatrickZM}
\eqn\anomdimatlargen{
\gamma_{n,0}^{(1)} \approx {\lambda n^{d-3} \over 2^{2+d}\pi^{d/2}\Gamma(d/2)},\qquad n\gg 1.
}
Finally, the large-$n$ limit of the Mean Field Theory OPE coefficients is \FitzpatrickDM
\eqn\mftOPEatlargen{
p_{n,0}^{(0)} \approx { 2^{2+{3d\over 2}} n^{4\Delta - {3 d\over 2}} \pi \Gamma(d/2)
\over
\Gamma(\Delta)^2 \Gamma(1-{d\over 2} + \Delta)^2
},\qquad n\gg 1.
}
Plugging {\correctlimitforblockspinzero, \anomdimatlargen, and \mftOPEatlargen} into {\correctiontofnearlightup} and approximating $\sum_{n=0}^\oo \to \int_0^\oo dn$, we obtain precisely the Witten diagram integral \witdiagro\ with $n=2\omega|\sin\phi|$.  This gives the familiar singularity
\eqn\singularitycomparisonOPEWittenDiagram{
{\prod_a k_a^{\Delta-1} \over (-ix)^{4\Delta-3}} \propto {1 \over |\sin\phi|} {1 \over \epsilon^{4\Delta-3}},
}
with the correct coefficient. For a four-point bulk interaction with $m$ derivatives we get $\gamma^{(1)}_{n,\ell} \propto n^{ d-3 + m} $ instead of \anomdimatlargen, resulting in a singularity $\epsilon^{-(4\Delta-3+m)}$.

Note that at finite $N$, we must exponentiate $\gamma_{n,\ell}$ again. With a sufficiently chaotic spectrum, we expect that all phases average out 
in \fnearlightup. Thus, at finite $N$,  we do not expect an extra enhancement to the singularity in the four-point function 
beyond the one present for the individual conformal blocks. 
For a similar reason, we do not expect that the single trace terms in \correctiontofnearlightup\ will contribute to the singularity since 
such terms have the dimensions already in the exponent, even to leading order in the $1/N$ expansion. 

Exponentiation gives another way to understand the regime of validity of the Witten diagram computation in a bulk gravity theory.   The sum over blocks \correctiontofnearlightup\ is reliable as long as $\gamma_{n,\ell}\ll 1$ so that we can expand $e^{-i\pi \gamma_{n,\ell}}\sim 1-i\pi \gamma_{n,\ell}+\dots$.  The fixed-angle amplitude in gravity grows as ${\cal A}\propto s$, leading to an anomalous dimension $\gamma_{n,\ell}^{(1)}\sim G_N n^{d-1}$.  This gives the condition $n\ll G_N^{1-d}$ or $\epsilon\gg \ell_{Pl}$, in agreement with the discussion in sections 3 and 4.

The relation between a bulk Witten diagram as an integral over $\omega$ and the conformal block expansion is easy to understand.  Conformal blocks are eigenfunctions of the quadratic Casimir $C$ of the conformal group, acting on the two initial (or final) operators.  In the bulk, $C$ becomes the squared total momentum plus the Casimir of the Lorentz group.  At high energies, this is just $C=(\omega k_1n_1+\omega k_2 n_2)^2=16\omega^2 |\sin\phi|^2$.  The Casimir for an operator of dimension $\Delta_{n,0}$ is $4n^2$ for $n\gg 1$.  Thus, inserting $\delta(n-2\omega|\sin\phi|)$ into \witdiagro\ gives the contribution of a single conformal block.\foot{For more general interactions, we can also project onto a specific angular momentum block by picking out individual partial waves in the scattering process.}

\subsec{The Regge limit and the bound on chaos}

When $\eps\to 0$, $\phi\to \pi$ with $|\pi-\phi|/\eps<1$ held fixed (or equivalently $z,\bar z \to 1$ with ${1-z \over 1-\bar z}$ fixed), the physical picture changes.  This is the so-called Regge limit, which in the bulk is controlled by high energy, fixed impact parameter scattering \refs{\CornalbaXK,\CornalbaXM\CornalbaZB\CornalbaFS-\CornalbaQF}. Recently, this kinematical regime was analyzed in the context of chaos in \refs{\ShenkerCWA,\MaldacenaWAA}.
As illustrated in \zequalzbregions, the physics of the Regge limit is different from that of the bulk-point singularity, even though we have $z\to \bar z$ in both cases. Here we simply want to compare and contrast the two regimes. 

 Following  the notation of \MaldacenaWAA, we  consider a time-ordered flat-space correlation function
\eqn\chaoscorr{
F(t ) \equiv \la {\rm T}\{V(x_1)V(x_2)W(x_3)W(x_4) \}\ra,
}
where we restrict $x_i\in {\Bbb R}^{1,1} \subset {\Bbb R}^{1,d-1}$ as follows
\eqn\coordchaos{
x_1^{\pm}=\pm 1,
~~~
x_2^\pm = \mp 1,
~~~
x_3^\pm = \pm e^{\sigma \pm t'},
~~~
x_4^\pm = \mp e^{\sigma \pm t'},
}
where $t = t' + i \pi/2$. The advantage of the variable $t$ is that $F(t)$ is real for real $t$ \MaldacenaWAA. We are interested in the values of 
$F(t)$ in the strip $ |{\rm Im}(t)| < \pi/2$. The above correlator is at the upper boundary ${\rm Im}(t)=\pi/2-\varepsilon$.
The Regge limit corresponds to $t'\to \infty$.  In this limit, the $V(x_1)V(x_2)$ OPE is no longer valid.  However, note that after a boost, $V(x_1)$ and $W(x_4)$ can be placed at time $-\pi/2$ on the Lorentzian cylinder, while $V(x_2)$ and $W(x_3)$ are approaching time $+\pi/2$ on the Lorentzian cylinder. This is equivalent to the configuration \lorcylfourp\ with $\phi\to \pi$.  Hence, we can safely approach this limit using the $V(x_1)W(x_4)$ OPE.
Doing the OPE in this channel we get the variables 
\eqn\rhopm{ 
\rho  \sim  e^{ - 2 \pi i } \left( 1 + 4 i e^{ -  \sigma - t' \over 2}  \right) ,~~~~~~~~~~~~~\bar \rho \sim  1 + 4 i e^{ \sigma - t' \over 2},  
} 
for large $t'$. The $e^{ - 2 \pi i }$ phase factor indicates the path of analytic continuation to get to the Regge regime. Note that we have $|\rho|,|\bar\rho|<1$ for all $-{3\pi \over 2}<{\rm Im}(t)<{\pi \over 2}$, and in particular for real $t$.  Swapping $t \leftrightarrow \bar t$ corresponds to exchanging points $3$ and $4$.  Thus, the line ${\rm Im}(t)=-{\pi  \over 2}$, where $\rho,\bar\rho$ are both real and less than 1, corresponds to the performing the $V(x_1)W(x_3)$ OPE in the Lorentzian correlator \chaoscorr\ with real $t'$, which is also convergent.

One point that these observations make clear is the following. There is no singularity when $\sigma\to 0$ (so that $z\to \bar z$) in the Regge regime.  This is because there exists an OPE channel where $|\rho|, |\bar \rho| < 1$ when $\sigma=0$ (with $t$ finite).  From \zequalzbregions, it is not surprising that $z=\bar z$ in the Regge regime has different physics from $z=\bar z$ in the bulk-point regime.

We now  consider the correlator 
\eqn\boundonchaosmodel{
{F(t) \over F_d} = 1 - {1 \over N^2}\left(Ae^{(j-1)t}+\dots\right) + O\left({1\over N^4}\right),\qquad t\gg 1.
}
where the ``$\dots$" represents subleading terms at large $t$.  The chaos bound  \MaldacenaWAA\ states that   
\eqn\boundstatement{
j \leq 2 ,~~~~~~~~~{\rm and } ~~~~~ A\geq 0 .
}
The quantity $j$ is called the Regge intercept, see, e.g., \CostaCB.
In theories with gravity duals we have $j=2$, with the Regge limit controlled by graviton exchange. 
In weakly coupled gauge theories, $j$ is slightly bigger than one. In the exact theory, 
the correlator should go to zero for large $t$. 

By looking at the OPE, we can perform an analysis similar to the one done in section~6.2 and relate the growth of the correlator to 
the dimensions of double-trace operators obtained from a gravity computation in the bulk. 
  For example, a contact interaction with amplitude $i{\cal A}=\lambda s^k+O(s^{k-1})$ with $k\geq 2$ gives
\eqn\contactinteractioncontribution{
\delta^{(1)} {F(t) \over F_d} = \lambda e^{(k-1)t}f(\sigma),\qquad t\gg 1,
}
where $\delta^{(1)}$ denotes the leading correction in the large $N$ expansion, and $f(\sigma)$ is a positive function of $\sigma$.   Assuming this contact term dominates at high energies, \boundstatement\ implies 
$k\leq 2$.  For the case $k=2$, corresponding to an interaction $\lambda(\partial\phi)^4$, we also find $\lambda<0$. This constraint was obtained in \AdamsSV\ for field theories in flat space, and more recently in \HartmanLFA\ for $AdS$.

In terms of anomalous dimensions of double-trace operators, the condition $j\leq 2$ is equivalent to the statement that $\gamma_{n,\ell}^{(1)}$ can grow no faster than $n^{d+1}$, when a finite number of spins contribute:
\eqn\boundchaos{
\lim_{n \to \infty,\ \ell \leq \ell_{{\rm max}} } \gamma_{n,\ell}^{(1)} \leq O(n^{d+1}) .
}
(And furthermore, the coefficient of $n^{d+1}$ must be negative.)
This uses the large $\Delta$ limit of the blocks at fixed $\ell$.
One can repeat the analogous exercise for corrections that correspond to exchange of particles in the bulk. The relevant limit of the blocks was considered in \refs{\CornalbaXK,\CornalbaXM\CornalbaZB\CornalbaFS-\CornalbaQF}. The result is the following bound:
\eqn\boundchaosB{
\lim_{n \to \infty,\ {\ell \over n}\,{\rm fixed} } \gamma^{(1)}_{n, \ell} \leq O(n^{2}) .
}
This bound is equivalent to the statement that the scattering phase  $\delta(s,b)$ does not grow faster than $s$ \CamanhoAPA .

 \newsec{ Singularities in 1+1-dimensional theories} 
        
In this section we continue to pursue the strategy of bounding Lorentzian correlators by Euclidean correlators, now making use of the special structure present in two dimensions. We first describe a quantization of the theory that makes manifest certain positivity properties of Virasoro conformal blocks. Working in this quantization, we prove the absence of bulk-point singularities in the four-point function nonperturbatively.

Before presenting the general proof, let us first remark on a simpler case. 
In rational CFTs it is easy to see that we will not get a singularity.  The reason is that the correlator is a sum of a 
{\it finite} number of products of holomorphic times antiholomorphic functions. Such a product can never give rise to a singularity at $z = \bar z$, which is a 
non-holomorphic condition. Now let us consider the general case.

In two dimensions, it is standard to define
\eqn\fourpointf{\eqalign{
{\cal F}(z, \bar z) &= \la {\cal O} (0) {\cal O} (z, \bar z) {\cal O} (1) {\cal O} (\infty)  \ra = \lim_{x_4\to \oo} x_4^{2 \Delta}\la {\cal O} (0) {\cal O} (z,\bar z) {\cal O} (1) {\cal O} (x_4)  \ra, \cr
f(z, \bar z) &= z^{\Delta} \bar z^{\Delta} {\cal F}(z, \bar z),
}}
where ${\cal F}(z,\bar z)$ can be expanded as a sum of Virasoro conformal blocks,
\eqn\confblocks{
{\cal F}(z, \bar z) = \sum_{h, \bar h} f_{\cO\cO\cO_{ h,\bar h}}^2 {\cal V}_{h}(z) {\cal V}_{\bar h}(\bar z).
}
The ${\cal V}_h(z)$ are complicated, but fortunately we will not need their detailed structure to obtain bounds on the Lorentzian cylinder. We can take a simpler route by quantizing the theory in the right way.

Our quantization will yield an expansion for the four-point function in the elliptic nome $q$, defined as
\eqn\definitionq{\eqalign{
q &= e^{i \pi \tau}={z \over 16}+\dots,\cr
\tau &= i {K(1-z) \over K(z)},\cr
K(z) &={1 \over 2}\int_0^1 {dt \over \sqrt{t(1-t)(1-zt)}},\cr
z &= {\theta_2(q)^4 \over \theta_3(q)^4},
}}
where $K$ is an elliptic integral of the first kind.  The Virasoro block ${\cal V}_ h(z)$ has a natural expression in terms of $q$, obtained by Zamolodchikov in \ZamolodchikovXX, 
\eqn\zamolodchikovblock{
{\cal V}_ h(z) = (16 q)^{ h-{c-1 \over 24}}(z(1-z))^{{c-1 \over 24}-\Delta}\theta_3(q)^{{c-1 \over 2}-8\Delta} H( h, q),
}
where $H( h,q)$ can be determined recursively 
\refs{\ZamolodchikovXX\ZamolodchikovIE-\ChangJTA}.  In Zamolodchikov's analysis, the prefactors above come from a semiclassical Liouville theory computation of the large-$ h$ limit of ${\cal V}_ h$.  We will give an alternative understanding of these factors using an appropriate quantization of the CFT.

First, let us describe the geometry underlying the $q$-variable.  The quantity $\tau$ is the modulus of a torus given by a double-cover of the Riemann sphere branched at $0,z,1,$ and $\oo$.  This torus is described by the equation
\eqn\ellipticeqn{
y^2 = x(z-x)(1-x),
}
where $x$ is a coordinate on the base ${\Bbb P}^1$.

What does a torus have to do with a four-point function on the Riemann sphere?  The answer is that the Riemann sphere can be thought of as the quotient of the torus by $\Bbb Z_2$ covering transformations $y \mapsto -y$.  That is, ${\Bbb P}^1 \cong T^2/{\Bbb Z}_2$.  Via this quotient, the sphere inherits a metric that is flat except for four conical defects at the fixed-points of $\Bbb Z_2$.  We will refer to the sphere with this metric as the ``pillow."

\ifig\TorusQuotient{The quotient $T^2/{\Bbb Z}_2$ gives the pillow, which has the topology of a sphere and a metric that is flat except at four conical defects.  On the left, we show the positions of the ${\Bbb Z}_2$ fixed points, which become conical defects on the pillow.  (The $u$-coordinate is defined below.)  The shaded region is a fundamental domain of ${\Bbb Z}_2$.  On the right, we show the result of the quotient and indicate the former $A$ and $B$ cycles of the torus, which become contractible $S^1$'s on the pillow separating pairs of conical defects.}
 {\epsfxsize5.6in \epsfbox{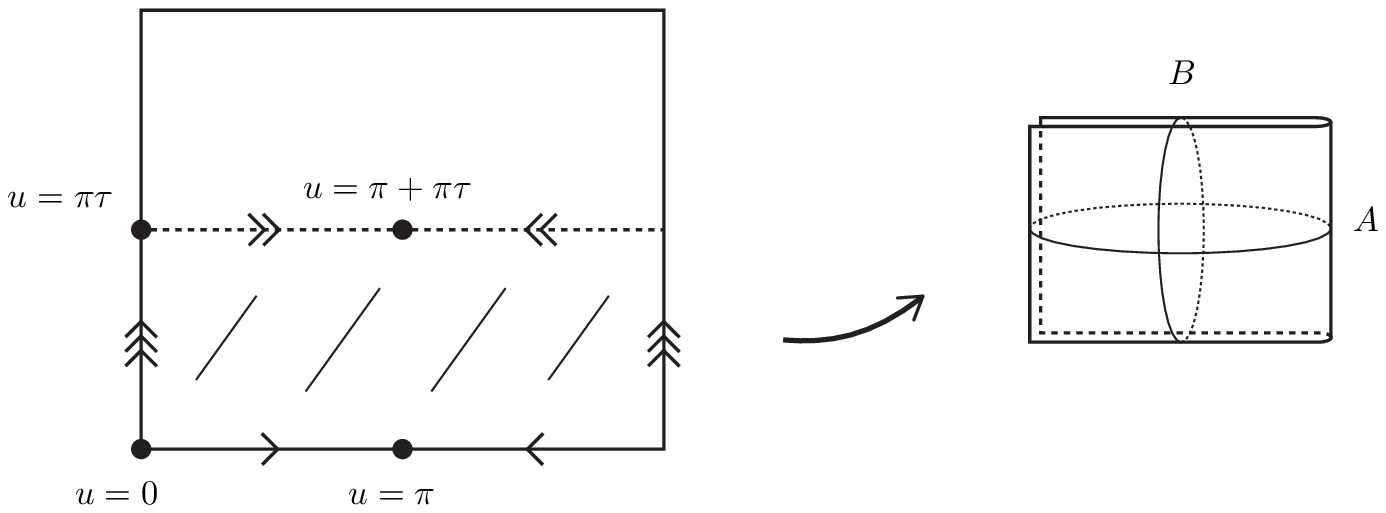} } 

Let us describe the quotient $T^2/{\Bbb Z}_2$ more explicitly, since it will be useful in the discussion that follows.  On the left-hand side of \TorusQuotient, we show the torus in the uniform coordinate $u$, where it is a rectangle with opposite sides identified.  In the $u$-coordinate, the ${\Bbb Z}_2$ acts as $u\mapsto -u$.  A cycle around $0$ and $z$ in the $x$-plane maps to the $A$-cycle of the torus (the horizontal direction), while a cycle encircling $z$ and $1$ maps to the $B$-cycle (the vertical direction).  Now, let us cut the torus into two sheets along a pair of $A$-cycles.  Taking one of the sheets (the bottom half of the torus in \TorusQuotient), we recover the sphere by re-gluing the cuts together.  The former $A$-cycle is now contractible --- it corresponds to an $S^1$ that separates pairs of conical defects.

Let us now return to our CFT four-point function ${\cal F}(z,\bar z)$.  The operators ${\cal O}$ lived at the fixed-points $0,z,1,\infty$ of this $\Bbb Z_2$ quotient, so on the pillow we have one operator at each conical defect.  The key idea to obtain a $q$-expansion is to quantize the CFT in the pillow geometry, with the former $A$-cycle of the torus as a spatial slice.  Let us normalize the $A$-cycle to have length $2\pi$, so the states associated with this spatial slice have the usual left- and right-moving Hamiltonians $L_0 - {c \over 24}, \bar L_0-{c\over 24}$ of the CFT on a cylinder.  These states evolve for half the length of the $B$-cycle because the pillow comes from half of the torus.  Consequently, the correlator will be an expansion in $e^{i\pi\tau( L_0-{c\over 24})-i\pi\bar\tau(\bar L_0-{c\over 24})}=q^{ L_0-{c \over 24}}\bar q^{\bar L_0-{c\over 24}}$.

Let's explore this idea in more detail.  The uniformizing coordinate on the torus $u$ is defined by
\eqn\flattoruscoordinate{
du = {1 \over \theta_3(q)^2} {dx \over y},
}
where $y$ satisfies {\ellipticeqn} and the prefactor $\theta_3(q)^{-2}=2\pi/(4K(z))$ comes from normalizing the $A$-cycle to have length $2\pi$. 
The ${\Bbb Z}_2$ acts as $u \to -u$. In the $u$-coordinate, the operators sit at the fixed points 
 $u_1=0, u_2=\pi, u_3=\pi(\tau+1),$ and $u_4=\pi \tau$.  We would like to compute the four-point correlator $\<\cO\cO\cO\cO\>$ by first performing a Weyl transformation to the uniform metric,
\eqn\weyltransform{
dx\,d\bar x \to e^{2\omega} dx\,d\bar x = du\,d\bar u.
}
Under this transformation, the correlator gets contributions both from the Weyl anomaly and from local rescaling near the operator insertions at $0,z,1$, $\oo$.  Both of these factors are infinite and must be regularized appropriately (see Appendix C for details), giving
\eqn\hindsight{\eqalign{
{\cal F}(z, \bar z) &= \Lambda(z)\Lambda(\bar z) g(q , \bar q),\cr
\Lambda(z) &\equiv \theta_3(q)^{{c\over 2}-8\Delta}(z(1-z))^{{c\over 24}-\Delta},\cr
g(q,\bar q) &\equiv \<\cO(u=0)\cO(u=\pi)\cO(u=\pi(\tau+1))\cO(u=\pi\tau)\>_{\rm pillow},
}}
where $g(q,\bar q)$ is an appropriately regularized four-point function in the pillow geometry with operators at the conical defects.\foot{This definition of $g(q,\bar q)$ is schematic because of  the need for regularization.  We define $g(q,\bar q)$ precisely in Appendix C.}  Note that the rescaling factor $\Lambda(z)$ gives precisely the $c$- and $\Delta$-dependent prefactors in Zamolodchikov's expression \zamolodchikovblock.

As discussed above, by quantizing the theory with the $A$-cycle as the spatial slice, we can write $g(q,\bar q)$ as a sum over states on the circle,
\eqn\gassumoverstates{\eqalign{
g(q,\bar q) &= \<\psi''|q^{L_0-{c \over 24}} \bar q^{\bar L_0 - {c\over 24}}|\psi''\>, \cr 
|\psi''\> &\equiv |\cO(u=0)\cO(u=\pi)\>_{\rm pillow},
}}
where $|\psi''\>$ is defined by cutting the path integral along an $A$-cycle just above the defects at $u=0,\pi$.\foot{The state $|\psi''\>$ is non-normalizable, but it can be made normalizable by a small amount of evolution in Euclidean time. The same is true of a boundary state or any state created by a local operator.}

Equivalently, we can write
\eqn\confblocksg{
g(q , \bar q) =  \sum_{ h, \bar  h} f_{\cO\cO\cO_{ h,\bar h}}^2 \tilde {\cal V}_{ h}(q) \tilde {\cal V}_{\bar  h}(\bar q),
}
where the modified blocks $\tilde {\cal V}_ h(q)$ are given by
\eqn\zamoblock{
\tilde {\cal V}_{ h}(q) = \Lambda(z)^{-1} {\cal V}_{ h}(z) = (16 q)^{ h - {c \over 24}} \prod_{k=1}^{\infty} (1 - q^{2 k})^{ - {1 \over 2}} H( h,q).
}
By interpreting $\tilde {\cal V}(q)$ as a sum over states on the pillow, it follows that $\tilde {\cal V}_{ h}(q)$ has an expansion with nonnegative coefficients whenever $c, h, \Delta$ have values appropriate for a unitary theory:
\eqn\exppos{
\tilde {\cal V}_{ h}(q) = \sum_{n=0}^{\infty} a_n q^{ h + n - {c \over 24}},~~~ a_n \geq 0.
}
This fact is non-obvious from the recursive definition of $H( h, q)$ \ZamolodchikovXX.

Considering our four-point function in the pillow geometry makes crossing symmetry look extremely similar to modular invariance of the torus partition function --- it is simply the statement that the partition function is unchanged under a $90^\circ$ rotation of the (Euclidean) spacetime manifold.  Instead of quantizing the theory with the $A$-cycle as a spatial slice, we could instead choose the $B$-cycle, which would lead to an expansion in the image of $q$ under a modular $S$-transformation.  More precisely, crossing symmetry of the four-point function
\eqn\crossingsymmetry{
{\cal F}(z,\bar z) = {\cal F}(1-z,1-\bar z)
}
implies that $g(q,\bar q)$ is a (non-holomorphic) modular form,\foot{Formulated in terms of $g(q,\bar q)$, the four-point function bootstrap \refs{\PolyakovGS\FerraraYT\MackJR\BelavinVU-\RattazziPE} is almost identical to the modular bootstrap \refs{\MooreUZ,\HellermanBU}.  The key differences are that we must allow for non-integral coefficients in the $q,\bar q$ expansion and that the vacuum character gets replaced by the conformal block for the identity operator.}
\eqn\modularinvariance{\eqalign{
 g(q , \bar q) &= \left(\sqrt{\tau \bar \tau} \right)^{{c \over 2} - 8 \Delta} g(\tilde q , \bar{\tilde{q}}),\cr
\tilde q &= e^{i\pi \tilde\tau} = e^{-i\pi/\tau}.
}}

\subsec{All Lorentzian singularities in $d=2$}

\ifig\Rhotoq{The map $\rho\mapsto q$ takes the interior of the unit circle to the shaded region on the right.  In particular, the boundary of the unit $\rho$-circle maps inside the unit $q$-circle, except for $\rho=\pm 1$, which map to $q=\pm 1$.}
 {\epsfxsize6.0in \epsfbox{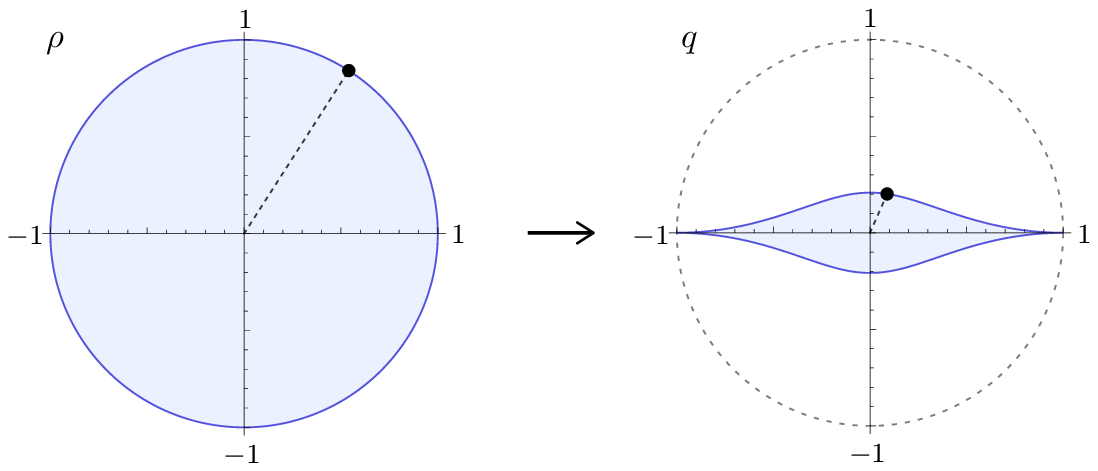} } 

On the Lorentzian cylinder, $\rho,\bar\rho$ take values on the unit circle.  In $q,\bar q$-coordinates, the unit $\rho$-circle gets mapped to the  shape shown in \Rhotoq\ --- {\it inside} the unit $q$-circle (aside from $q=\rho=\pm 1$).  The correlator $g(q,\bar q)$ is finite here, since it is given by a series in $q,\bar q$ with positive coefficients, and this series converges for real $q\in[0,1)$.  It follows that the four-point function is completely finite on the Lorentzian cylinder (aside from when $\rho,\bar\rho=\pm 1$), and in particular there is no bulk-point singularity.

\ifig\Sigmavsphi{The value of $\sigma(\phi)=-\log|q|$ for angles $\phi\in[0,2\pi]$.  It is positive everywhere apart from $\phi\in \pi \Bbb Z$.}
 {\epsfxsize7.0in \epsfbox{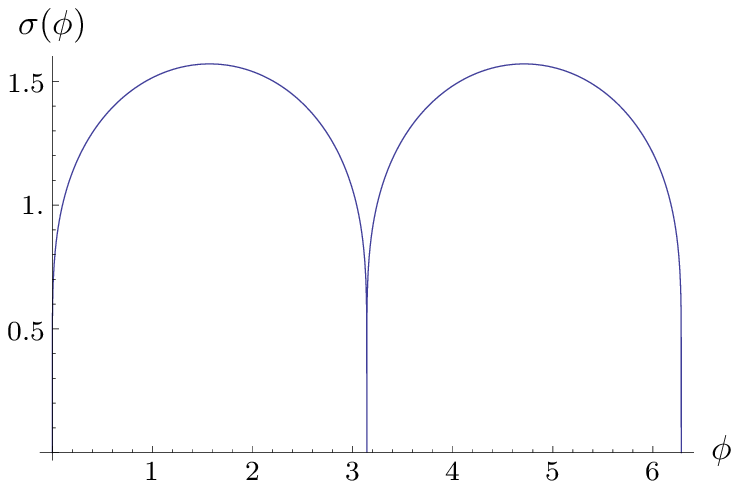} } 

Let us see this more explicitly. On the Lorentzian cylinder, our four-point function becomes
\eqn\confblocksgOPELor{
g(q,\bar q) = g(e^{- \sigma(\tau_L+\phi) + i \theta(\tau_L+\phi)}, e^{- \sigma(\tau_L-\phi) + i \theta(\tau_L-\phi)}), 
}
where
\eqn\sigmatheta{\eqalign{
\log q\left(z=\csc^2(\phi/2)\right) \equiv -\sigma(\phi)+i\theta(\phi).
}}
We plot the function $\sigma(\phi)$ in \Sigmavsphi.  Importantly for us, it is positive everywhere aside from $\phi=0,\pi$.

We can now repeat the argument of the previous section, this time using $q$-quantization. Starting from {\gassumoverstates}, the Cauchy-Schwarz inequality implies
\eqn\cauchyschwartzq{\eqalign{
|g(q,\bar q)| &\leq \<\psi''| |q|^{ L_0-{c\over 24}}|\bar q|^{ \bar L_0-{c\over 24}}|\psi''\>
= g(e^{-\sigma(\tau_L+\phi)}, e^{-\sigma(\tau_L-\phi)}) \leq g(e^{- \sigma_{min}}, e^{- \sigma_{min}}), \cr
\sigma_{min} &= {\rm min} [\sigma(\tau_L+\phi), \sigma(\tau_L - \phi)].
}}
Thus, we have bounded the Lorentzian correlator by the Euclidean one evaluated at $q = \bar q = e^{- \sigma_{min}}$. Note that
\eqn\zamomapprop{\eqalign{
\sigma(n\pi \pm \phi) &=  \sigma(\phi),\qquad n\in {\Bbb Z}
}}
so that the fundamental domain for $\sigma$ is $[0, {\pi \over 2}]$. The Euclidean correlation function is only singular when $\sigma_{min} = 0$. In this way we just proved that the only singularities of the Lorentzian four-point correlation function are light-cone singularities, which occur at 
\eqn\lightcone{
|\tau_{L}| = |\pi n \pm \phi|.
}

The bulk-point configuration, which corresponds to $\tau_{L} = \pi$ and generic $\phi$, is non-singular. More precisely,
\eqn\boundong{
|g(q,\bar q)|_{\rm bulk-point} \leq g(e^{-\sigma(\phi)},e^{-\sigma(\phi)}).
}
The latter is clearly finite for $\phi\neq 0,\pi$, since it is strictly increasing as $\sigma(\phi)\to 0$, and approaches the (finite) value dictated by crossing symmetry when $\sigma(\phi)$ is sufficiently small but nonzero.  Re-expressing {\boundong} in terms of the four-point function $f(z,\bar z)$, we have

\eqn\boundfinal{\eqalign{
|f_{\rm bulk-point} (\phi)| &\leq \left|{z_L^\Delta \Lambda(z_L) \over z_E^\Delta \Lambda(z_E)}\right|^2 f(z_E, z_E),\cr
z_L &= \csc^2\left({\phi \over 2}\right),\cr
z_E &= {\theta_2(e^{-\sigma(\phi)})^4 \over \theta_3(e^{-\sigma(\phi)})^4}.\cr
}}
As an example, when $\phi = {\pi \over 2}$, {\boundfinal} reads
\eqn\result{
|f_{\rm bulk-point} (\pi /2 )| \leq  2^{c / 8 - 4 \Delta}f \left(12\sqrt 2 - 16 , 12\sqrt 2 - 16 \right).
}
Note that $12\sqrt 2 - 16\approx 0.97$, so we expect the correlator on the right-hand side to be well-approximated by the unit operator in the other channel.  In particular, the $c$-dependence of the bound \result\ comes primarily from the prefactor $2^{c/8}$.  The fact that the bound grows faster than any power of $c$ is consistent, for any $\Delta$, with the expectation that a gravity correlator grows like $f_{\rm bulk-point}^{\rm gravity}(\phi)\sim c^{4\Delta-2}$ as $c\to \oo$ (\bumpcartoon).\foot{It may be possible to prove a stronger bound by using the full structure of Virasoro blocks, together with crossing symmetry, instead of just the $q$-expansion.}

\subsec{Other analytic continuations}

Moving in the time direction on the Lorentzian cylinder corresponds to repeatedly circling the origin $\rho,\bar \rho=0$.  It is also interesting to consider analytic continuations around the other singular points $\rho=\pm 1$ (equivalently $z=1,\infty$).  In Lorentzian signature, such continuations can be interpreted in terms of two operators crossing each others' light cones, and are needed to calculate non-time-ordered correlators in various quantizations of the theory.  We will see that arbitrary analytic continuations of this type correspond to moving around {\it inside} the $q,\bar q$ unit discs, so that the OPE expansion \gassumoverstates\ remains convergent.

This fact is easiest to understand in terms of the modulus $\tau$.  First note that analytic continuation around $z=0$ corresponds to the ${\rm PSL}(2,\Bbb Z)$ transformation $T^2:\tau\mapsto \tau+2$.  Since crossing symmetry $z\leftrightarrow 1-z$ is a modular $S$-transformation, continuation around $z=1$ corresponds to $S T^2 S:\tau \mapsto {\tau \over 1-2\tau}$.  We do not need to separately consider the cycle around $z=\infty$, since it is linearly dependent with cycles around $z=0,1$. Together, $T^2$ and $ST^2 S$ generate the principal congruence subgroup $\bar \Gamma(2)\subset {\rm PSL}(2,\Bbb Z)$, which clearly preserves the upper half-plane and hence the unit $q$-disc.  Succinctly, $\tau$ provides a uniformization of the universal cover of the three-punctured sphere.

\ifig\analyticcontinuationdomains{The map $z\to q$ takes the universal cover of the sphere with punctures at $z=0,1,\infty$ to the interior of the unit $q$-disc.  On the left, we show paths between the punctures in different colors.  We imagine drawing these paths on every sheet of the universal cover.  On the right, we show the images of these paths in the unit $q$-disc.  Analytic continuation around punctures on the left corresponds to moving around inside the unit $q$-disc on the right.  A dense set of points on the boundary of the $q$-disc corresponds to approaching a puncture on some sheet of the universal cover.}
 {\epsfxsize5.5in \epsfbox{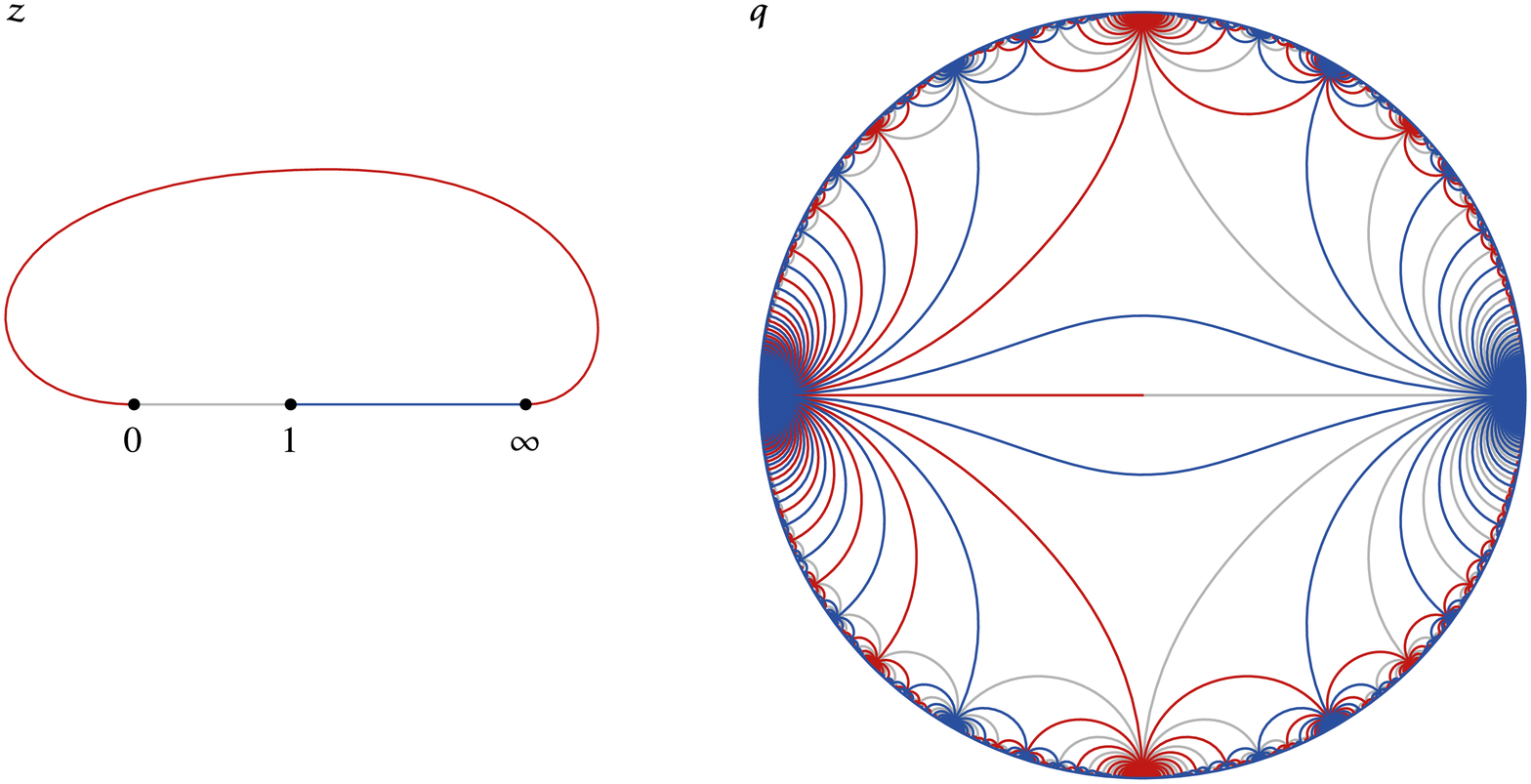} } 

We can now map out all possible analytic continuations of our four-point function as follows.  In the $z$-plane, let us draw cuts between each pair of branch points, $\{\infty,0\}$, $\{0,1\}$, $\{1,\infty\}$, colored red, gray, and blue, respectively.  Only two cuts are necessary for keeping track of the branch structure, but we include all three so as not to break the symmetry between $0,1,\infty$.  Encircling a branch point thus corresponds to crossing two cuts of different colors.  In the $q$-plane, the cut from $1$ to $\infty$ maps to the boundary of the shaded region in \Rhotoq\ (this locus corresponds to the Lorentzian cylinder configuration discussed in section~6), while the cuts between $\{z=\infty,z=0\}$ and $\{z=0,z=1\}$ map to segments on the real line between $\{q=-1,q=0\}$ and $\{q=0,q=1\}$ respectively.  Now acting with $\bar\Gamma(2)$ on $\tau$, we can produce the images of these cuts under continuation around the different branch points.  The result is depicted in \analyticcontinuationdomains.  

Every region accessible by analytic continuation has spikes that touch the boundary of the unit $q$-disk.  These spikes correspond to light-cone singularities as $z\to 0,1,\infty$ on some sheet of the branched-cover of the three-punctured sphere.   (These are left-moving light-cone singularities, since we are only discussing the holomorphic variable $z$.  By additionally analytically continuing in $\bar z$, we can also explore right-moving light cones as well as combinations of both right- and left-moving light cones.)   In fact, a dense set of points on the unit $q$-circle  correspond to some light-cone singularity after analytic continuation.  They are arranged in a fractal pattern at rational angles.  Because of this fractal of singularities at $|q|=1$, we expect it should not be possible to analytically continue the four-point function outside the unit $q$-circle.\foot{A simple example of a function that cannot be analytically continued outside the unit $q$-circle is $\sum_{n\geq 0} q^{n!}$, which is super-exponentially convergent for $|q|<1$, but divergent on the unit circle at every rational angle.}  

It is remarkable that, using $q$-quantization, a single OPE expansion is sufficient to cover every possible analytic continuation of the four-point function $g(q,\bar q)$.  By contrast, in higher dimensions where only the $\rho$-variable is available (without assuming additional symmetries), the OPE has only a finite radius of convergence.\foot{This radius of convergence can be extended by using additional information about the spectrum and OPE coefficients to partially (or completely) resum the expansion, as we did with Virasoro symmetry in this section.}

   \newsec{Conclusions}

In this paper we have analyzed some Lorentzian singularities of correlators. 
We showed that singularities of weakly coupled local quantum field theories are at the
location of Landau diagrams. These Landau diagrams are a purely geometric construction that is 
theory independent. At each order in perturbation theory, there is a finite number of possible 
diagrams. It would be interesting to derive general formulas for the location of these
singularities, since this could have useful applications for symbology. Of course, these remarks also
hold for the usual momentum-space Landau diagram locations. 

 We considered particular $(d+2)$-point correlation functions and argued that they have singularities (perturbatively in the $1/N$ expansion) at a 
 codimension-one hypersurface in the space of cross-ratios  given by $\det X_a^I=0$. This singularity arises from a particular point in the bulk. 
 Indeed, it is a signature of a local bulk theory,
 acting as a microscope for that bulk point.  In principle, one can imagine defining 
 bulk correlators of $n$ points by taking $n(d+2)$ points and grouping them in $n$ groups of $d+2$ so that
 we approach the bulk-point singularity in each group, associated to $n$ different bulk points. 
 In this limit, the correlator will be related to the insertion of
 $n$ operators in the bulk after we factor out pieces from each singularity. This is only a sketch since we would also 
 have to subtract contributions from graviton exchanges between the various bulk null lines. 
 
 We showed that in 1+1 or 2+1 dimensions, these bulk-point singularities 
 cannot be reproduced by weakly-coupled theories. It would be nice to see whether this is true in 
 3+1 dimensions. We suspect that it should be true  for generic configurations with zero
 determinant, but the method we used in lower dimensions was no longer applicable, since there are special configurations
 that do lead to such a singularity (see Appendix C).

 After reviewing and slightly extending the discussion of \OkudaYM ,  we  argued that finite $\alpha'$ effects should remove the singularity
 from planar correlators. It is interesting that the emergence of bulk-point singularities is something that could be 
 seen directly by looking at planar correlators, since these might be computed using integrability in the not so distant future! 
  
 We noted that instanton effects, curiously, both at weak and strong coupling, give rise to a bulk-point singularity. This is due to a couple of  simple facts: the moduli space being $AdS$ and the correlators factorizing in the presence of an instanton. 
 In principle, this instanton discussion is irrelevant for the emergence of the singularity in bulk perturbation theory. However, it is 
 tempting to imagine that there could be an underlying mechanism that uses a similar idea. As a vague idea, one would suggest ``fractional instantons''
 whose action would be divided by $N$ and thus important at strong coupling. Such configurations were discussed in, e.g., \PoppitzNZ. 
 More physically, one would like to argue that the sphere diagram in planar gauge theory has some zero modes corresponding to conformal transformations
 and nothing else. When we attach the external lines, we get a picture similar to the instanton discussion. It would be nice to make these ideas concrete.  Note that in the twistor string theory, the spacetime interactions also arise from D-instanton contributions \WittenNN.

The picture for parton evolution at strong coupling is that it is very rapid, with momentum becoming rapidly 
spread over very many low energy partons, which fill the whole spatial region
 within  the light cone of the operator insertion
\HattaKN, see also  \RobertsISA. The bulk-point singularity arises when all these partons interact coherently, 
each carrying an infinitesimal amount of momentum. In this case we are not
 obeying the Landau rules, which hold only 
in perturbation theory, where parton evolution only leads to splitting into a finite number of partons. 
 
 We have noted that at finite $G_N$, we do not expect bulk-point singularities. In 1+1 dimensions, we proved this using the
 full power of the conformal group. In that case, the only true singularities are light-cone singularities.

 It would also be nice to relate the emergence of bulk-point singularities to the spectrum of the theory. 
 In other words, one expects \HeemskerkPN\ that as the dimension of the lightest single trace higher spin ($S>2$) operator, $\Delta_*$, becomes
 large, then the theory should become local. 
 The authors of \HeemskerkPN\ showed that the only solutions to crossing symmetry in this situation correspond to local-like interactions, in 
 the sense that they can be described by bulk interactions of the form $ \lambda_n \phi^2  \partial^{ 2 n} \phi^2$, with various contractions of derivatives. 
 However, we expect more to be true.  We expect
  that the coefficient of such interactions should be suppressed for
   $n>1$ as $ |\lambda_n | \leq \lambda_*/\Delta_*^{2n } $. We have been unable to prove this conjecture.\foot{For an argument for 
   the graviton three-point function see \CamanhoAPA .}
   We propose  a corresponding conjecture in flat-space physics. Namely we consider a  tree-level amplitude (containing only poles)   that respects causality 
   so that it grows less rapidly than $s^2$ for large $s$. When we perform a low energy expansion of the amplitude, we will get terms that are polynomial 
   in the Mandelstam variables, $s$, $t$, $u$.  
   Then we expect that higher derivative corrections to the amplitude should be suppressed 
    by  the inverse mass of new particles. In other words, a term whose amplitude goes as ${\cal A} \sim s^{ 2 + n}$ in
   the large $s$ (fixed $t$) region should be suppressed by $ 1/M_*^{ 2 n} $, where $M_*$ is the mass of the lightest higher spin particle. 
   We prove a weak version of this flat space conjecture in appendix~A, using a slight variation of the method in \AdamsSV. 
   It would be nice to prove a stronger version of the flat space conjecture. 
   Mellin space looks like the best tool  to study these issues, since the Mellin amplitudes have analytic properties similar to string tree-level amplitudes \refs{\PenedonesUE,\MackMI,\FitzpatrickIA}.
    The bulk-point singularity arises from a Mellin amplitude that is polynomial in $s$ and $t$. A true local theory governed by Einstein gravity  
   would have higher order polynomials suppressed by inverse powers of $1/\Delta_*$. 
   It seems that a proof of these statements (or a corrected version of them) should be feasible.     

\bigbreak\bigskip           
\noindent {\bf Acknowledgements } 
\par\nobreak\medskip\nobreak

We thank N.~Arkani-Hamed, B.~Burrington,  T.~Hartman, D.~Kosower,  J.~Penedones, M.~Spradlin, D.~Stanford, G.~Turiaci, P.~Vieira, I.~Zadeh  and A.~Zamolodchikov for discussions. J.M. and D.S.D. are supported in part by U.S. Department of Energy grant
DE-SC0009988.  D.S.D. is also supported by a William D. Loughlin Membership at the Institute for Advanced Study. A.Z. is supported in part by U.S. Department of Energy grant DE-SC0007870.

\appendix{A}{Bounds on higher derivative interactions}

Let us assume we have an amplitude $A(s,t)$ that is meromorphic and with Regge behavior at infinity, 

\eqn\higheng{ 
|A(s,t) | \leq |s|^2 ,~~~~~~~s ~~{\rm large} ,~~~~~t\leq  0 
}
 for large $s$ (in any direction of the complex plane) and fixed $ t\leq 0$ (negative $t$ is spacelike $t$). 
The rationale for imposing this condition is that we want the amplitude to respect causality  in both the $u$ and $s$ channels. 
Let us also assume that the first massive state appears at $M_s$. 

We can now apply an argument similar to the one in \AdamsSV . 
First let us do exactly what they do. 
Namely, we consider $\tilde A(s,t) = A(s,t)  - $poles, where the poles are the low energy poles of particles with spin less than two. 
The subtracted amplitude $\tilde A(s,t)$ continues to obey  the high energy behavior \higheng . 
Because we subtracted the  poles (including the one at $t=0$), $\tilde A$ has a power series expansion around 
$s,~t=0$. Let us first set  
 $t=0$ in $\tilde A $. 

We now consider the integral 
\eqn\contint{ 
 c_{ 2 n} =  \oint { ds \over 2 \pi i } {  \tilde A(s,t=0) \over s^{ 2 n+1 } } = { 2 \over \pi } \int_{cuts, s>0} d s { s \sigma(s) \over s^{ 2 n+1}  }.
}
We can neglect the contribution at infinity as long as  $n>1$. 
For a meromorphic function, the sum over cuts is simply a sum over delta functions $i \pi \delta( s - M_k^2)$. In other words, $\sigma(s)$ contains such 
$\delta $ functions. 
Now we therefore end up with an expression of the rough form 
\eqn\roughf{ 
c_{2n} = { 2 \over \pi } \int^\infty_{M_{min}^2}  { d s \over s} { \sigma(s) \over s^{2 n -1} }, 
}
where the integral is really a sum over $\delta $ functions. It would be nice to assume that the right-hand side was finite for 
$n=1$. But this does not follow from our assumptions. In fact, it could be divergent. On the other hand it would be consistent with 
our assumptions to say that the integral on the right-hand side is convergent for $n =1 + \epsilon$. This implies
\eqn\boudnoth{ 
c_{2n} \leq      { c_{ 2 + 2 \epsilon}   \over (M_{min}^2)^{ 2 n -2 - 2 \epsilon} },
}
where $c_{2 + 2 \epsilon} $ is defined to be the right hand side of \roughf\ for $n=1+\epsilon$. 
 For the case of a theory where the integral for $n=1$ is finite, as was 
 considered in \AdamsSV, one can set $\epsilon=0$ in 
 \boudnoth . We can also saturate the bound by classically integrating out a massive scalar field of mass $M_{min}$.

All this discussion was for $t=0$. We can now consider non-zero $t$. In this case, we obtain a similar expression with 
\eqn\exprdu{ 
c_m(t) = \oint { ds \over 2 \pi i } {  \tilde A(s,t ) \over s^{ m+1 } } = \sum_{k}   { 1 \over (M^2_k)^{ m +1} } { A_{12, M} A_{M,34} } + 
 (-1)^m \sum_{k}   { 1 \over (M^2_k)^{ m +1} } { A_{14, M} A_{M,32} }, 
}
where the first sum contains the poles in the $s$ channel and the second contains the poles in the $u$ channel. 
We also have a sum over spins implicit in these expressions:
\eqn\sgar{ 
 \sum_{\rm spins} { A_{12, M} A_{M,34} }  = \hat C_l^{{d-3 \over 2}}( \cos \theta ) R_k,
 }
 where $\hat C_l^\nu(\cos\theta)={\Gamma(2\nu)l! \over \Gamma(\ell+2\nu)}C_l^\nu(\cos\theta)$ is a normalized Gegenbauer polynomial. (In four dimensions, it is a Legendre polynomial.) 
 Here, $R_k$ is the same as the left-hand side when $t=0$, which is the residue that appeared in \contint . Now, an important point is that for real $\theta $, $|\hat C^{\nu}_l(\cos \theta) |\leq 1$ and for $\theta =0$ (or $t=0$), it is equal to one. 
This is because
 \eqn\cospco{ 
 \hat C^{d-3\over 2}_l(\cos \theta) = ( \hat k_1)^l   .  ( \hat k_3 )^l   ,~~~~~~~ \sin^2{ \theta \over 2} = { - t \over s } ,
 }
 where $\hat k_1$ and $\hat k_3$ are unit vectors in the direction of the center of mass frame along the momenta of the particles 1 and 3. 
 $(\hat k_1)^l$ is a symmetrized,  traceless, and unit-normalized  combination of $l$ powers of $\hat k_1$. This is maximal when $\hat k_3$ is in the direction of 
 $\hat k_1$. 
 To ensure $\theta$ is real, we may demand $- M_{min}^2 \leq  t \leq 0$, since then $s,t$ take values possible in a physical scattering process for each pole. 
 Therefore, in the above expressions, we can  bound $ { A_{12, M} A_{M,34} }$ by their  value at $t=0$, allowing us to apply our previous argument.
 
Thus we find that 
 \eqn\fesgc{ 
  |c_m(t)| \leq { c_{ 2 + 2 \epsilon}  \over ( M_{min}^2)^{ m-2 - 2 \epsilon} }  ,~~~~~~~ - M_{min}^2 \leq  t\leq 0 ,~~~~~~~m> 2.
  }

It seems clear that with extra assumptions we might be able to do better. Another possible assumption is to demand that
for $t< 0 $, the power for large $s$ is strictly less than two in \higheng . It might be possible also that we can limit more strongly the corrections
to the gravitational effective action rather than generic corrections to scalar fields. 
Assuming ${\cal N}=8$ supersymmetry, we can do better, but not as well as we expected. 

\appendix{B}{Limits of conformal blocks and MFT OPE coefficients}

Conformal blocks in $d$-dimensions are eigenfunctions of the Casimir operator for the conformal group $SO(d,2)$ \DolanHV. Solving this equation in the large $\Delta$ limit gives \KosTGA
\eqn\largedimensionblock{
g_{\Delta,\ell}(r,\phi) = {\ell! \over (d-2)_\ell}{r^\Delta C^{d/2-1}_\ell(\cos\phi) \over (1-r^2)^{d/2-1} \sqrt{(1+r^2)^2 - 4r^2 \cos^2 \phi}} \qquad (\Delta \gg 1),
}
where $\rho=re^{i\phi}$ is defined in {\rhovariabledefinition} and  
 $C^{d/2-1}_\ell(\cos\phi)$ is a Gegenbauer polynomial.  The above expression is valid in the limit $\Delta \gg 1$ with $r$ fixed.  Here, we have normalized the block so that the leading term of $g_{\Delta,\ell}(r,\phi=0)$ at small $r$ is $r^{\Delta}$.

We also need the mixed limit $\Delta \to \oo$, $r=e^{-\epsilon}\to 1$ with the product $t\equiv \Delta \epsilon$ fixed.  Let us define $g_{\Delta, \ell}(e^{-t/\Delta},\phi)\equiv f_{\Delta,\ell}(t,\phi)$.  Taking the leading terms in the Casimir equation for $f_{\Delta,\ell}$ in the large $\Delta$ limit, we find
\eqn\mixedcasimireq{
\left(t {\partial^2 \over \partial t^2} + (d-2) {\partial \over \partial t} - t\right) f_{\Delta,\ell}(t,\phi) = 0.
}
This has solution
\eqn\mixedcasimirsoln{
f_{\Delta,\ell}(t,\phi) = t^{3-d\over 2} K_{d-3 \over 2}(t) j_{\Delta,\ell}(\phi),
}
where $K_{d-3\over 2}(t)$ is a Bessel function of the first kind.  The function $j_{\Delta,\ell}(\phi)$ can be fixed by demanding that the limit $t\to \oo$ of $f_{\Delta,\ell}(t,\phi)$ correctly reproduces the $r \to 1$ limit of {\largedimensionblock}.  This gives
\eqn\mixedlimitsolution{
g_{\Delta,\ell}(e^{-\epsilon},\phi) = 
{2^{1-d\over 2}\ell! \over \sqrt \pi (d-2)_\ell} {C_\ell^{d/2-1}(\cos \phi) \over |\sin\phi|} \sqrt \Delta\, \epsilon^{3-d\over 2} K_{d-3\over 2}(\Delta \epsilon) \qquad (\Delta \gg 1, \Delta\epsilon\ {\rm fixed}).
}

Several different normalizations of the conformal blocks are present in the literature.  In our normalization, the Mean Field Theory OPE coefficients for double-trace operators $\cO_{n,\ell}\equiv \cO \partial^{2n}\partial^{\mu_1}\cdots\partial^{\mu_\ell}\cO$ are given by \FitzpatrickDM
\eqn\mftOPEcoefficients{\eqalign{
\bar p_{n,\ell} &\equiv f_{\cO\cO\cO_{n,\ell}}^2|_{MFT}\cr
&=  {
4^{2\Delta+2n+\ell}(1+(-1)^\ell) (2h-2)_\ell (\Delta)_{\ell+n}^2(\Delta+1-h)_n^2
\over
\ell! n! (h-1)_\ell (h+\ell)_n (2\Delta+1-2h+n)_n(2\Delta-h+\ell+n)_n(2\Delta-1+\ell+2n)_\ell
},
}}
where $h=d/2$.

\appendix{C}{A Landau diagram on $R \times S^3$}

Here we present an example of a  set of points in four dimensions, on $R \times S^3$, such that we can 
draw a Landau diagram on $ R \times S^3$.  At $\tau =   \pi/2$ (the final time) we have two 
points on the north and south pole of $S^3$. Then at $\tau = - \pi/2$ (the initial time) we consider four points 
that are on the equatorial $S^2$ inside $S^3$. 
Two of the points are on opposite sides of a circle at $\theta_0$ and two are  on opposite sides of the circle at $\pi -\theta_0$. 
More explicitly, we have the following points on $S^2$:
\eqn\poitns{ 
 {\rm top}_\pm = ( \pm \sin \theta_0 , 0, \cos \theta_0 ) ,~~~~~~~~ {\rm bottom}_\pm= ( \pm \sin \theta_0 \cos \phi , \pm \sin \theta_0 \sin \phi , - \cos \theta_0 ) .
 }
 We can now have two lines that start from top two points and meet at the north pole of $S^2$ 
 at time $\theta_0$. They  send lines along the great circle that contains
 the bottom two points. These lines  travel for a time $\pi/2 -\theta_0$. After this time, they   meet a line coming from one of the bottom points that 
 is coming along the same great circle and will collide with it. This will happen after a total time of $\pi/2$. So at this time, they  produce 
 the lines going to  the
 north and south poles of the $S^3$. The existence of this diagram suggests that there is a qualitatively new entry for the symbol of the three-loop contribution to the
 six-point function. 
 
 This is a special configuration, but it shows that the strategy we used to prove that there are no boundary Landau diagrams in $d=2,3$ does not work here. 
  It would be nice to find out whether a completely generic configuration of six points with $\det X =0$ can or cannot
 have a Landau diagram purely on the boundary.

\appendix{D}{Transformation to the pillow metric}

Consider a four-point function on the plane with coordinate $x$,
\eqn\flatspacecorrelator{
\<\cO_1(x=0)\cO_2(x=z)\cO_3(x=1)\cO_4(x=\oo)\>,
}
where the operators $\cO_i$ have conformal weights $\delta_i,\bar\delta_i$.  The pillow metric is given by $du\,d\bar u$, where $u$ satisfies
\eqn\uniformcoordinate{\eqalign{
du &= {1 \over \theta_3(q)^2} {dx \over y},\cr
y^2 &= x(z-x)(1-x).
}}
Under the Weyl transformation
\eqn\weyltransformtopillow{
dx\,d\bar x \to e^{2\omega} dx\,d\bar x = \left|{du \over dx}\right|^2 dx\,d\bar x,
}
the correlator {\flatspacecorrelator} gets contributions from local rescaling near each operator insertion and also the Weyl anomaly.  Because of the singular nature of the map $x\mapsto u$ near the operator insertions, both of these contributions must be evaluated with some care.  We will address each one in turn.  However, let us first make some preliminary remarks about the uniformizing coordinate $u$.

We choose branch cuts for $y$ to run along $(0,z)$ and $(1,\oo)$.  With these cuts, the plane maps to half of the torus $u\in[0,2\pi]+\tau[0,\pi]$, with the operators mapping as follows:
\eqn\newoperatorpositions{\eqalign{
\cO_1(x=0) &\to \cO_1(u=0), \cr
\cO_2(x=z) &\to \cO_2(u=\pi), \cr
\cO_3(x=1) &\to \cO_3(u=\pi+\pi\tau), \cr
\cO_4(x=\oo) &\to \cO_4(u=\pi \tau).
}}
(We have not yet kept track of the rescaling of the operators due to the change of local coordinate.)

The segment $u\in (0,\pi)$ on the pillow corresponds to moving along the top of the branch cut between $x=0$ and $x=z$.  Meanwhile, the segment $u\in(\pi,2\pi)$ corresponds to moving back below the same branch cut on the $x$-plane.  Since the theory on the sphere has no actual cut, these paths on the pillow should be identified
\eqn\pillowidentification{
u=t \quad\sim\quad u=2\pi - t, ~~~ t\in(0,\pi).
}
Similarly for the other cut,
\eqn\pillowidentificationtwo{
u=\pi\tau+t \quad\sim\quad u=\pi\tau+2\pi - t, ~~~ t\in(0,\pi).
}

Notice that locally near each operator insertion, the map $x\mapsto u$ looks like a square-root and a rescaling, with the branch cut re-identified to create a conical defect.  In the following subsections, we will examine more closely the behavior of operators and partition functions under these sorts of maps.  For simplicity, we will sometimes assume the operators $\cO_i$ are purely holomorphic ($\bar\delta_i=0$), restoring non-holomorphic dependence at the end.

\subsec{Rescaling of local operators at branch points}

Consider the behavior of $\cO(x=0)$ under a square-root map
\eqn\squarerootmap{
x ~~~\mapsto ~~~ \xi = 2a \sqrt x.
}
We take the branch cut along the positive real $x$-axis, so the positive and negative real $\xi$-axes should be identified to create a conical defect.  Since our map is singular at $x=0$, we should define $\cO(\xi=0)$ in terms of a limit of operators at nonsingular points.  We have
\eqn\operatoratbranchpoint{\eqalign{
\cO(x=0) &= \lim_{\epsilon\to 0} \cO(x=\epsilon) \cr
&= \lim_{\epsilon\to 0} \left({a \over \sqrt \epsilon}\right)^\delta \cO(\xi=2a \sqrt \epsilon) \cr
&= a^{2\delta} \left[\lim_{\sigma\to 0} \left({2\over \sigma}\right)^\delta \cO(\xi=\sigma)\right],
}}
where we have redefined $2a\sqrt\epsilon \equiv \sigma$.

This suggests that the quantity in brackets,
\eqn\regularizedoperator{
\cO^{(*)}(\xi=0) \equiv \lim_{\sigma\to 0} \left({2 \over \sigma}\right)^\delta \cO(\xi=\sigma),
}
is the correct definition of a regularized operator at a conical defect in the $\xi$ coordinate.
Our calculation above now reads
\eqn\regularizedoperatorsummary{
\cO(x=0) = a^{2\delta} \cO^{(*)}(\xi=0).
}

By writing the map $x\mapsto u$ locally in the form {\squarerootmap} near each operator insertion, we can now use {\regularizedoperatorsummary} to relate operators at branch points in the $x$-plane to regularized operators at conical defects on the pillow,
\eqn\pillowoperators{\eqalign{
\cO_1(x=0) &= \theta_3(q)^{-4\delta_1}z^{-\delta_1}\cO_1^{(*)}(u=0),\cr
\cO_2(x=z) &= \theta_3(q)^{-4\delta_2}(z(1-z))^{-\delta_2}\cO_2^{(*)}(u=\pi),\cr
\cO_3(x=1) &= \theta_3(q)^{-4\delta_3}(1-z)^{-\delta_3}\cO_3^{(*)}(u=\pi+\pi\tau),\cr
\cO_4(x=\oo) &= \theta_3(q)^{-4\delta_4}\cO_4^{(*)}(u=\pi\tau).
}}
(As usual, the operator at infinity $\cO_4(x=\oo)$ is defined by $\cO_4(w=0)$, where $w=1/x$ is a local coordinate near $\oo$.)

\subsec{The Weyl anomaly}

The Weyl anomaly for a rescaling $\delta\to e^{2\omega}\delta$ is given by
\eqn\starweylanom{\eqalign{
\cal A &\equiv \log Z[e^{2 \omega} \delta] - \log Z[\delta] 
=  {c \over 24 \pi} \int d^2\sigma\, \delta^{a b}\pa_{a} \omega \pa_{b} \omega,
}}
where $x = \sigma^1 + i \sigma^2$, $\bar x = \sigma^1 - i \sigma^2$, and $d^2\sigma=d\sigma^1d\sigma^2$.

As a warmup, let us compute ${\cal A}$ for the square-root mapping {\squarerootmap}.  The coordinate $\xi$ defines a metric
\eqn\sqrmap{\eqalign{
d\xi d\bar\xi &= e^{2\omega} dx d\bar x, ~~~
\omega ={1 \over 2} \log {|a|^2 \over |x|}.
}}
Plugging $\omega$ into {\starweylanom}, we find a logarithmic divergence at $x=0$.  This phenomenon is familiar from the plane-to-cylinder map (where $\omega$ differs from {\sqrmap} by a factor of $2$).  There, the anomaly contributes to a divergence in the partition function on the infinite cylinder.  This divergence has a simple physical interpretation: it comes from the Casimir energy of the theory on the circle, integrated along the infinite length of the cylinder.  A simple way to regulate this infinity is to instead consider a Weyl transformation to the finite-length cylinder.  Equivalently, we can modify $\omega$ inside small circles around $x=0$ and $x=\oo$ so that it is everywhere nonsingular, see, e.g., \refs{\SonodaI,\SonodaII}.

We can adopt the same procedure here.  Let us modify $\omega$ to be
\eqn\regularized{\eqalign{
\omega &= \cases{
  {1 \over 2} \log {|a|^2 \over |x|}      &\quad if $|\xi|>\epsilon$, \cr
  {1 \over 2} \log {|a|^2 \over |\epsilon/2a|^2} &\quad if $|\xi|\leq \epsilon$. \cr
 }
}}
The anomaly contribution close to $x=0$ is now
\eqn\closetobranch{
{\cal A} \sim 
 {c \over 24 \pi} {2 \pi \over 4} \int_{({\eps \over 2 a})^2} {d r\over r} = - {c \over 24} \log \eps + {\rm finite}.
}
The infinite piece $-{c\over 24}\log\eps$ can be subtracted off to define a regularized partition function in the $d\xi d\bar\xi$ metric.

Let us now return to the pillow Weyl transformation,
\eqn\weylforus{\eqalign{
\omega &= -\log |\theta_3 (q)|^2 - {1 \over 2} \log |y|^2 \cr
 &= - \log |\theta_3 (q)|^2 - {1 \over 2} \log |x (z - x) (1 - x)|.
}}
As before, we regulate ${\cal A}$ by modifying the pillow metric inside small circles of radius $\epsilon$ around the points $u=0,\pi,\pi\tau,\pi+\pi\tau$.  It is important that we modify the metric in the same way near each of the conical defects.  Suppose instead we were to choose circles of different radii $\epsilon_i$ around the points $u_i$.  Then the regularized pillow would no longer be invariant under reflection in the ${\rm Im}(u)$ direction when ${\rm Re}(\tau)=0$.  Consequently, the pillow four-point function may no longer be reflection-positive.  For this reason, we should choose each circle to have the same radius $\epsilon$ in the $u$-coordinate.  These circles then map to different size circles in the $x$-coordinate with radii
\eqn\ircutoffs{\eqalign{
r_1 &= \left( { \theta_3 (q)^2 \sqrt{z} \over 2} \right)^2 \eps^2, \cr
r_2 &=  \left( { \theta_3 (q)^2 \sqrt{z (1-z)} \over 2} \right)^2 \eps^2, \cr
r_3 &=  \left( { \theta_3 (q)^2 \sqrt{1-z} \over 2} \right)^2 \eps^2, \cr
r_4 &=  \left( { \theta_3 (q)^2  \over 2} \right)^{-2} \eps^{-2}.
}}

The divergent part of the anomaly near $x=0,z,1$ is again given by {\closetobranch}.  A similar computation near the point at infinity gives the divergent piece $- 9 {c \over 24} \log \eps$.  Thus, we can define the regularized anomaly contribution
\eqn\reganom{
{\cal A}^{*} =\lim_{\eps \to 0}  \left( \log Z[e^{2 \omega} \delta] - \log Z[\delta] + {c \over 2} \log \eps \right).
}

We are finally ready to compute ${\cal A}^*$ for the transformation to the pillow metric.  We have
\eqn\anomcontr{\eqalign{
{\cal A} &= 
{c \over 24\pi} {1 \over 4} \int_R d^2 \sigma\, \partial^a \log |y|^2\, \partial_a \log|y|^2
\cr
&= {c \over 24\pi} {1 \over 4} \oint_{\partial R} \log|y|^2 \partial_a \log|y|^2 dn^a,
}}
where we have used the divergence theorem together with the fact that $\log|y|^2$ is harmonic.  Locally near the points $1,2,$ and $3$, $\log |y|^2$ has the form $\log |x-x_i|+b_i(x)$, where $b_i(x)$ is slowly varying.  Hence,
\eqn\nearcircle{
\int_{C_i} \log|y|^2 \partial_a \log|y|^2 dn^a \approx \log|x_i+r_i|^2 \int_{C_i} {1 \over r_i} (-r_i d\theta_i) = -2\pi \log|x_i+r_i|^2,
}
where we have discarded terms that vanish as $r_i\to 0$.  The circle at infinity contributes similarly, with an additional factor of $-3$.  Hence, the anomaly is
\eqn\fullanomaly{\eqalign{
{\cal A} &= {c \over 24 \pi} {1 \over 4} \left(-2\pi\right)\left(\sum_{i=1}^3 \log|y(x_i+r_i)|^2 - 3 \log|y(r_4)|^2 \right)\cr
&=- {c \over 48 } \left( 24 \log \eps + 48 \log |\theta_3 (q)| + 4 \log |z (1-z)| - 24 \log 2 \right).
}}
The regularized anomaly is
\eqn\fullregularizedanomaly{
{\cal A}^* = -c \log |\theta_3(q)|-{c \over 12} \log|z(1-z)|,
}
where we have absorbed constant pieces into a redefinition of $\epsilon$.

\subsec{Putting everything together}

Combining the results of the previous subsections, and writing only the holomorphic half of the transformation law for brevity, we have
\eqn\regularizedpillow{\eqalign{
\<\cO_1(x=0)\cO_2(x=z)&\cO_3(x=1)\cO_4(x=\oo)\>_{R^2}\cr
&=
\theta_3(q)^{{c \over 2}-4(\delta_1+\delta_2+\delta_3+\delta_4)} z^{{c \over 24}-\delta_1-\delta_2} (1-z)^{{c \over 24}-\delta_2-\delta_3}\cr
&\phantom{=}\times \<\cO_1^{(*)}(u=0)\cO_2^{(*)}(u=\pi)\cO_3^{(*)}(u=\pi+\pi\tau)\cO_4^{(*)}(u=\pi\tau)\>^{(*)}_{\rm pillow}.
}}
Here, the regularized correlator on the pillow is defined by combining the divergent part of the Weyl anomaly with the (divergent) partition function in the pillow metric to get a finite quantity,
\eqn\reabsorbdivergence{
\<\dots\>_{\rm pillow}^{(*)} \equiv e^{{c \over 2}\log \epsilon} \<\dots\>_{\rm pillow}.
}
We refer to the regularized four-point function in the main text as $g(q,\bar q)$.  Note that our regularization procedure does not spoil reflection positivity in the case that $\cO_3=\cO_2^\dagger$, $\cO_4=\cO_1^\dagger$ and ${\rm Re}(\tau)=0$ since we simply rescale the reflection-positive pillow correlator by a positive constant.

  \listrefs 
  
  \bye